\documentclass[prx,reprint,superscriptaddress,nofootinbib,longbibliography,floatfix]{revtex4-2}
\usepackage{amsmath,amssymb,graphicx,booktabs,braket,hyperref}
\usepackage[ruled,linesnumbered]{algorithm2e}
\hypersetup{hidelinks,pdftitle={Conserved and relaxing modes enable polynomial-success coherent Carleman lattice Boltzmann evolution},pdfauthor={Hua-Dong Yao, Sauro Succi}}
\graphicspath{{figures/}}
\makeatletter
\newcommand{\raggedcells}{\let\raggedcells@apr\@arrayparboxrestore\def\@arrayparboxrestore{\raggedcells@apr\raggedright\let\\\tabularnewline}}
\makeatother
\newcommand{\Ma}{\mathrm{Ma}}
\newcommand{\Lc}{\mathcal{L}}
\newcommand{\Qc}{\mathcal{Q}}
\newcommand{\Kc}{\mathcal{K}}
\newcommand{\Cc}{\mathcal{C}}
\newcommand{\Tr}{\mathrm{T}}
\newtheorem{theorem}{Theorem}
\newtheorem{corollary}{Corollary}
\begin{document}
\title{Conserved and relaxing modes enable polynomial-success coherent Carleman lattice Boltzmann evolution}
\author{Hua-Dong Yao}\email{huadong.yao@chalmers.se}
\affiliation{Chalmers University of Technology, Gothenburg, Sweden}
\author{Sauro Succi}
\affiliation{Fondazione Istituto Italiano di Tecnologia, Center for Life Nano-Neuroscience at La Sapienza, Rome, Italy}
\affiliation{Department of Physics, Harvard University, Cambridge, Massachusetts 02138, USA}
\date{\today}

\begin{abstract}
Carleman linearization represents nonlinear lattice Boltzmann collision dynamics in a form amenable to quantum block encoding, but low postselection probabilities in existing population-space encodings make coherent multi-step evolution exponentially unlikely to succeed. We present a quantum Carleman lattice Boltzmann scheme based on the multi-relaxation-time collision operator in a weighted Hermite moment basis. In this basis, the lifted collision operator separates into small coupled blocks whose norm exceeds unity because of the coupling between conserved and relaxing modes. This block structure enables a norm-attaining block encoding with a single-step success probability close to unity. Crucially, scaling the pair sector for a prescribed evolution horizon makes the combined encoding and sector-selection overhead grow polynomially rather than exponentially with the number of steps. For a hundred-step evolution, the resulting complete-run success probability exceeds that of population-space encodings by orders of magnitude. Over horizons of several tens of steps, the truncated dynamics reproduces the nonlinearly generated modes of a benchmark flow to within a few percent. We verify the combined preparation, evolution and readout protocol as a single unitary circuit and give a complete gate-level resource estimate for extracting a nonlinear observable of the benchmark flow from a coherent Carleman evolution. The block structure of conserved and relaxing modes holds equally on three-dimensional lattices and provides a constructive route to norm-attaining encodings for lattice Boltzmann models with quadratic equilibria.
\end{abstract}
\maketitle
\section{Introduction}\label{sec:intro}
Quantum computers hold the prospect of information processing capabilities with no known efficient classical counterpart \cite{NielsenChuang2010,Preskill2018}. A pure state of $n$ qubits is described by $2^n$ complex amplitudes, so that a quantum circuit acts on a state space of exponential dimension. Coherence and entanglement, the resources on which quantum computation rests, are fragile against environmental perturbations, so that present devices run circuits of limited depth before errors accumulate \cite{Preskill2018}, and the tendency of quantum systems to behave classically at large scales limits the coherent volume that can be exploited \cite{SucciSanavioLove2026}. Fault-tolerant architectures suppress logical errors at a substantial overhead in physical qubits and in the cost of non-Clifford gates \cite{Fowler2012}. These constraints determine the quantities by which a quantum algorithm is assessed, namely the number of logical qubits, the number of non-Clifford gates and the depth of the circuit that has to be run coherently. An amplitude-encoded state holds a field in the amplitudes of the state vector, and a non-unitary linear map of such a state is implemented as a block of a larger unitary and recovered by postselection, which succeeds with a probability below one \cite{GSLW2019}. A failed postselection discards the run, so that a low success probability multiplies the number of repetitions and exposes the full depth of the circuit to noise in each of them.

The difficulty is sharper for classical fluid dynamics. The coherent evolution of the quantum circuit model is linear and unitary, whereas the equations of fluid dynamics are nonlinear and dissipative \cite{SucciSanavioLove2026}. Carleman linearization maps a polynomial system onto an infinite linear hierarchy of tensor powers and truncates it, and \citet{Liu2021} showed that the truncation converges when dissipation dominates nonlinearity. \citet{Gaitan2020} applied a quantum algorithm for ordinary differential equations to the flow through a nozzle, and \citet{Bharadwaj2023} developed a hybrid quantum linear-systems workflow for Poiseuille and Couette flows. The lattice Boltzmann (LB) equation is a natural starting point for this route because its two substeps separate the difficulties. Streaming is a permutation of the populations and hence unitary, while the nonlinearity and the dissipation are confined to the local collision \cite{SucciEPL2023}. \citet{Mezzacapo2015} proposed a quantum simulator of the LB equation with a heralded collision, \citet{Todorova2020} constructed circuits for the collisionless streaming step, \citet{Budinski2021} built a full circuit for the advection-diffusion equation, and \citet{Wang2025} obtained a linear representation of the collision from a node-level ensemble description of a lattice gas.

Applied to the LB equation, the Carleman lift replaces the populations $f$ by the tensor powers $(f, f\otimes f, \dots)$ and truncates the hierarchy \cite{ItaniSucci2022}. The Carleman lattice Boltzmann (CLB) method truncated at the second tensor power reproduces the LB dynamics of a two-dimensional Kolmogorov flow at Reynolds numbers between 10 and 100 \cite{SanavioSucci2024}. Its quantum implementation is limited by the block encoding of the lifted collision. With sparse-access oracles the D2Q9 relaxation matrix has sparsity 81 and succeeds with probability $6\times10^{-5}$ per step \cite{SanavioSimonSucci2025}. \citet{BastidaZamora2026} keep the Carleman variables in population space, store the pair sector in coordinates relative to the first site, and run several steps at a per-step success probability of order $10^{-2}$, with a total probability that decays as $p^T$. \citet{Jennings2026} give an end-to-end analysis of the Carleman route through a history-state linear system, identify the convergence of the linearization, the time stepping, the condition number and the data extraction as the bottlenecks, bound the possible speedup over classical evolution by $\mathrm{Re}^{3D/8}$ in $D$ dimensions, and find that a modest quantum advantage may be preserved for selected observables in the high-error-tolerance regime. \citet{Demirdjian2026} address the loading of the Carleman operator with a linear combination of non-unitaries whose number of terms scales as the square of the Carleman order times the square of the number of velocities. \citet{KhanSucciYao2026} take a complementary route and realize the diagonal multi-relaxation-time (MRT) relaxation of the moments deterministically as a completely positive trace-preserving map on an expectation-level encoding, leaving the nonlinear equilibrium to a Carleman lift in a hybrid pipeline. The three Carleman routes of \citet{SanavioThreeRoutes2024} place these constructions in context. Classical MRT schemes collide in moment space and stream in population space \cite{LallemandLuo2000}, and applying this factorization to the Carleman-lifted operator on the amplitude-encoded state is the starting point of the present work, introduced in a preliminary version \cite{YaoSucci2026}.

Despite the body of work reviewed above, the total success probability of a coherent multi-step run decays exponentially with the number of steps for the population-space encodings, and this postselection overhead remains the main obstacle to the route. The cost of such a run has moreover been assessed through the success probability of one step, and the two are not the same quantity. A scaling of the pair sector raises the step probability toward one while it reduces the weight of the output sector, so that the relevant quantity is the cost of a specified observable at a specified accuracy over a specified horizon. Which property of the collision sets this cost, whether the truncated dynamics is accurate over the horizon on which the cost is favorable, and how the resulting resources compare with matched quantum constructions and with efficient classical evolutions of the same problem, are open questions. A block encoding can at best attain the operator norm of the stage it encodes, and the singular-value construction of \citet{BastidaZamora2026} does so for the plain population-space stage, whose norm exceeds 9 at $\omega = 1.5$. It has not been shown which representation brings that norm close to one, what structure of the collision fixes its value, and how the resulting encoding scales with the number of steps. Nor has a complete resource estimate of one nonlinear observable extracted from a coherent Carleman lattice Boltzmann run been given.

The present study addresses these questions for the D2Q9 lattice with MRT collision in the weighted Hermite moment basis. The central claim is that the conservation and relaxation structure of the collision, exposed in the weighted Hermite representation, fixes the operator norm of the scaled level-2 collision stage at $1 + c/\lambda^2 + O(\lambda^{-4})$ and admits a block encoding that attains it with three $2\times2$ blocks, and that this encoding, combined with sector scaling and coherent output recovery, controls the postselection overhead of a multi-step run. Nonlinear-observable validation and resource accounting then establish the benefit and its present limits. The contribution is threefold. First, a structure theorem for the scaled collision stage (Sec.~\ref{sec:structure}) locates its norm in the coupling from the conserved momenta into the relaxing second-order moments, with an explicit block encoding that attains this norm in the weighted Hermite representation, verified by statevector simulation together with its $1/\lambda^2$ scaling, and with the same structure on the D3Q19 and D3Q27 lattices. Second, the telescoped complete-run probability of a coherent run with output recovery (Sec.~\ref{sec:coherent}) exposes the trade-off between the per-step normalization and the final sector selection, and optimizing the pair-sector scale for each specified horizon gives polynomial $1/T^2$ and $1/T$ success laws in the number of steps $T$ for the two structured encodings, with the coupled block gaining a further factor linear in $T$. The same section gives the rest-shifted encoding, an explicit preparation circuit for few-mode initial data with exact amplification, and a validated protocol with deferred measurement, amplification and overlap readout. Third, two nonlinearly generated observables of a nondegenerate two-mode flow and their accuracy window are established (Sec.~\ref{sec:accuracy}), and a complete instance-specific resource estimate of their extraction, from state preparation to readout, is given against matched quantum constructions and efficient classical evolutions of the same problem (Sec.~\ref{sec:resources}). Section~\ref{sec:problem} defines the computational task and the kinetic structure, Sec.~\ref{sec:scope} states the generality and the limitations of the results and concludes, and the appendices give the closure analysis that fixes the state, the representations of the collision stage, the proof of the theorem, the quantum construction, the validation and error analysis, and the resource derivations.

\section{Computational task and kinetic structure}\label{sec:problem}
This section defines the lattice Boltzmann model, the weighted moment representation, the level-2 lift and the observable-extraction task that the later sections validate and cost. The D2Q9 lattice is used in lattice units, with velocities $c_i$ ($i = 0,\dots,8$), weights $w_i$ and sound speed $c_s^2 = 1/3$, on a periodic box of $N = L^2$ sites. The populations $f_i(x,t)$ define the density $\rho = \sum_i f_i$ and the momentum $J = \sum_i c_if_i$, and collision relaxes them toward the equilibrium
\begin{equation}
f_i^{\rm eq} = w_i\left[\rho + 3\,c_i\!\cdot\!J + \tfrac92(c_i\!\cdot\!J)^2 - \tfrac32J\!\cdot\!J\right],
\label{eq:feq}
\end{equation}
in which the density has been set to one inside the quadratic term, the weakly compressible form of \citet{SanavioSucci2024} and \citet{BastidaZamora2026}, which makes the equilibrium exactly quadratic in the populations, $f^{\rm eq} = E_1f + E_2(f\otimes f)$. Collision is performed in moment space. With $M$ a $9\times9$ matrix whose rows define nine moments $m = Mf$, the first three being $\rho$, $J_x$ and $J_y$ up to normalization, the MRT collision relaxes each moment at its own rate,
\begin{equation}
m_k' = (1-\omega_k)\,m_k + \omega_k\,m_k^{\rm eq}, \qquad \omega_k\in(0,2),
\label{eq:mrt}
\end{equation}
and leaves the conserved moments unchanged. The shear viscosity is $\nu = c_s^2(1/\omega_\nu - 1/2)$, with $\omega_\nu$ the rate of the two shear moments, $\omega_e$ that of the trace and $\omega_q$, $\omega_\varepsilon$ those of the three ghost moments. Equal rates give the Bhatnagar-Gross-Krook (BGK) model. Written back in population space, the per-site update and the lattice-wide step read
\begin{equation}
f' = Lf + Q(f\otimes f), \qquad F(t+1) = \Lc F + \Qc(F\otimes F),
\label{eq:step}
\end{equation}
with $L = M^{-1}(I-\Omega)M + M^{-1}\Omega ME_1$, $Q = M^{-1}\Omega ME_2$, $\Omega = \mathrm{diag}(\omega_k)$, $F\in\mathbb{R}^{9N}$ the population vector, $\Lc = S(I_N\otimes L)$, $S$ the streaming permutation $f_i(x + c_i, t+1) = f_i'(x,t)$, and $\Qc = S(I_N\otimes Q)$ acting on same-site pairs only.

Gram-Schmidt orthonormalization of the monomials $1$, $c_x$, $c_y$, $c_x^2+c_y^2$, $c_x^2-c_y^2$, $c_xc_y$, $c_xc_y^2$, $c_x^2c_y$ and $c_x^2c_y^2$ in the weighted inner product $\langle a,b\rangle_w = \sum_iw_ia_ib_i$ gives the tensor Hermite basis of D2Q9 up to a rotation of the trace and deviator, $M = H$ with $HWH^\Tr = I$ and $W = \mathrm{diag}(w)$. In this basis the linear part of the equilibrium map is $HE_1H^{-1} = \mathrm{diag}(1,1,1,0,\dots,0)$, so that the level-1 collision block is the diagonal
\begin{equation}
C_1 = \mathrm{diag}(1,1,1,1-\omega_3,\dots,1-\omega_8),
\label{eq:c1}
\end{equation}
and the quadratic part of the equilibrium enters only the three second-order moments $e$, $p_{xx}$ and $p_{xy}$. The map from the weighted populations $g_i = f_i/\sqrt{w_i}$ to the moments, $U = HW^{1/2}$, is orthogonal, and the rest state $g_i = \sqrt{w_i}$ has unit norm. The amplitude encoding stores $g$, and the norm of a state is the weighted norm $\|F\|_g = \|F_g\|$, $F_g = (I_N\otimes W^{-1/2})F$, which equals the Euclidean norm of the moments because $U$ is orthogonal.

Carleman linearization lifts $F$ to $(F, F\otimes F, \dots)$ and truncates the hierarchy. At level two the state $(F, F\otimes F)$ has dimension $9N + 81N^2$, and one step is the block upper triangular map
\begin{equation}
\begin{aligned}
\Cc &= \begin{pmatrix}\Lc & \Qc\\ 0 & \Lc\otimes\Lc\end{pmatrix},\\
\Cc_g &= \mathcal{W}^{-1/2}\Cc\,\mathcal{W}^{1/2} = \mathcal{S}\,\mathcal{T}^\Tr\Kc\,\mathcal{T},
\end{aligned}
\label{eq:lift}
\end{equation}
with $\mathcal{W} = (I_N\otimes W)\oplus(I_N\otimes W)^{\otimes2}$, $\mathcal{S} = S\oplus(S\otimes S)$ and $\mathcal{T} = (I_N\otimes U)\oplus(I_N\otimes U)^{\otimes2}$. In this factorization $\mathcal{S}$ is a permutation, $\mathcal{T}$ is orthogonal and $\Kc$ is the lifted collision in moment space, block upper triangular with $I_N\otimes C_1$ and $(I_N\otimes C_1)^{\otimes2}$ on the diagonal and the quadratic coupling above it. The arrangement, collision in moment space and streaming in population space on the lifted state, is the representation split \cite{YaoSucci2026}. The level-2 hierarchy does not close locally under streaming, which generates cross-site terms associated with the work of pressure, advection and stress on the momentum flux (Appendix~\ref{app:closure}), so the state remains the full pair array, and the moment lift is an orthogonal change of basis of the weighted population lift with identical dynamics and truncation error. Two properties of the lift are used throughout. Because $(\Lc\otimes\Lc)(G\otimes G) = (\Lc G)\otimes(\Lc G)$, the pair sector of a product initial state remains a product, and the level-2 dynamics reduces to
\begin{equation}
F(t+1) = \Lc F(t) + \Qc\big(G(t)\otimes G(t)\big), \qquad G(t) = \Lc^tF_0,
\label{eq:product}
\end{equation}
with $G$ the linear LB evolution of the initial state, at $O(N)$ storage and work per step. The pair sector may also be scaled, and along a trajectory the scaled state is
\begin{equation}
\psi_\lambda(t) = \big(F(t),\ \lambda\,G(t)\otimes G(t)\big),
\label{eq:state}
\end{equation}
in the weighted encoding, which divides $\Qc$ by $\lambda$ in Eq.~(\ref{eq:lift}) and leaves the dynamics unchanged. A block encoding of a matrix $A$ with subnormalization $\alpha \ge \|A\|_2$ applied to a state $\psi$ succeeds with probability
\begin{equation}
p = \frac{\|A\psi\|^2}{\alpha^2\|\psi\|^2} = \frac{r}{\alpha^2},
\label{eq:p}
\end{equation}
with $r$ the norm ratio of the map on the state and $r_\lambda$ its value on $\psi_\lambda$. Errors of a truncated evolution against the nonlinear LB reference are measured on the velocity,
\begin{equation}
e_u = \frac{\|u - u_{\rm ref}\|}{\|u_{\rm ref}\|},
\label{eq:eu}
\end{equation}
with the norm over the lattice, and on individual observables as defined below.

The test problem is a two-mode flow whose modes interact. The initial momentum field derives from the streamfunction $\psi_0 = U_0\kappa^{-1}[\sin\kappa x\sin\kappa y + A_2\sin(2\kappa x + \phi)\sin\kappa y]$, with $\kappa = 2\pi/L$, $A_2 = 0.6$ and $\phi = 0.3$, so that
\begin{equation}
\begin{aligned}
J_x &= U_0\left[\sin\kappa x\cos\kappa y + A_2\sin(2\kappa x + \phi)\cos\kappa y\right],\\
J_y &= -U_0\left[\cos\kappa x\sin\kappa y + 2A_2\cos(2\kappa x + \phi)\sin\kappa y\right],
\end{aligned}
\label{eq:twomode}
\end{equation}
with unit density. The Mach number is $\Ma = U_0/c_s$, the advective time $t_{\rm adv} = 1/(\kappa U_0)$ and the Reynolds number $\mathrm{Re} = U_0L/\nu$. The two modes have wavevectors $(1,\pm1)$ and $(2,\pm1)$, and the quadratic term generates the sums and differences of these wavevectors, which are absent from the initial momentum. Two of the generated modes are the observables. The Fourier component
\begin{equation}
\hat{J}_a(k) = \frac1N\sum_xJ_a(x)\,e^{-ik\cdot x}
\label{eq:jhat}
\end{equation}
at $k = \kappa(1,0)$ is purely longitudinal, its transverse part vanishing by the symmetry of the field, and is the acoustic mode that the interaction of the two vortical modes excites. It grows to its peak near $0.4t_{\rm adv}$ and oscillates at the acoustic frequency $c_s\kappa$ together with the density mode $\hat\rho(k)$. The transverse part $\hat{J}_\perp(k) = \hat k_\perp\!\cdot\!\hat J(k)$ at $k = \kappa(1,2)$ is the vortical mode that the same interaction generates. It grows smoothly to its peak near $0.76t_{\rm adv}$ without oscillation. The primary task is to estimate the magnitude $|\hat{J}_x(1,0)|$ at the time of its peak to relative precision $\epsilon_{\rm rel}$ with a confidence of 95\%, the horizon being $T = t_{\rm peak}$, 40 steps on the $32\times32$ lattice at $\Ma = 0.087$, and the same task for $|\hat{J}_\perp(1,2)|$ at its later peak, $T = 77$, is costed alongside; both horizons are specified inputs of the task. The initial populations of the task are the linearized equilibrium $f_i = w_i(1 + 3c_i\!\cdot\!J)$ of the field of Eq.~(\ref{eq:twomode}), the state that the preparation circuit of Sec.~\ref{sec:coherent} builds, and the LB reference of the task starts from the same populations; the observable validation of Sec.~\ref{sec:accuracy} and the resource accounting of Sec.~\ref{sec:resources} use this one initial state. The parameter study of the accuracy window uses the quadratic equilibrium of Eq.~(\ref{eq:feq}) instead, which differs from the linearized one by 8.3\% of the weighted flow norm at $t = 0$ and changes the peak of the acoustic observable by 0.64\%, as robustness evidence. On the amplitude-encoded state the observable is the overlap with the reference state
\begin{equation}
\begin{aligned}
\ket{\phi_a(k)} &= \frac{1}{\sqrt{Nc_s^2}}\sum_{x,i}e^{ik\cdot x}\,c_{ia}\sqrt{w_i}\,\ket{x,i},\\
\braket{\phi_a(k)|\Phi} &= \frac{\sqrt N\,\hat{J}_a(k)}{c_s\,\|F\|_g},
\end{aligned}
\label{eq:overlap}
\end{equation}
with $\ket{\Phi}$ the normalized level-1 sector, since $\sum_ic_{ia}\sqrt{w_i}\,g_i(x) = J_a(x)$ and $\sum_iw_ic_{ia}c_{ib} = c_s^2\delta_{ab}$, and the transverse component is the corresponding combination of the two reference states. In the coherent protocol of Sec.~\ref{sec:coherent} the observable is recovered from one measured amplitude and classical quantities computed from the initial data alone, without the norm of the evolved state. Two errors are distinguished throughout. The kinetic truncation error is the error of the level-2 dynamics against the nonlinear LB reference started from the same populations, measured as the relative error of the observable at its peak for the task and as the velocity error of Eq.~(\ref{eq:eu}) at fixed fractions of $t_{\rm adv}$ for the window study. The estimation precision $\epsilon_{\rm rel}$ is the statistical error of the quantum readout. The discretization error of the LB scheme itself against the Navier-Stokes solution, second order in the lattice spacing, is a diagnostic of the reference and enters neither (Appendix~\ref{app:validation}).

\section{A collision encoding determined by conserved and relaxing modes}\label{sec:structure}
This section shows that the norm of the scaled collision stage, and hence the success probability of its optimal block encoding, is fixed by the way the conserved momenta feed the relaxing second-order moments, separates this contribution from those of the metric and the basis, and gives the encoding that attains the norm. The level-1 block $C_1$ of Eq.~(\ref{eq:c1}) is diagonal with entries of modulus at most one, and the pair block $C_1\otimes C_1$ is diagonal with products of these entries. The only off-diagonal part of $\Kc$ is the coupling from same-site products of conserved momenta into the level-1 sector, per site the $3\times4$ block
\begin{equation}
C_{\rm site} = \frac{1}{2\lambda}\,\mathrm{diag}(\omega_e,\omega_\nu,\omega_\nu)\begin{pmatrix}1&1&0&0\\1&-1&0&0\\0&0&1&1\end{pmatrix}
\label{eq:csite}
\end{equation}
from the pair components $(m_1m_1, m_2m_2, m_1m_2, m_2m_1)$ to the trace, the deviator and the shear moment. Its singular values are $\omega_e/(\sqrt2\lambda)$ and $\omega_\nu/(\sqrt2\lambda)$ (twice), and a two-term linear combination of unitaries (LCU) of the diagonal and the coupling has the subnormalization
\begin{equation}
\alpha_{\rm L} = 1 + \frac{a}{\lambda}, \qquad a = \frac{\omega_{\max}}{\sqrt2}, \qquad \omega_{\max} = \max(\omega_e,\omega_\nu).
\label{eq:alcu}
\end{equation}
The operator norm of the stage is smaller, and the difference is the subject of the following statement.

\begin{theorem}[Block structure]\label{th:structure}
Let the per-site collision in a moment basis satisfy the following. (i) The conserved moments and the relaxing moments are orthogonal in a positive metric in which the level-1 collision is the diagonal $C_1$ with entries in $[-1,1]$, unity on the conserved moments and $|1 - \omega_s| < 1$ for every relaxing moment $s$ coupled by the quadratic term. (ii) The quadratic part of the equilibrium depends only on products of conserved moments and enters only relaxing moments, with $\rho = 1$ inside the quadratic term. (iii) The rows of the coupling matrix from the conserved-pair slots into the relaxing moments are mutually orthogonal, so that the coupling has the form $\Sigma V^\Tr$ with $V$ orthogonal on the pair slots and $\Sigma$ diagonal on the coupled moments. (iv) Streaming is an isometry of the metric. Then, with the pair sector scaled by $\lambda$ and stored in relative coordinates, the rotation $W_n = V^\Tr$ of the same-site conserved-pair slots is such that the collision stage is the direct sum of a diagonal with entries in $[-1,1]$ and one $2\times2$ block per coupled relaxing moment $s$,
\begin{equation}
B_s = \begin{pmatrix}1-\omega_s & \sigma_s/\lambda\\ 0 & 1\end{pmatrix},
\label{eq:block}
\end{equation}
where $\sigma_s/\lambda$ is the singular value of the coupling into $s$. The norm of the stage is $\|\Kc_\lambda\|_2 = \max_s\sigma_{\max}(B_s)$, with
\begin{equation}
\begin{aligned}
\sigma_{\max}(B_s) &= 1 + \frac{c_s}{\lambda^2} + \frac{c_{4,s}}{\lambda^4} + O(\lambda^{-6}),\\
c_s &= \frac{\sigma_s^2}{2\omega_s(2-\omega_s)} .
\end{aligned}
\label{eq:sigma}
\end{equation}
The expansion is for $\lambda\to\infty$ at fixed interior relaxation rates, $0 < \omega_s < 2$, and is not uniform as $\omega_s\to2^-$. The exact singular value of $B_s$ is used for finite-$\lambda$ costs.
\end{theorem}

\begin{figure*}[t]\centering\includegraphics[width=\textwidth]{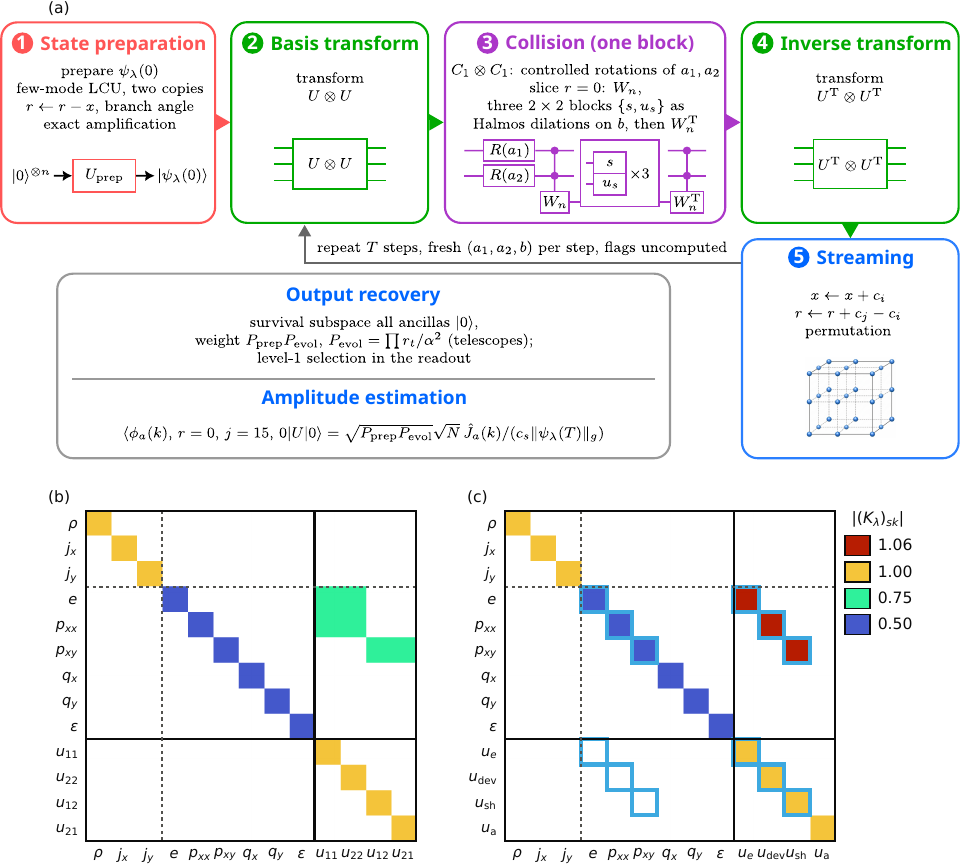}
\caption{The algorithm and the structure of its collision stage. (a) The five stages of the protocol, from the preparation of the scaled state through one step on the level-2 register in relative coordinates to the streaming, with the output recovery and the amplitude that is estimated. (b) One-site level-2 collision stage of the Hermite MRT scheme on the nine moments and the four same-site conserved-pair slots at $\lambda = 1$ and $\omega = 1.5$; in the product slots $(m_1m_1, m_2m_2, m_1m_2, m_2m_1)$ the coupling fills a $3\times4$ block. (c) After the rotation to $(u_e, u_{\rm dev}, u_{\rm sh}, u_{\rm a})$ the stage is a diagonal plus three $2\times2$ blocks on $\{e, u_e\}$, $\{p_{xx}, u_{\rm dev}\}$ and $\{p_{xy}, u_{\rm sh}\}$ (frames), and the antisymmetric slot decouples. The dotted lines separate the conserved from the relaxing moments.}\label{fig:structure}\end{figure*}

For D2Q9 the natural rotation is $u_e = (m_1m_1 + m_2m_2)/\sqrt2$, $u_{\rm dev} = (m_1m_1 - m_2m_2)/\sqrt2$, $u_{\rm sh} = (m_1m_2 + m_2m_1)/\sqrt2$ and $u_{\rm a} = (m_1m_2 - m_2m_1)/\sqrt2$, the antisymmetric slot $u_{\rm a}$ decouples, and $\sigma_s = \omega_s/\sqrt2$ for $s = e$, $p_{xx}$, $p_{xy}$, so that
\begin{equation}
c_s = \frac{\omega_s}{4(2-\omega_s)}, \qquad c_{4,s} = -\frac{\omega_s(3\omega_s^2 - 6\omega_s + 4)}{32(2-\omega_s)^3},
\label{eq:c}
\end{equation}
by the series expansion of the largest singular value of Eq.~(\ref{eq:block}) in $1/\lambda$ (Appendix~\ref{app:proof}). Figures~\ref{fig:structure}(b) and (c) show the one-site stage before and after the rotation. The proof rests on two observations, that the conserved-pair slots are left unchanged by the collision, and that under assumption (iii) the singular value decomposition of the coupling rotates only its input space, so that the coupled part of the stage on the pair $\{s, u_s\}$ is exactly Eq.~(\ref{eq:block}) and every other entry is diagonal. Without (iii) a general coupling would also rotate the output space, which mixes relaxing moments with different rates, and the stage would not split into independent blocks. Assumption (iii) holds for D2Q9 because the three rows of Eq.~(\ref{eq:csite}) are orthogonal, and it is the property that makes the block structure a statement about the conserved and relaxing modes rather than about the lattice. The numerical check on the $2\times2$-lattice stage (dimension 1332) gives a direct-sum residual of $5\times10^{-16}$ after the rotation, and the norm $\max_s\sigma_{\max}(B_s)$ agrees with the recorded $\|\Kc_\lambda\|_2$ over 85 rate and scale combinations to $7\times10^{-16}$. The expansion overestimates the norm near $\omega = 2$, where $c_s$ diverges, 1.098 against the exact 1.051 at $\omega = 1.95$ and $\lambda = 10$, and Eq.~(\ref{eq:block}) is used exactly in all costs below.

\begin{corollary}[BGK and 3D lattices]\label{cor:bgk}
For equal rates $\omega_s = \omega$ the three blocks coincide and $\|\Kc_\lambda\|_2 = \sigma_{\max}(B)$ with $c = \omega/[4(2-\omega)]$. For the D3Q19 and D3Q27 lattices with the quadratic equilibrium of Eq.~(\ref{eq:feq}) in their weighted Hermite bases, the coupling acts from the six symmetric conserved-momentum products, which span the symmetric subspace of the nine ordered momentum-pair slots with orthonormal directions $\ket{aa}$ and $(\ket{ab} + \ket{ba})/\sqrt2$ for $a < b$, into the six second-order moments along six orthogonal directions with $\sigma_s = \omega_s/\sqrt2$, so that the stage is a diagonal plus six blocks of the form of Eq.~(\ref{eq:block}) and its norm is the same function of $\omega$ and $\lambda$ as on D2Q9.
\end{corollary}

The corollary was checked numerically on the three lattices at $\omega = 1.5$. The linear equilibrium map $HE_1H^{-1}$ is diagonal on the conserved moments, the coupling from products of non-conserved moments vanishes to $10^{-15}$, and $\|\Kc_\lambda\|_2$ is 1.504861, 1.076738, 1.007436, 1.000833 and 1.000075 at $\lambda = 1$, 3, 10, 30 and 100 on all three lattices. The physical content of the theorem is that the cost is set by the relaxing moments that the conserved momenta feed, three in two dimensions and six in three, and not by the number of velocities or by the amount of dissipation. Figure~\ref{fig:normalization} compares the two subnormalizations. At $\omega = 1.5$ and $\lambda = 10$ the two-term LCU has $\alpha_{\rm L} - 1 = 0.106$ and the block structure gives $\|\Kc_\lambda\|_2 - 1 = 0.0074$, and the ratio grows with $\lambda$ because one is first and the other second order in $1/\lambda$.

\begin{figure*}[t]\centering\includegraphics[width=0.65\textwidth]{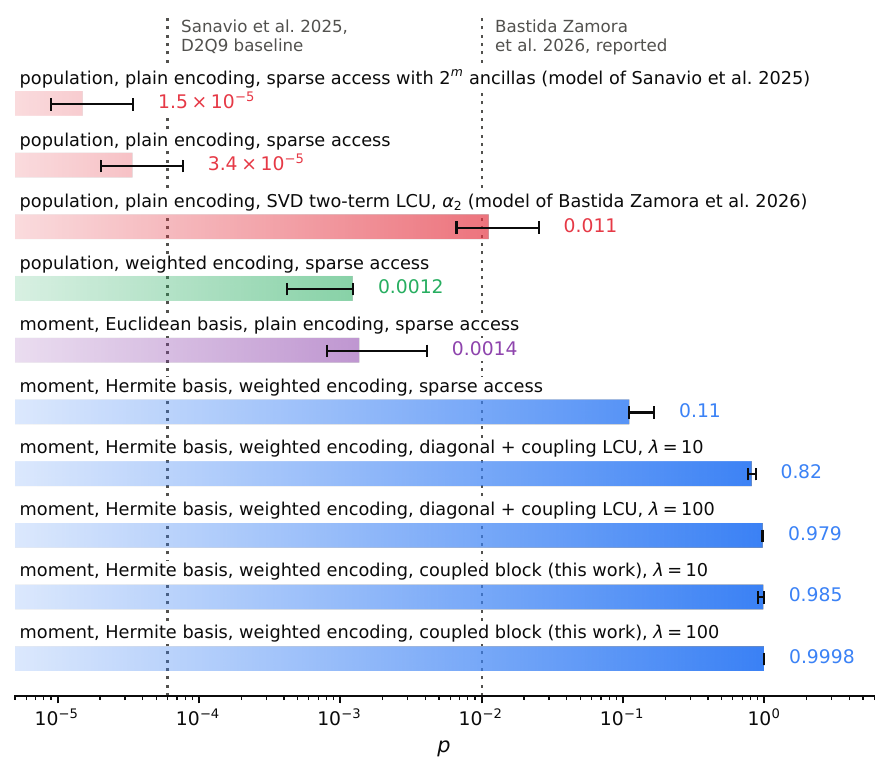}
\caption{Single-step success probability $p$ of the collision stage by representation and block-encoding construction, on the $32\times32$ Taylor-Green state at $\Ma = 0.087$ and $\omega = 1.5$ (bars, whose values are printed at the right); the horizontal line through each bar spans the range of $p$ over $\omega = 1.0$ to $1.95$. The fill of each bar darkens with $p$ on the same scale in every row, reaching full opacity at $p = 1$. The dotted vertical lines are the D2Q9 baseline $6\times10^{-5}$ of \citet{SanavioSimonSucci2025} and the value of order $10^{-2}$ reported by \citet{BastidaZamora2026} at $N = 16$, whose construction is modeled as the optimal subnormalization of the population-space stage. The last two rows are the coupled-block encoding of this work.}\label{fig:ladder}\end{figure*}

\begin{figure*}[t]\centering\includegraphics[width=0.9\textwidth]{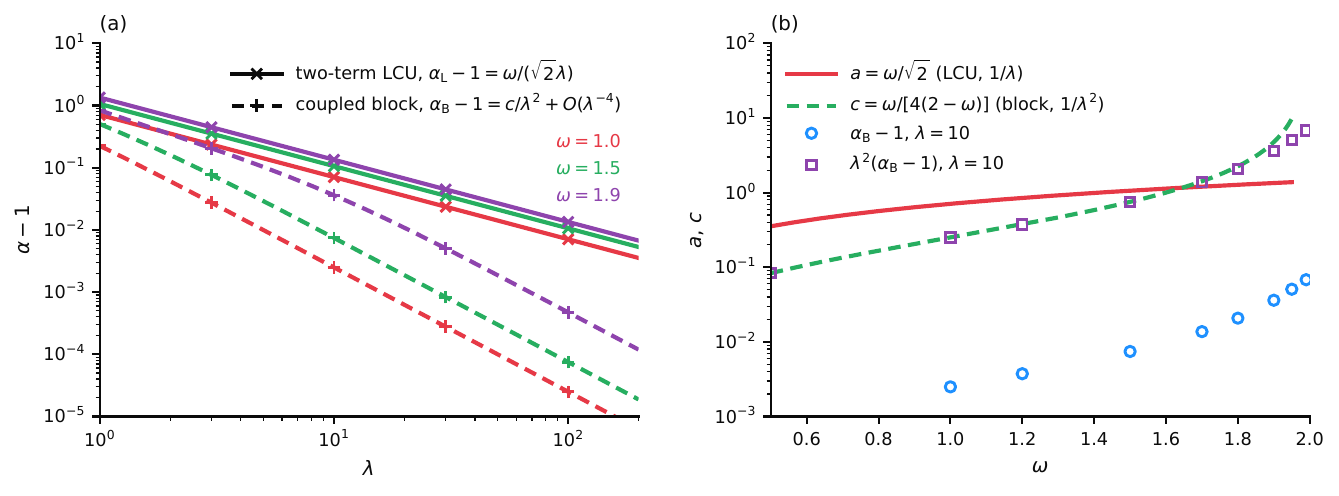}
\caption{Normalization cost of the scaled collision stage. (a) $\alpha - 1$ against the pair-sector scale $\lambda$ for the two-term LCU of Eq.~(\ref{eq:alcu}) (solid lines, crosses at the recorded values) and for the operator norm of Theorem~\ref{th:structure} (dashed lines, plus signs), at three BGK rates. (b) The coefficients $a$ of $1/\lambda$ and $c$ of $1/\lambda^2$ against the rate, with the recorded $\|\Kc_\lambda\|_2 - 1$ at $\lambda = 10$ and its product with $\lambda^2$, which follows $c$ up to $\omega = 1.9$ and falls below it as $\omega\to2$.}\label{fig:normalization}\end{figure*}

Figure~\ref{fig:ladder} places the coupled-block encoding in the sequence of representations and constructions of the collision stage, evaluated on one state, the $32\times32$ Taylor-Green vortex at $\Ma = 0.087$ and $\omega = 1.5$, with the range over the BGK rates 1.0 to 1.95 as whiskers. The three contributions to the success probability are separate. In population space the collision is dense, with sparsity 81 and 90 in rows and columns and entries up to 2.0, and the Euclidean metric of the plain encoding is not the one in which relaxation is a contraction, so that the sparse-access construction gives $1.5\times10^{-5}$ with the seven-ancilla subnormalization of \citet{SanavioSimonSucci2025} and $3.4\times10^{-5}$ with $\alpha_s = \sqrt{s_rs_c}\max|a_{ij}|$, and the two-term SVD construction of \citet{BastidaZamora2026}, modeled as the optimal subnormalization $\sigma_{\max} = 9.4$ of the population stage, gives $1.1\times10^{-2}$. The weighted encoding $f/\sqrt w$ changes the metric and moves the singular values of the stage into $[0.25, 1.5]$, which raises the sparse-access value to $1.2\times10^{-3}$. The Euclidean moment basis with plain encoding changes the sparsity to four and nine but not the singular values and gives $1.4\times10^{-3}$. The weighted Hermite representation, with sparsity three, entries at most one and $\alpha_s = 3$, gives $0.11$ with the same generic construction. The structured constructions then act on the coupling alone, $0.82$ and $0.98$ for the two-term LCU at $\lambda = 10$ and $100$, and $0.985$ and $0.9998$ for the coupled block at the same scales. The metric, the basis and the encoding of the coupling thus contribute factors of order $10^2$, $10^2$ and $10$, the first through the singular values and the second through the sparsity, and the last of them is the subject of Theorem~\ref{th:structure}. The representations themselves, their sparsity patterns and the density of the assembled full-step matrix in either basis are given in Appendix~\ref{app:representations}.

A block encoding that reaches the operator norm is obtained from the block structure directly. Each $2\times2$ block divided by $\alpha = \|\Kc_\lambda\|_2$ is a contraction $A_s$, and its Halmos dilation
\begin{equation}
V_s = \begin{pmatrix}A_s & \sqrt{I - A_sA_s^\Tr}\\ \sqrt{I - A_s^\Tr A_s} & -A_s^\Tr\end{pmatrix}
\label{eq:halmos}
\end{equation}
is a unitary on the block and one ancilla \cite{Halmos1950}. After the rotation $W_n$ the stage is $\tilde\Kc_\lambda = (\bigoplus_s B_s)\oplus D_0$, where $D_0$ contains all uncoupled diagonal and padding sectors. The assembled unitary $\tilde U$ is chosen so that
\begin{equation}
\bra{0_a}\tilde U\ket{0_a} = \Big(\bigoplus_s\frac{B_s}{\alpha}\Big)\oplus\frac{D_0}{\alpha} = \frac{\tilde\Kc_\lambda}{\alpha},
\label{eq:assembled}
\end{equation}
with $\alpha = \|\Kc_\lambda\|_2$, on the encoded physical subspace, the span of the register states that the lifted state occupies, and conjugating by $W_n$ gives the block encoding of $\Kc_\lambda/\alpha$ on that subspace. The diagonal part with entries of modulus at most one is block encoded by uniformly controlled rotations of one ancilla per moment register, as in the LCU construction, and the three dilations act on the pairs $\{s, u_s\}$, addressed by the second moment register on the slice $r = 0$ of the relative coordinate, controlled by the flag of that slice. The rotation $W_n$ and its inverse are applied on the same slice before and after the dilations. The whole stage is then one block of a unitary with subnormalization $\|\Kc_\lambda\|_2$ and no linear combination of terms, at $4n + 15$ qubits for $L = 2^n$, one more than the LCU circuit (Appendix~\ref{app:circuit}). The circuit was verified by statevector simulation on the $4\times4$ lattice (23 qubits) against the classical lift. The postselected state equals $\Kc_\lambda\psi/\alpha$ in both sectors to $2\times10^{-14}$, and the success probability measured as the postselected norm fraction equals $r_\lambda/\alpha^2$ to six digits, with $p = 0.984965$, $0.965744$ and $0.980375$ over three consecutive steps at $\lambda = 10$ where the LCU gives $0.8171$, $0.8012$ and $0.8133$, $0.999518$ at $\lambda = 100$ against $0.9788$, $0.862257$ at $\lambda = 3$ against $0.5456$, and $0.980263$ for the MRT rates $\omega_\nu/\omega_e/\omega_q/\omega_\varepsilon = 1.3/1.6/1.1/1.8$ at $\lambda = 10$ against $0.8068$ (Fig.~\ref{fig:validation}a). The remaining loss $1 - r_\lambda$, between $3\times10^{-4}$ and $2\times10^{-2}$ over these steps, is the non-equilibrium norm that the relaxation removes from the scaled state, which the linear evolution generates at the lattice scale on this coarse grid and which is at most $2\times10^{-3}$ per step on the $32\times32$ two-mode run at $\Ma = 0.173$.

The physical part of the loss is the same in every representation and is separated from the encoding by the norm balance of one collision. Along a trajectory the pair sector evolves by $C_1\otimes C_1$, so its norm ratio in a collision is exactly $(1-\ell)^2$ with $\ell = \sum_{k\ge3}[1 - (1-\omega_k)^2]\phi_k$, the fraction that $C_1$ removes from $G$, and $\phi_k$ the norm fraction of moment $k$ of $G$. In the level-1 sector the ghost moments relax as $(1-\omega_k)m_k$ and the three second-order moments as $(1-\omega_k)m_k + \omega_k\mu_k$, with $\mu_k$ the quadratic equilibrium moment evaluated on $G$ that the coupling supplies, so that the squared norm of the level-1 sector changes by
\begin{equation}
\begin{aligned}
\Delta_1 = &-\sum_{k\ge3}\big[1-(1-\omega_k)^2\big]\|m_k - \mu_k\|^2\\
&- 2\sum_{k=3}^{5}\omega_k\langle\mu_k, m_k - \mu_k\rangle,
\end{aligned}
\label{eq:balance}
\end{equation}
with $\mu_k = 0$ for the ghost moments. The first sum is the non-equilibrium norm relaxed in the step. The second is a cross term between the equilibrium and the non-equilibrium parts of the second-order moments, which has no definite sign, so the level-1 norm change is not the relaxed norm alone. On the trajectories considered the cross term is at most $10^{-3}$ of the norm, stays below the relaxed norm on the unperturbed Taylor-Green runs at rates up to 1.7, exceeds it in a few steps at the rates 1.9 and 1.95 where both are below $10^{-5}$ of the norm, and is comparable to the relaxed norm in a perturbed run and in the transient of a linearized-equilibrium start, whose second-order moments vanish so that the coupling first raises the level-1 norm by $\sum_k\omega_k^2\|\mu_k\|^2$ before relaxation removes anything. Dissipation thus reduces the success probability in proportion to the relaxed non-equilibrium norm, up to the cross term, and this norm is of order $(\mathrm{Kn}\,\Ma)^2$ for a resolved flow, with Kn the ratio of the relaxation length to the flow scale. The factor $1/\alpha^2$ carries the remainder, and Theorem~\ref{th:structure} fixes it.

\section{From collision success to coherent output recovery}\label{sec:coherent}
This section derives the success probability of a $T$-step run followed by output recovery, shows how telescoping reduces the multi-step problem to a global normalization-selection balance, shows that optimizing the scale for the specified horizon yields polynomial rather than exponential success scaling for the structured encodings, explains the rest shift, and states the preparation, evolution and readout protocol under which amplitude amplification applies. For a fixed scale the subnormalization $\alpha$ is the same in every step and step $t$ succeeds with probability $r_t/\alpha^2$. The transforms and the streaming preserve the norm, so $r_t$ is the ratio of the squared norms of $\psi_\lambda$ before and after the step, and the product over the steps telescopes. The product $P_{\rm evol} = \prod_{t=0}^{T-1}r_t/\alpha^2 = \|\psi_\lambda(T)\|_g^2/(\alpha^{2T}\|\psi_\lambda(0)\|_g^2)$ is the probability that the encoded evolution survives, the weight of the branch in which every encoding ancilla returns to zero, before any selection on the state itself. A run followed by the selection of the level-1 sector, which holds the nonlinear solution, succeeds with
\begin{equation}
\begin{aligned}
P &= \prod_{t=0}^{T-1}\frac{r_t}{\alpha^2}\,\frac{\|F(T)\|_g^2}{\|F(T)\|_g^2 + \lambda^2\|G(T)\|_g^4}\\
&= \frac{\|F(T)\|_g^2}{\alpha^{2T}\big(\|F_0\|_g^2 + \lambda^2\|F_0\|_g^4\big)},
\end{aligned}
\label{eq:total}
\end{equation}
an identity for the level-2 dynamics, verified along every trajectory of this study to $10^{-15}$. The first factor of the first line is $P_{\rm evol}$ and the second the selection weight of the level-1 sector. The norm of the pair sector cancels between the survival and the selection factors, so the pair-sector loss of a run does not enter $P$. The level-1 factor is bounded through mass conservation, $\|F(T)\|_g^2 \ge \sum_x\rho(x)^2 \ge N$ for unit mean density, while $\|F_0\|_g^2 = N(1 + \epsilon_0)$ defines the initial flow fraction $\epsilon_0$ of the norm, of order $\Ma^2$. With $\epsilon_0\to0$ and $\|F(T)\|_g^2 = N$ the near-rest model
\begin{equation}
P(\lambda) = \frac{\alpha(\lambda)^{-2T}}{1 + \lambda^2N}
\label{eq:model}
\end{equation}
follows. At any fixed finite $\lambda$ with nonzero coupling, $\alpha(\lambda) > 1$ and $P(\lambda)$ still decays exponentially with $T$. The polynomial laws below are the horizon-optimized envelope obtained by choosing $\lambda = \lambda(T)$ separately for the specified simulation horizon. The maximum of Eq.~(\ref{eq:model}) over $\lambda$ depends on the encoding through $\alpha(\lambda)$. For the LCU, $\alpha = 1 + a/\lambda$, the stationary condition is $N\lambda^3 - aN(T-1)\lambda^2 - aT = 0$, $\lambda = aT$ approximates the root to relative order $1/T$, and $P^\ast_{\rm L} \to e^{-2}/(a^2T^2N)$. For the block encoding, $\alpha = 1 + c/\lambda^2 + O(\lambda^{-4})$, the same argument gives $\lambda^2 = 2cT$ and
\begin{equation}
P^\ast_{\rm B} \to \frac{e^{-1}}{2cTN}, \qquad \frac{P^\ast_{\rm B}}{P^\ast_{\rm L}} \to \frac{ea^2}{2c}\,T = e\,\omega(2-\omega)\,T,
\label{eq:gain}
\end{equation}
for BGK rates, both asymptotic in $T$ for the near-rest model; at finite $T$ the optimum is taken numerically from Eq.~(\ref{eq:model}) with the exact singular value, and the actual probability of a run is Eq.~(\ref{eq:total}). The cost of the run thus falls from $T^2$ to $T$ and the gain grows linearly with the number of steps.

\begin{figure*}[t]\centering\includegraphics[width=\textwidth]{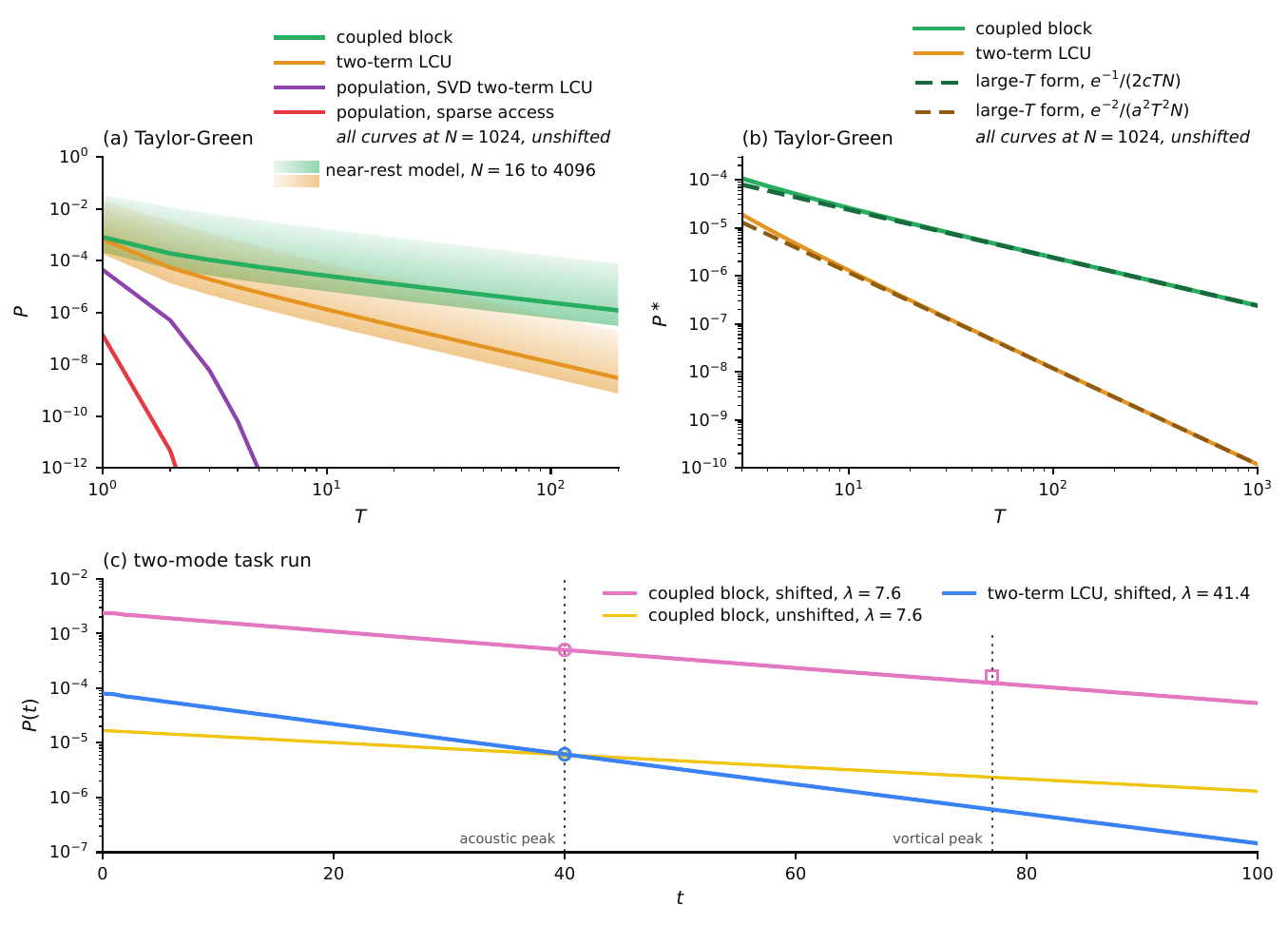}
\caption{Success of a complete run with output selection. (a) $P$ of Eq.~(\ref{eq:total}) against the number of steps on the $32\times32$ Taylor-Green trajectory at $\Ma = 0.087$ and $\omega = 1.5$ for the population constructions with sparse access ($\alpha = 171$) and with the two-term SVD ($\alpha = 9.39$), unscaled as published, and for the two structured encodings of the weighted Hermite stage at the scale optimal for each horizon; the shaded bands are the near-rest model of the two structured encodings between $N = 16$ and $4096$. (b) Finite-$T$ optimum $P^\ast$ of the near-rest model for the two encodings at $\omega = 1.5$ and $N = 1024$, with the large-$T$ forms (dashed); the values of Eq.~(\ref{eq:total}) along the Taylor-Green trajectory at the optimal scale are within 0.4 to 0.7\% of the model over 10 to 200 steps and are not drawn. (c) Running product of the step factors along the task run of Sec.~\ref{sec:problem} in the shifted encoding at the scales optimal for the acoustic horizon $T = 40$, for the block encoding, the LCU and the unshifted block encoding at its own optimal scale; the circles mark $P$ at the acoustic horizon and the square the block value at the vortical horizon $T = 77$ at its own scale.}\label{fig:joint}\end{figure*}

The comparison with the specified population constructions is made on one trajectory. Figure~\ref{fig:joint}a evaluates Eq.~(\ref{eq:total}) for every construction of Table~\ref{tab:compare} on the $32\times32$ Taylor-Green run of Fig.~\ref{fig:ladder}, with the norm ratios in the metric of each encoding and the population constructions unscaled, as published. The sparse-access construction and the two-term SVD construction of the population stage lose the factors $r_t/171^2$ and $r_t/9.39^2$ at every step and reach $8.8\times10^{-48}$ and $1.4\times10^{-22}$ after ten steps. The powers $(6\times10^{-5})^T$ and $(10^{-2})^T$ of the published step probabilities lie above these curves by the selection weight of about $4\times10^{-3}$ and, for the sparse-access construction, by the ratio $(6\times10^{-5}/3.4\times10^{-5})^T$ of the published to the reproduced step probability, and are not drawn. The two structured encodings of the weighted Hermite stage, each at the scale optimal for its horizon, reach $1.3\times10^{-6}$ and $2.6\times10^{-5}$ after ten steps and $1.2\times10^{-8}$ and $2.4\times10^{-6}$ after a hundred. The exponential loss of the population constructions is thus replaced by a polynomial one, and the coupled block adds the factor of Eq.~(\ref{eq:gain}). The near-rest model between $N = 16$ and $4096$ (shaded bands) shows the lattice-size dependence, and the trajectory curves lie inside the bands because the flow norm decays along the run. Figure~\ref{fig:joint}b shows the two finite-$T$ optima of Eq.~(\ref{eq:model}) against $T$ at $\omega = 1.5$ and $N = 1024$ with the asymptotic forms; the trajectory values are not drawn, because $P$ of Eq.~(\ref{eq:total}) at the optimal scale is 0.993 to 0.996 of the near-rest model over 10 to 200 steps for both encodings and the two curves would coincide. The ratio $P^\ast_{\rm B}/P^\ast_{\rm L}$ at $N = 4096$ is 20, 61 and 204 at $T = 10$, 30 and 100 for $\omega = 1.5$, and 7.3, 17.5 and 54 at $\omega = 1.9$, where the expansion of the norm is less accurate and the asymptotic scale $\lambda^2 = 2cT$ reaches 83 to 99.6\% of the optimum. At $\omega\le1.5$ and $T\ge30$ it reaches at least 99.8\%. These probabilities describe the survival of the run, not its accuracy. The horizon over which the truncated dynamics is accurate is the subject of Sec.~\ref{sec:accuracy}, and $1/P$ is not the runtime, since each use of the run costs the $T$-step circuit of Appendix~\ref{app:circuit}.

The selection weight $1/(1 + \lambda^2N)$ is set by the norm of the rest state, which carries no information about the flow. This factor is removed by encoding the deviation from rest. With $w$ the rest populations and $g = f - w$, the level-1 collision leaves $w$ unchanged, $Lw = w$, and the quadratic part of the equilibrium is bilinear in the momentum with unit density inside, so that
\begin{equation}
\begin{aligned}
Q(w\otimes g) &= Q(g\otimes w) = 0,\\
f' - w &= L(f - w) + Q\big((f-w)\otimes(f-w)\big),
\end{aligned}
\label{eq:shift}
\end{equation}
exactly. Streaming is a permutation that maps $w$ onto itself. The level-2 lift of the shifted state $(g, \lambda\,g\otimes g)$ therefore has the same dynamics, the same truncation and the same circuits as the unshifted one, and reproduces $F - w$ to better than $10^{-9}$ of the flow norm along every trajectory of this study. The same shift underlies the formulation of \citet{Jennings2026}. Its effect on the cost is twofold. The selection weight becomes $1/(1 + \lambda^2N\epsilon_0)$, because $\|g_0\|_g^2 = N\epsilon_0$, and the near-rest optimization applies with $N\epsilon_0$ in place of $N$, so that the optimal scale and the cost are those of a lattice smaller by the factor $\epsilon_0$. The overlap of Eq.~(\ref{eq:overlap}) with the reference state grows by $1/\sqrt{\epsilon_0}$ at $t = 0$, and along a run by the inverse square root of the flow-norm ratio in addition, from $1.1\times10^{-3}$ to $1.7\times10^{-2}$ at the peak of the observable at $\Ma = 0.087$ (Fig.~\ref{fig:observable}c), because the norm of the level-1 sector no longer contains the rest state. The price is that the level-1 factor of Eq.~(\ref{eq:total}) is no longer bounded through mass conservation and becomes the physical decay of the flow norm, $\|g(T)\|_g^2/\|g_0\|_g^2$, which is 0.58 at the observable horizon of the reference case. Figure~\ref{fig:joint}c shows the running product of the step factors along the task run, from the linearized start, for the two encodings in the shifted form and for the unshifted block encoding at its own optimal scale, whose probability at the horizon is $6.1\times10^{-6}$ against $5.0\times10^{-4}$ shifted.

The amplified cost requires the run to be one unitary. The protocol is $U = U_{\rm evol}U_{\rm prep}$ (Algorithm~\ref{alg:step}), where $U_{\rm prep}$ prepares $\psi_\lambda(0)$ and $U_{\rm evol}$ applies the $T$ steps with a fresh triple of encoding ancillas per step, the flag ancillas being uncomputed and reused. The good subspace of the protocol is the survival subspace, the one in which all preparation and encoding ancillas are zero, and its weight is $P_{\rm prep}P_{\rm evol}$, with $P_{\rm prep}$ the success of the preparation. This weight carries no selection on the state and is not the $P$ of Eq.~(\ref{eq:total}). The level-1 selection enters through the reference state of the readout below. For a few-mode initial state the preparation is explicit. The shifted linearized equilibrium $g_0 = 3w_ic_i\!\cdot\!J$ of a field with $M$ Fourier terms, the initial state of the task of Sec.~\ref{sec:problem}, is, in the weighted encoding, a linear combination of $M$ orthonormal product states $\ket{e_k}\otimes\ket{v_a}$ of a plane wave on the site register and a fixed velocity vector on the moment register, and is prepared by a linear combination of unitaries with $\lceil\log_2M\rceil$ ancillas, one controlled basis-state write per term, two quantum Fourier transforms and a success probability $p_1 = (\|c\|_2/\|c\|_1)^2$ \cite{ChildsWiebe2012}. The pair sector is built from two copies of this preparation on the two registers, a subtraction of the site coordinates that brings the second copy into relative coordinates, and a branch ancilla whose angle is compensated for the success of the second copy (Appendix~\ref{app:circuit}). The preparation success is $P_{\rm prep}\approx p_1^2$ because the pair branch passes two LCU postselections, $4.3\times10^{-3}$ for the 16-term field, and it is made deterministic by exact amplitude amplification of $U_{\rm prep}$ alone \cite{BrassardHoyer2002}. An auxiliary qubit is rotated so that the good weight becomes $\sin^2[\pi/(2(2k+1))]$ with $k = \lceil\pi/(4\theta_{\rm p}) - 1/2\rceil$ and $\sin\theta_{\rm p} = \sqrt{P_{\rm prep}}$, and $k$ rounds of $-U_{\rm prep}S_0U_{\rm prep}^\dagger S_{\rm good}$ end exactly on the good state, at the cost of $2k + 1$ preparation circuits per use of $U$, twelve rounds for the task.

\begin{figure*}[t]\centering\includegraphics[width=0.9\textwidth]{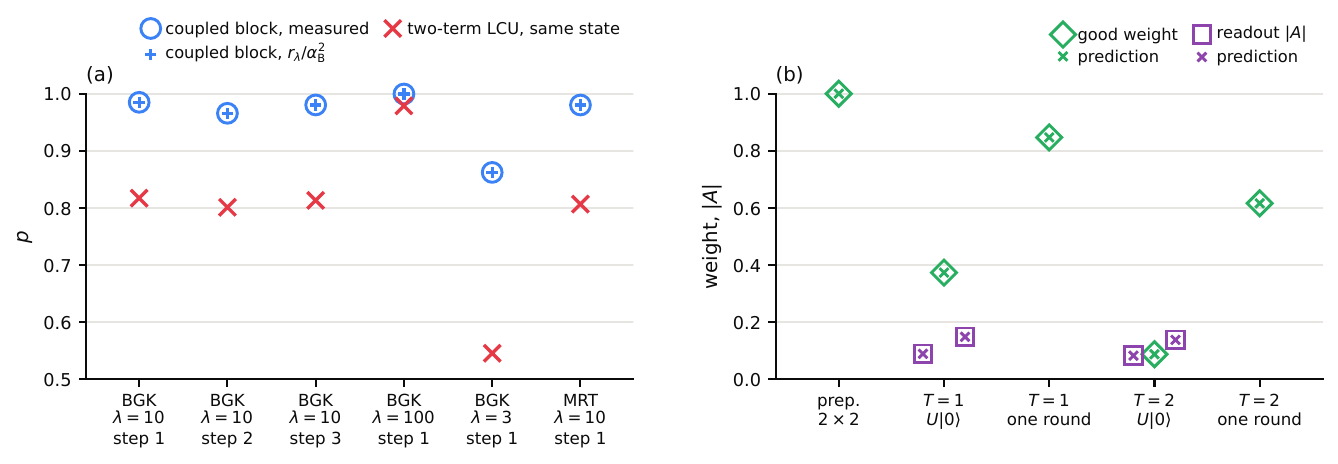}
\caption{Verification of the quantum construction. (a) Success probability $p$ of one collision stage of the coupled-block circuit on the $4\times4$ Taylor-Green state at $\Ma = 0.17$ (23 qubits), measured as the postselected norm fraction against $r_\lambda/\alpha_{\rm B}^2$ from the classical state, and the value the same state gives with the two-term LCU. (b) Weight of the survival subspace after the exact amplification of the $2\times2$ preparation, after $U\ket{0}$ and after one amplification round of the one- and two-step protocols on the $2\times2$ lattice at $\lambda = 3$ and $\omega = 1.5$, and the readout amplitudes $|A|$ of the two momentum components, each against its prediction.}\label{fig:validation}\end{figure*}

The construction was verified by statevector simulation. The level-1 state is prepared with fidelity one on the $4\times4$ (12 terms, 12 qubits) and $8\times8$ (16 terms, 14 qubits) lattices at the predicted success 0.106 and 0.066, and the full $\psi_\lambda(0)$ with fidelity one at the predicted success, 0.429 on the $2\times2$ lattice for $\lambda = 3$ (15 qubits) and 0.0120 on the $4\times4$ lattice for $\lambda = 10$ (25 qubits). The exact amplification of the $2\times2$ preparation, one round after the auxiliary rotation to the weight 0.250, brings the good weight to $1 - 4\times10^{-14}$ at fidelity one (16 qubits). On the $2\times2$ lattice the two-mode field of Sec.~\ref{sec:problem} vanishes ($\kappa = \pi$), so the protocol checks use the field $J_x = U_0\cos\pi y$, $J_y = 0.6U_0\cos\pi x$ with $U_0 = 0.05$ and read its initially present modes (Supplemental Material). The small circuit validates the composition and the readout identity. The nonlinear generation of the target modes is validated classically in Sec.~\ref{sec:accuracy}. A run of the complete protocol, one step at $\lambda = 3$ on 22 qubits, leaves the good branch of $U\ket{0}$ equal to $\psi_\lambda(1)$ with fidelity one at the weight $0.3735 = 0.4295\times0.8698$ predicted from the preparation success $P_{\rm prep} = 0.4295$ and the survival $P_{\rm evol} = 0.8698$ of the one-step evolution, and one round of amplitude amplification, $-US_0U^\dagger S_{\rm good}$ with the reflections applied exactly, raises the weight to $\sin^23\theta = 0.8470$ with the good branch unchanged (Fig.~\ref{fig:validation}b). The two-step protocol on 25 qubits behaves in the same way, with the good weight $0.0878 = 0.4295\times0.2045$, the two-step $P_{\rm evol}$ being 0.2045, raised to $0.6162 = \sin^23\theta$ by one round at fidelity one. The output recovery was checked on the same statevectors. The amplitude of $U\ket0$ on the product of the reference state of Eq.~(\ref{eq:overlap}), the level-1 slot and the zero ancillas is the quantity that amplitude estimation measures, and for the two momentum components of the $2\times2$ field it equals $\sqrt{P_{\rm prep}P}\,\sqrt N\hat{J}_a(k)/(c_s\|g(T)\|_g)$, with $P$ the level-1-selected probability of Eq.~(\ref{eq:total}) and equivalently $\sqrt{P_{\rm prep}P_{\rm evol}}\,\sqrt N\hat{J}_a(k)/(c_s\|\psi_\lambda(T)\|_g)$, from the classical state to $3\times10^{-15}$, $0.08927$ and $0.1488$ after one step and $-0.08291$ and $-0.1382$ after two (Supplemental Material), so that the chain from preparation through evolution, sector selection and overlap is validated as one unitary.

With $U = U_{\rm evol}U_{\rm prep}$ available as one circuit and the preparation exact, the good branch of $U\ket0$ is $\psi_\lambda(T)/(\alpha^T\|\psi_\lambda(0)\|_g)$, with $\psi_\lambda(T)$ the state after $T$ full steps including the transforms and the streaming, so the amplitude of $U\ket0$ on the product of the reference state of Eq.~(\ref{eq:overlap}), the level-1 slot and the zero ancillas is
\begin{equation}
A = \frac{\sqrt N\,\hat{J}_a(k;T)}{c_s\,\alpha^T\,\|\psi_\lambda(0)\|_g},\qquad \|\psi_\lambda(0)\|_g^2 = \|g_0\|_g^2 + \lambda^2\|g_0\|_g^4,
\label{eq:A}
\end{equation}
with $\hat{J}_a(k;T)$ the Fourier coefficient of the level-2 state at the horizon. Neither the norm of the evolved state nor the run probability enters: the product $\sqrt{P}\,a$ of the square root of the run probability and the overlap of Eq.~(\ref{eq:overlap}) is the same number by the telescoping identity, and the observable is recovered from an estimate of $|A|$ as $|\hat{J}_a(k;T)| = c_s\alpha^T\|\psi_\lambda(0)\|_g|A|/\sqrt N$ with $\alpha$ the exact norm of the scaled collision stage and $\|\psi_\lambda(0)\|_g$ computed from the prepared initial state. Equation~(\ref{eq:A}) reproduces the measured amplitudes of the $2\times2$ protocol above to $3\times10^{-15}$ from the initial norm alone, and a run of the exactly amplified preparation followed by one step on 23 qubits, with the reflections of the amplification built as circuits, leaves $U\ket0$ with the survival weight $P_{\rm evol} = 0.8698$ of the evolution alone and the amplitudes $0.1362$ and $0.2270$ of Eq.~(\ref{eq:A}) to $10^{-14}$ (Supplemental Material).

The magnitude $|A|$ is estimated by amplitude estimation \cite{BrassardHoyer2002} on the state $U\ket0$ with the good subspace spanned by the reference product state, whose reflection $S_\chi$ is the preparation $V$ of that state, a reflection about $\ket0$ and $V^\dagger$. The three weights are nested: the survival weight on the zero ancillas, the $P$ of Eq.~(\ref{eq:total}) once the level-1 sector is selected as well, and the estimated $|A|^2$ on the reference product state. With resolution parameter $M = 2^m$, canonical phase-estimation amplitude estimation uses $M - 1$ controlled applications of $Q = -US_0U^\dagger S_\chi$ after the initial preparation $U$, and the estimate $\tilde a$ of $a = |A|^2$ satisfies $|\tilde a - a| \le 2\pi\sqrt{a(1-a)}/M + \pi^2/M^2$ with probability at least $8/\pi^2$, so that $|\tilde A| = \sqrt{\tilde a}$ has relative error at most $\epsilon_{\rm rel}$ whenever
\begin{equation}
\frac{2\pi}{M|A|} + \frac{\pi^2}{M^2|A|^2} \le 2\epsilon_{\rm rel} - \epsilon_{\rm rel}^2,
\label{eq:M}
\end{equation}
and $M$ is the smallest power of two that satisfies Eq.~(\ref{eq:M}), $M \approx 3.5/(\epsilon_{\rm rel}|A|)$ before rounding, with $m = \log_2M$ phase qubits. Since $|A|$ enters Eq.~(\ref{eq:M}), the estimation length is set from the benchmark amplitude of the classical level-2 calculation of the same instance; the counts that follow are therefore instance-dependent resource estimates calibrated with that amplitude, and no stopping rule for an output that is not known in advance is implemented or costed here. The $8/\pi^2$ guarantee is raised to the confidence level by the median of $r$ independent runs, which fails only if at least $\lceil r/2\rceil$ runs fail; $r = 7$ gives a failure probability of $0.028$, while $r = 5$ gives $0.050$. Each application of $Q$ uses $U$ once and $U^\dagger$ once, and only the two reflections carry the control of the phase register, since $UU^\dagger$ is the identity on the branch that skips them. The task therefore needs
\begin{equation}
N_U = r\,(2M - 1) \approx 14M \approx \frac{49}{\epsilon_{\rm rel}|A|}
\label{eq:NU}
\end{equation}
applications of $U$ or $U^\dagger$ before the rounding of $M$, against $1.96^2(1 - |A|^2)/(4\epsilon_{\rm rel}^2|A|^2)$ incoherent repetitions of preparation, evolution and measurement of the projector onto the reference product state for the same precision and confidence, a normal approximation of the binomial sample, and without the amplification of the preparation the number of applications would be larger by $1/\sqrt{P_{\rm prep}}$. This is the cost that Sec.~\ref{sec:resources} evaluates.

\section{Nonlinear observables and accuracy window}\label{sec:accuracy}
\begin{figure*}[t]\centering\includegraphics[width=\textwidth]{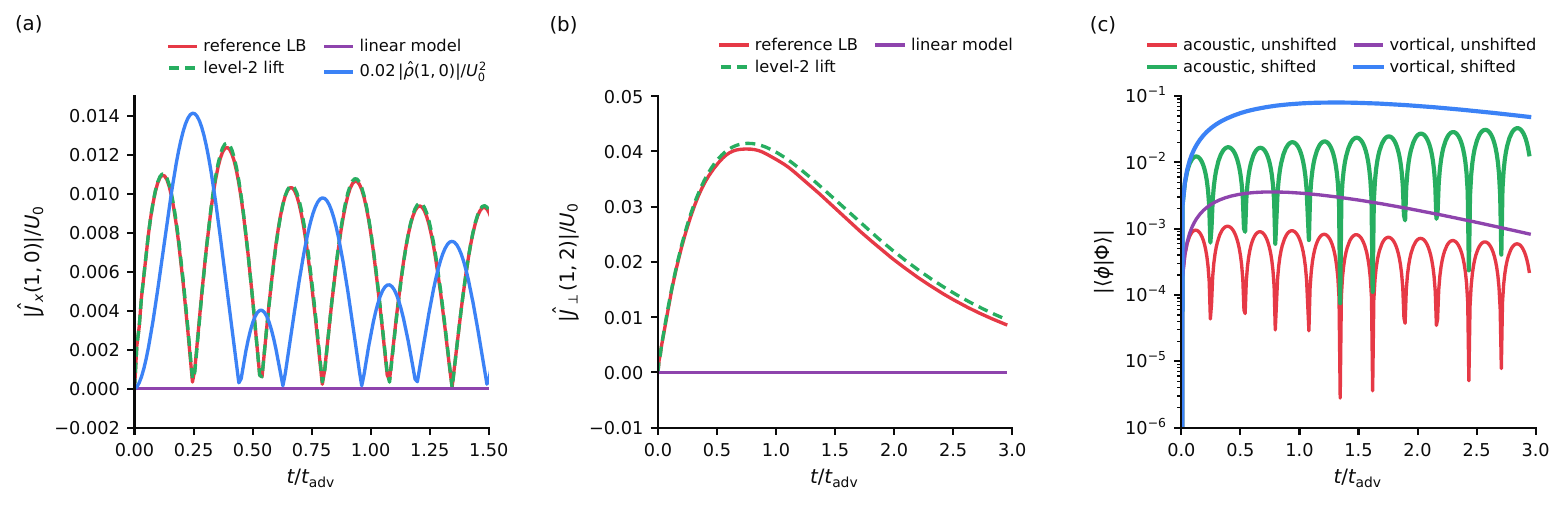}
\caption{Nonlinearly generated observables on the $32\times32$ two-mode flow at $\Ma = 0.087$ and $\omega = 1.5$ from the linearized start. (a) The acoustic mode $|\hat{J}_x(1,0)|/U_0$ for the LB reference, the level-2 lift and the linear model, with the density mode $|\hat\rho(1,0)|$ scaled to the same panel. (b) The vortical mode $|\hat{J}_\perp(1,2)|/U_0$ for the same three evolutions. (c) Overlap amplitudes of the two modes with their reference states along the run in the unshifted and the shifted encodings.}\label{fig:observable}\end{figure*}

This section establishes which physically meaningful signals the level-2 dynamics recovers and the linear approximation misses, over which horizon it is accurate on a flow whose modes interact, and why the error behaves as it does. The two-mode flow of Eq.~(\ref{eq:twomode}) is run from the linearized start of Sec.~\ref{sec:problem} on the $32\times32$ lattice at $\Ma = 0.087$ and $\omega = 1.5$ ($\mathrm{Re} = 28.8$), and the momentum modes generated by the quadratic term are decomposed into their longitudinal part along $k$ and their transverse part along $k_\perp$ (Fig.~\ref{fig:observable}). The $(1,0)$ mode is longitudinal to rounding, its transverse part being below $10^{-15}$ by the symmetry of the field, and it is acoustic. It is seeded at order $\Ma^2$ by the equilibrium of the first collision, grows to $0.0124\,U_0$ at $t = 40 = 0.39t_{\rm adv}$ through the pressure and advective work of the two vortical modes on each other, and oscillates with a period of 55.0 steps. The same period is measured at $\Ma = 0.020$ and $0.173$, 55.6 steps in both, where the advective time is 443 and 51 steps, and it becomes 27.4 and 110.2 steps on the $16^2$ and $64^2$ lattices at $\mathrm{Re} = 28.8$, so the oscillation follows the acoustic period $2\pi/(c_s\kappa) = 55.4$ steps of this wavevector and not the flow, and the density mode $\hat\rho(1,0)$ oscillates with it. It is the sound that the vortical interaction radiates, an aeroacoustic source term in the sense of Lighthill, resolved here as a discrete mode of the periodic box. The $(1,2)$ mode has a transverse part that grows smoothly to $0.040\,U_0$ at $t = 77 = 0.76t_{\rm adv}$ and decays viscously without oscillation, the vortical mode that the interaction feeds. Its longitudinal part is a small acoustic transient that peaks at $t = 5$. The linear model, run from the same linearized start, leaves both modes empty, because its translation-invariant evolution cannot populate a wavevector that carries no initial momentum, so its error at the peaks is one. The level-2 lift reproduces the acoustic peak with a relative error of $1.9\times10^{-2}$ and the vortical peak with $2.6\times10^{-2}$, where the relative error of a mode is the modulus of the difference of the complex Fourier coefficients of the lift and the reference divided by the modulus of the reference coefficient, an upper bound on the error of the estimated magnitude, and at fixed physics the errors are $1.8\times10^{-2}, 1.9\times10^{-2}, 1.9\times10^{-2}$ for the acoustic mode and $2.3\times10^{-2}, 2.6\times10^{-2}, 2.7\times10^{-2}$ for the vortical mode on $16^2$, $32^2$ and $64^2$ at $\mathrm{Re} = 28.8$, resolution independent, and $2.2\times10^{-2}, 2.2\times10^{-2}$ for the acoustic mode on $32^2$ and $64^2$ at $\mathrm{Re} = 100$. The acoustic error scales as $1.0\times10^{-3}, 1.9\times10^{-2}, 7.2\times10^{-2}$ over $\Ma = 0.020$, $0.087$ and $0.173$ and the vortical one as $1.4\times10^{-3}, 2.6\times10^{-2}, 8.1\times10^{-2}$, close to $\Ma^2$, except that the vortical error grows somewhat more slowly between the two larger Mach numbers. Both generated modes have even transverse index, which cubic products of the initial modes, of odd transverse index, cannot populate, so the dropped cubic term, of absolute order $\Ma^3$, does not feed them directly. The $\Ma^2$ scaling of their relative error follows the $\Ma^2$ truncation error of the level-2 state and is a property of these observables, not a generic rule for every generated mode. At $\mathrm{Re} = 100$ the vortical peak moves to $2.1t_{\rm adv}$ and its level-2 error there is $1.7\times10^{-1}$, outside the accuracy window established next, whereas the acoustic peak at $0.4t_{\rm adv}$ stays inside it. Figure~\ref{fig:observable}c shows the overlap amplitudes of Eq.~(\ref{eq:overlap}) in the two encodings along the run. The shifted values, $1.7\times10^{-2}$ for the acoustic mode at its peak and $7.0\times10^{-2}$ for the vortical mode at its later peak, are the conditional overlaps of Eq.~(\ref{eq:overlap}). The readout of Sec.~\ref{sec:coherent} estimates the combined amplitude $A$ of Eq.~(\ref{eq:A}), the product of this overlap with the square root of the run probability.

\begin{figure*}[t]\centering\includegraphics[width=\textwidth]{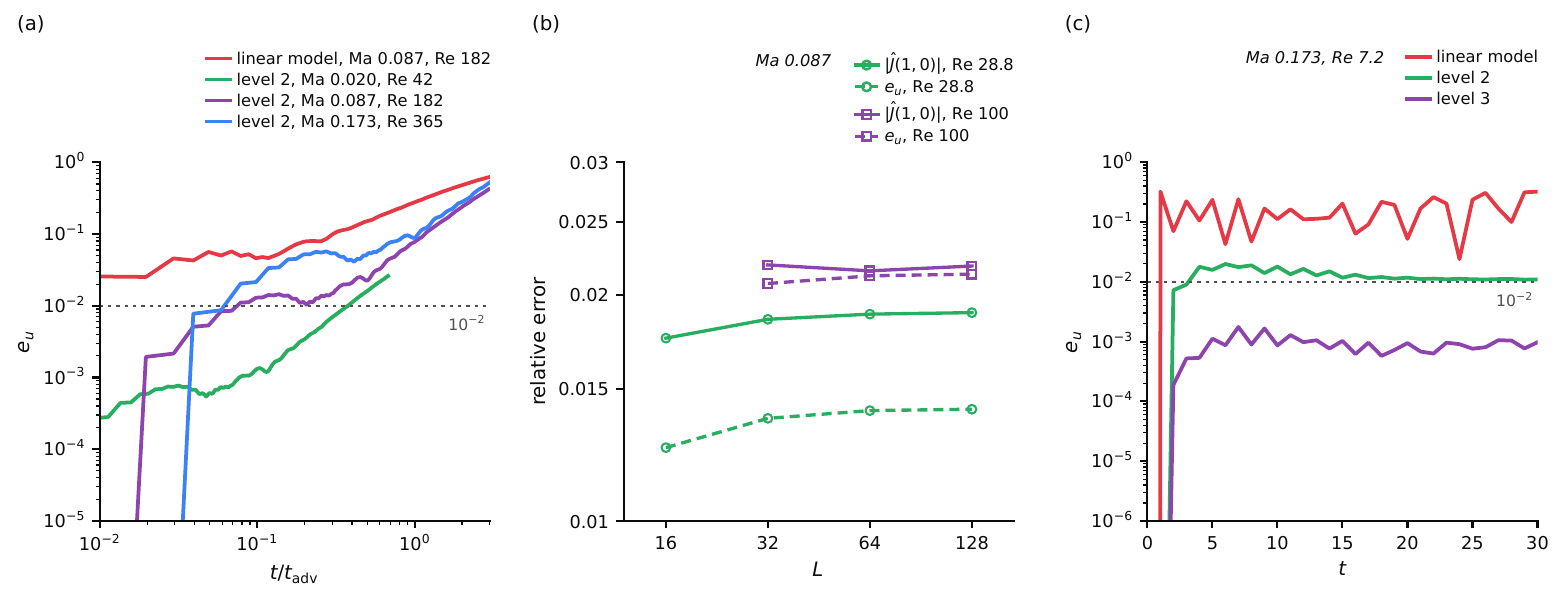}
\caption{Accuracy window of the level-2 lift on the two-mode flow. (a) Velocity error against $t/t_{\rm adv}$ on the $32\times32$ lattice at $\omega = 1.9$ for the linear model at $\Ma = 0.087$ and the level-2 lift at three Mach numbers; the dotted line in (a) and (c) marks $10^{-2}$. (b) Fixed-physics refinement at $\Ma = 0.087$, relative error of $|\hat{J}(1,0)|$ at its peak and $e_u$ at $t_{\rm adv}/2$ against the lattice side at $\mathrm{Re} = 28.8$ and $100$. (c) Level-3 spot check on the $4\times4$ lattice at $\Ma = 0.173$ and $\omega = 1.5$ for the linear model and the level-2 and level-3 lifts.}\label{fig:window}\end{figure*}

The accuracy window is established on the wider parameter study with the quadratic-equilibrium start. The two-mode flow of Eq.~(\ref{eq:twomode}) was run on the $32\times32$ lattice at $\Ma = 0.020$, $0.087$ and $0.173$ with BGK rates $\omega = 1.0$, $1.5$ and $1.9$ and with the MRT set of Sec.~\ref{sec:structure}, and at fixed physics, $\Ma = 0.087$ with $\mathrm{Re} = 28.8$ on $16^2$ to $128^2$ and $\mathrm{Re} = 100$ on $32^2$ to $128^2$, for 150 to 1200 steps, against the nonlinear LB reference and the linear model $G(t)$ of Eq.~(\ref{eq:product}). Figure~\ref{fig:window}a shows the velocity error at $\omega = 1.9$. At early times the level-2 error is of order $\Ma^2$, as the dropped cubic term of the lift implies, and the fit of the level-2 error over the Mach numbers of the level-3 check runs on the $4\times4$ and $6\times6$ lattices gives $e_u(t=2)\propto\Ma^{2.00}$ and $\Ma^{2.02}$. The error then grows with the accumulated nonlinear interaction. At $\Ma = 0.087$ and $\omega = 1.5$ it levels off on the advective time scale, at $3.3\times10^{-2}$ after one advective time and $5.6\times10^{-2}$ after three, and at $\omega = 1.9$ it continues to grow, reaching $0.42$ after three advective times. At $t = t_{\rm adv}/2$ it is $1.6\times10^{-2}$, $2.2\times10^{-2}$ and $5.0\times10^{-2}$ at the three Mach numbers, where the linear model has $0.12$ to $0.15$, and at $\Ma = 0.087$ it is $1.3$ to $1.4\times10^{-2}$ on the four lattices at $\mathrm{Re} = 28.8$ and $4.7\times10^{-3}$ to $2.2\times10^{-2}$ across the rates on $32^2$, increasing with $\omega$. The window in which $e_u < 10^{-2}$ ends at $0.08$ to $0.11t_{\rm adv}$ for $\Ma\ge0.087$ at $\omega\ge1.3$ and for $\Ma = 0.173$ at every rate, at $1.9t_{\rm adv}$ for $\Ma = 0.087$ at $\omega = 1.0$ where the viscous decay removes the flow first, at $0.38t_{\rm adv}$ for $\Ma = 0.020$ at $\omega = 1.9$, and is not left within $0.68t_{\rm adv}$ at $\Ma = 0.020$ for $\omega\le1.5$. This behavior indicates that the truncation error is controlled by the advective time and not by the resolution. Figure~\ref{fig:window}b confirms this at fixed physics. The relative error of the observable at its peak is $1.75$, $1.85$, $1.88$ and $1.89\times10^{-2}$ on $16^2$ to $128^2$ at $\mathrm{Re} = 28.8$, and $2.19$, $2.15$ and $2.18\times10^{-2}$ on $32^2$ to $128^2$ at $\mathrm{Re} = 100$, so that refinement at fixed Reynolds and Mach numbers changes neither the window nor the error at the horizon. Level-3 truncation extends the window. On the $4\times4$ lattice at $\Ma = 0.173$ the level-2 error exceeds $10^{-2}$ at $t = 4$ and the level-3 error stays below it over 30 steps (Fig.~\ref{fig:window}c), with $e_u(t=2)\propto\Ma^{4.00}$ on $4\times4$ and $\Ma^{3.01}$ on $6\times6$, at a state dimension that grows as $N^3$.

The regime in which the level-2 lift fails is also identified. At $\Ma = 0.173$ and $\omega = 1.9$ the velocity error exceeds $10^{-1}$ at $t = 54 = 1.1t_{\rm adv}$, and on the Kolmogorov flow of Appendix~\ref{app:closure} at the same rate and Mach number the level-2 velocity error grows to order unity over 300 steps against the decaying reference, because the cubic term dropped by the truncation is no longer small against the viscous damping at this Mach number and viscosity. The favorable horizon of Secs.~\ref{sec:structure} and \ref{sec:coherent} is thus the horizon of a resolved, weakly compressible flow up to a fraction of the advective time, on which the dynamics is nonlinear in the sense that the observable is generated by mode coupling.

The growth and the saturation follow from the recurrence of the error. With $F_{\rm ref}$ the LB solution, whose update is exactly $\Lc F + \Qc(F\otimes F)$ for the equilibrium of Eq.~(\ref{eq:feq}), and $F_2$ the level-2 solution of Eq.~(\ref{eq:product}), the error $e_t = F_{\rm ref}(t) - F_2(t)$ obeys
\begin{equation}
e_{t+1} = \Lc e_t + S_t, \qquad S_t = \Qc\big(F_{\rm ref}\otimes F_{\rm ref} - G\otimes G\big),
\label{eq:recurrence}
\end{equation}
and by bilinearity and Eq.~(\ref{eq:shift}) the source is $S_t = \Qc(\delta_t\otimes(F_{\rm ref} - w)) + \Qc((G - w)\otimes\delta_t)$ with $\delta_t = F_{\rm ref} - G = e_t + d_t$, where $d_t = F_2 - G$ is the nonlinear correction that the lift carries. In the weighted norm $\|\Lc\|_g\le1$, so $\|e_t\|_g \le \sum_{k<t}\|S_k\|_g$ and $\|S_t\|_g\le2\|Q\|_g\|\delta_t\|_g\max(\|F_{\rm ref} - w\|_g, \|G - w\|_g)$, a source of order $\Ma$ times the departure of the nonlinear solution from the linear one. Along the reference run the source is dominated by $d_t$ rather than by $e_t$, by a factor of 3 to 18 over the first hundred steps at $\Ma = 0.087$ (2 to 18 over the run), and the ratio $\|S_t\|_g/(\|\delta_t\|_g\|{\rm flow}\|_g)$ stays between $0.01$ and $0.07$ against the bound $2\|Q\|_g = 2.1$, so the error does not feed itself and its growth follows the nonlinear correction, which is of order $\Ma^2$ at early times and saturates when $\delta_t$ saturates on the advective time scale. The summed bound is loose by a factor that grows from 5.5 at $t = 10$ to 120 at $t = 300$ because the homogeneous part of Eq.~(\ref{eq:recurrence}) is dissipative, and a bound that uses this dissipation is left open (Appendix~\ref{app:validation}).

\section{Resources and competing approaches}\label{sec:resources}
This section accounts for the preparation, evolution, amplification, estimation, synthesis and classical recovery of the task of Sec.~\ref{sec:problem} as a ledger, identifies the dominant costs, and compares the result with matched quantum constructions and with the classical evolutions of the same problem. The ledger separates three categories: operations implemented as circuits and verified by statevector simulation (the coupled-block step on $4\times4$, the preparation on $2\times2$ and $4\times4$, its exact amplification and the complete chain with the readout amplitude on $2\times2$); operations counted analytically from their operator lists with the rules of Appendix~\ref{app:circuit} (a $k$-controlled gate at $2(k-1)$ Toffolis with $k - 1$ clean workspace qubits, or at $8(k-3)$ Toffolis with one workspace qubit in any state for the reflections on the full register; uniformly controlled rotations, generic four-qubit unitaries and the Halmos dilations by the cited decompositions; every generic single-qubit gate as three $z$-rotations); and the synthesis of the $z$-rotations at the leading-order ancilla-free cost $3\log_2(1/\delta_r)$ T gates of \citet{RossSelinger2016}, with a Toffoli at 7 T gates. The repeat-until-success synthesis of \citet{Bocharov2015}, $1.15\log_2(1/\delta_r) + 9.2$ T gates in expectation with one ancilla and a mid-circuit measurement with feed-forward per rotation, is recorded in the Supplemental Material as a conditional alternative, valid only if measurement and feed-forward are available inside the coherent estimation circuit. The statevector verification certifies the operators the circuits implement, not the counts. Excluded are error correction, routing and classical control; the preparation of the reference state, the workspace and the phase register of the estimation are included.

\begin{figure*}[t]\centering\includegraphics[width=\textwidth]{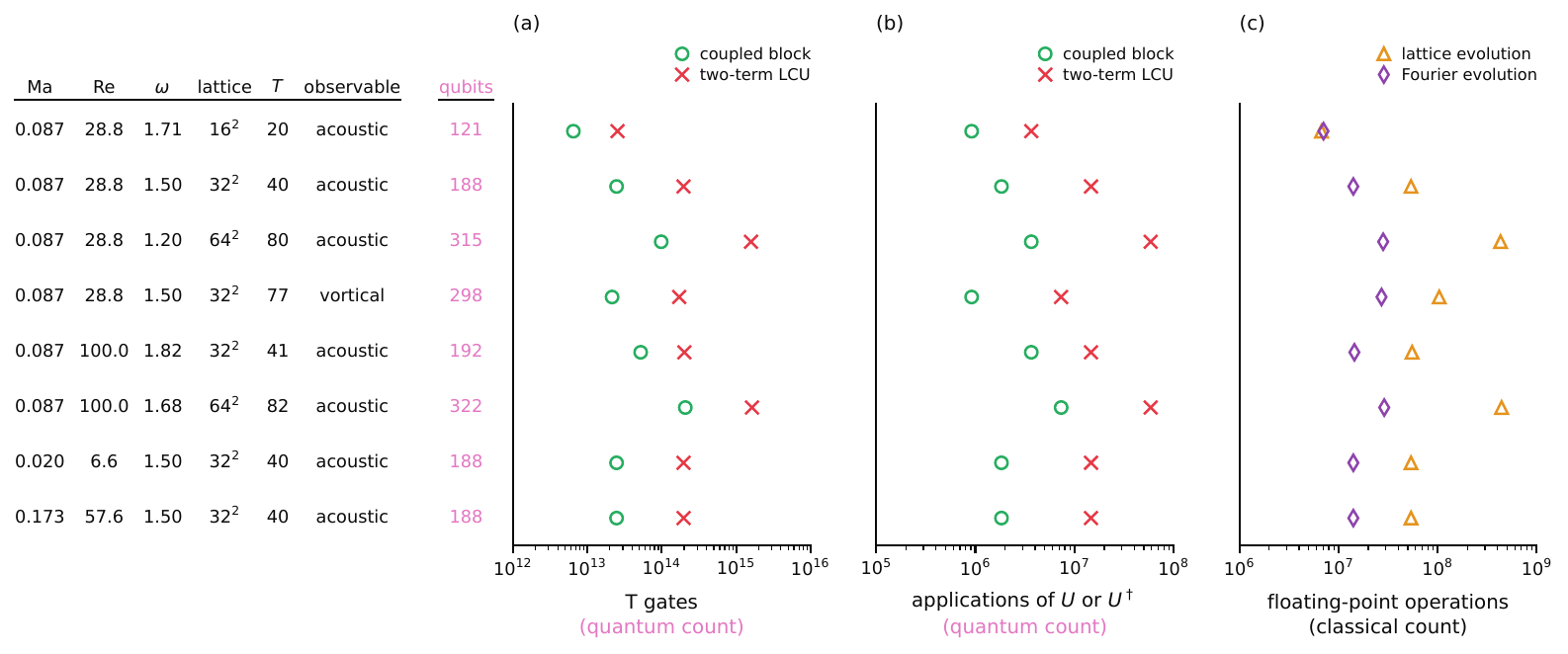}
\caption{Resources of the observable task at 10\% estimation precision and 95\% confidence for the seven acoustic cases and the $32\times32$ vortical case; the two remaining vortical cases are in the Supplemental Material. The table gives the parameters of each case and the number of logical qubits of the coupled-block encoding, which the LCU exceeds by two to four. (a) T-gate counts of the two encodings with the ancilla-free synthesis, a quantum count. (b) Number of applications of $U$ or $U^\dagger$ for the two encodings. (c) Floating-point operation counts of the implemented classical lattice and Fourier evolutions of the same problem to the same horizon, a classical count in a different unit; the ratio of the two counts is not a runtime.}\label{fig:resources}\end{figure*}

The error budget of the task has three parts, and the accuracy statement is made at the level of the observable. The kinetic truncation error of the level-2 dynamics at the horizon is $1.9$\% for the acoustic and $2.6$\% for the vortical observable at $\Ma = 0.087$ (Sec.~\ref{sec:accuracy}). It is a property of the classical level-2 model and is not reduced by the quantum resources. The estimation precision is $\epsilon_{\rm rel} = 0.1$ of the magnitude with a confidence of 95\%, from Eq.~(\ref{eq:M}) and the median of seven runs. The synthesis error is budgeted at a tenth of $\epsilon_{\rm rel}|A|$ per run, divided among the rotations of that run, so that it is a tenth of the estimation error. If the three parts add, the estimate lies within about 13 to 14\% of the nonlinear LB value with 95\% confidence, dominated by the estimation precision. The discretization error of the LB reference itself against the Navier-Stokes solution, $3\times10^{-4}$ in the viscosity on $32^2$ at $\omega = 1.5$ (Appendix~\ref{app:validation}), is a diagnostic of the reference and is not part of this budget. Table~\ref{tab:resources} lists the representative cases and Fig.~\ref{fig:resources} plots all of them. The full sweep over both observables, both encodings and both precisions, with the conditional synthesis alternative, is in the Supplemental Material, and the ledger of the two reference cases is in Appendix~\ref{app:resources}. For the acoustic observable on $32\times32$ at $\Ma = 0.087$ and $\mathrm{Re} = 28.8$, the optimal scale in the shifted encoding is $\lambda = 7.6$, the norm of the prepared state is $\|\psi_\lambda(0)\|_g = 55.5$, the joint success at $T = 40$ is $P = 5.0\times10^{-4}$, the combined amplitude of Eq.~(\ref{eq:A}) is $|A| = 3.8\times10^{-4}$, and Eq.~(\ref{eq:M}) gives $M = 2^{17}$, that is $M - 1$ applications of $Q$ per run, seven runs, and $N_U = 1.8\times10^{6}$ applications of $U$ or $U^\dagger$ on 188 logical qubits, 162 in the register of $U$, nine in the clean workspace pool and 17 in the phase register. Each application of $U$ has $1.5\times10^{5}$ Toffolis (1932 in the streaming and 726 in the collision stage per step, $4.7\times10^{4}$ in the amplified preparation against 1734 for one preparation, the preparation success $P_{\rm prep} = 4.3\times10^{-3}$ being amplified in twelve rounds) and $7.9\times10^{4}$ $z$-rotations, of which $6.0\times10^4$ come from the generic single-qubit gates of the moment transforms and the dilations. Each application of $Q$ adds two reflections of $1272$ Toffolis and the reference-state preparation and its inverse. The rotations of one run are synthesized at $\delta_r = 1.8\times10^{-16}$, or 157 T gates each. The total is $2.5\times10^{13}$ T gates, of which the rotations contribute $2.3\times10^{13}$ and the Toffolis $2.0\times10^{12}$. The conditional repeat-until-success count is $1.2\times10^{13}$. Without the amplification of the preparation the number of applications would be $2.8\times10^{7}$, and the incoherent alternative needs $6.7\times10^{8}$ repetitions. The two-term LCU at its own optimum, $\lambda = 41$, has $P = 6.2\times10^{-6}$, a factor of 81 below the block value, a combined amplitude nine times smaller, and needs eight times as many applications of $U$ and $2.0\times10^{14}$ T gates, a factor of 8.0 more, because the estimation cost scales as $1/|A|$. For the coupled-block encoding, the count at $\epsilon_{\rm rel} = 0.02$ is $1.1\times10^{14}$ T gates. The vortical observable, with its horizon $T = 77$ and its 2.4 times larger combined amplitude, needs $9.2\times10^{5}$ applications on 298 qubits and $2.1\times10^{13}$ T gates, fewer than the acoustic one because the larger amplitude outweighs the longer horizon. Over the seven cases and both observables the count ranges from $5.2\times10^{12}$ to $2.1\times10^{14}$ T gates on 121 to 539 qubits, growing with the lattice through $T\propto L$ at fixed physics and through the streaming count, and with the Reynolds number through the smaller step probability at larger $\omega$. The rotations dominate the T count in every case, and among the Toffolis the streaming stage dominates from $L = 8$ onward (Appendix~\ref{app:circuit}).

\begin{table*}[t]\centering
\caption{Quantum constructions of the lattice Boltzmann collision and of a run under matched assumptions. The status of each entry is derived here (D), reproduced here on the same operator and states (R) or reported in the cited work (L). $n = \log_2L$, $T$ the number of steps, $N$ the number of sites, $a = \omega_{\max}/\sqrt2$, $c = \omega/[4(2-\omega)]$ and $\alpha_C$ the Carleman order.}\label{tab:compare}
\scriptsize\setlength{\tabcolsep}{3pt}
\raggedcells\begin{tabular}{p{22mm}p{27mm}p{28mm}p{29mm}p{30mm}p{30mm}}\toprule
 & population CLB, sparse access \cite{SanavioSimonSucci2025} & population CLB, local \cite{BastidaZamora2026} & MRT-CLB, two-term LCU \cite{YaoSucci2026} & MRT-CLB, coupled block (this work) & history-state QLSA \cite{Jennings2026}\\\midrule
state and access & amplitude, $9N + 81N^2$, sparse-access oracles (L) & amplitude, pair sector in relative coordinates (L) & amplitude, relative coordinates, pair sector scaled by $\lambda$ (D) & the same, rest-shifted $g = f - w$ (D) & history state of the shifted variables over all steps (L)\\
collision block & $\alpha = 256$ with seven ancillas (L); $\alpha_s = 171$ (R) & SVD two-term LCU, $\alpha = \sigma_{\max}$ of the population stage, 9.4 at $\omega = 1.5$ (R) & diagonal plus coupling, $\alpha = 1 + a/\lambda$ (D) & three Halmos dilations, $\alpha = \|\Kc_\lambda\|_2 = 1 + c/\lambda^2 + O(\lambda^{-4})$ (D) & linear system, condition number growing with Re (L)\\
$p$ per step & $6\times10^{-5}$ (L); $1.5$ to $3.4\times10^{-5}$ (R) & order $10^{-2}$ at $N = 16$ (L); $1.1\times10^{-2}$ (R) & 0.82 ($\lambda = 10$), 0.98 ($\lambda = 100$), verified (D) & 0.985 ($\lambda = 10$), 0.9995 ($\lambda = 100$), verified (D) & not step based\\
$T$ steps with output & $\prod r_t/\alpha^2\times$ selection, $\approx(3\times10^{-5})^T$ (R) & $\approx(10^{-2})^T$ (R) & $\to e^{-2}/(a^2T^2N)$, shifted $N\to N\epsilon_0$ (D) & $\to e^{-1}/(2cTN)$, shifted $N\to N\epsilon_0$ (D) & not step based; speedup over classical evolution bounded above by $\mathrm{Re}^{3D/8}$, quantum query complexity estimated numerically as $\mathrm{Re}^{1.936}q_M$ in 2D with $q_M = O(\mathrm{Re}^{3/8})$ for the drag (L)\\
qubits & $\lceil\log_2D\rceil + 7$, $D = 9N + 81N^2$ (L) & $4n + 10$ plus LCU ancillas (L) & $4n + 14$ (D) & $4n + 15$, plus $3(T-1) + 2n_a + 2$ coherent, $2n - 1$ workspace and $\log_2M$ phase qubits (D) & 722 at Carleman order 10 (L)\\
gates per step or query & $O(\log^2N)$ (L) & $O(Q^3 + \log^2N)$ (L) & $36n^2 + 204n + 12$ Toffolis streaming, $8n + 194$ collision (D) & the same streaming, $8n + 686$ collision (D) & $10^6$ to $10^8$ per query (L)\\
preparation & not included & not included & explicit LCU for few-mode data, $(\|c\|_2/\|c\|_1)^2$, exact amplification (D) & the same, verified as circuits with the readout of Eq.~(\ref{eq:A}) (D) & LCNU loading of \citet{Demirdjian2026}, $O(\alpha_C^2Q^2)$ terms (L)\\
\bottomrule\end{tabular}
\end{table*}

\begin{table*}[t]\centering
\caption{Representative cases of the observable task in the shifted encoding with the coupled-block collision and the amplified preparation, and one case with the two-term LCU. $T$ is the horizon at the peak of the observable, $\lambda$ the optimal scale, $e_{\rm trunc}$ the relative kinetic truncation error of the observable at the peak, $\epsilon_{\rm rel}$ the estimation precision at 95\% confidence, $P$ the joint success probability of Eq.~(\ref{eq:total}) with the shifted norms, $|A|$ the combined amplitude of Eq.~(\ref{eq:A}), $M$ the resolution parameter of the estimation, with $M - 1$ applications of $Q$ per run, and the uses of $U$ count applications of $U$ or $U^\dagger$ over the seven runs; the T-gate count uses the ancilla-free synthesis rule. All cases start from the linearized equilibrium of the two-mode field.}\label{tab:resources}
\footnotesize\setlength{\tabcolsep}{3.5pt}\begin{tabular}{llcccccccccc}\toprule
case & observable & $T$ & $\lambda$ & $e_{\rm trunc}$ & $\epsilon_{\rm rel}$ & $P$ & $|A|$ & $M$ & uses of $U$ & qubits & T gates\\\midrule
Re 28.8, $16^2$ & $|\hat J_x(1,0)|$ & 20 & 7.2 & $1.8\times10^{-2}$ & 0.1 & $2.0\times10^{-3}$ & $7.2\times10^{-4}$ & $2^{16}$ & $9.2\times10^{5}$ & 121 & $6.5\times10^{12}$\\
Re 28.8, $32^2$ & $|\hat J_x(1,0)|$ & 40 & 7.6 & $1.9\times10^{-2}$ & 0.1 & $5.0\times10^{-4}$ & $3.8\times10^{-4}$ & $2^{17}$ & $1.8\times10^{6}$ & 188 & $2.5\times10^{13}$\\
Re 28.8, $64^2$ & $|\hat J_x(1,0)|$ & 80 & 7.7 & $1.9\times10^{-2}$ & 0.1 & $1.2\times10^{-4}$ & $1.9\times10^{-4}$ & $2^{18}$ & $3.7\times10^{6}$ & 315 & $9.8\times10^{13}$\\
Re 28.8, $32^2$ & $|\hat J_\perp(1,2)|$ & 77 & 10.6 & $2.6\times10^{-2}$ & 0.1 & $1.6\times10^{-4}$ & $8.9\times10^{-4}$ & $2^{16}$ & $9.2\times10^{5}$ & 298 & $2.1\times10^{13}$\\
Re 100, $32^2$ & $|\hat J_x(1,0)|$ & 41 & 13.7 & $2.2\times10^{-2}$ & 0.1 & $2.2\times10^{-4}$ & $2.1\times10^{-4}$ & $2^{18}$ & $3.7\times10^{6}$ & 192 & $5.2\times10^{13}$\\
Re 28.8, $32^2$ (LCU) & $|\hat J_x(1,0)|$ & 40 & 41.4 & $1.9\times10^{-2}$ & 0.1 & $6.2\times10^{-6}$ & $4.2\times10^{-5}$ & $2^{20}$ & $1.5\times10^{7}$ & 191 & $2.0\times10^{14}$\\
\bottomrule\end{tabular}

\end{table*}

The classical baselines are evaluated for the same problem from the same linearized initial state to the same horizons, and their operation counts are derived from the implemented methods (Appendix~\ref{app:resources}). The product property of Eq.~(\ref{eq:product}) evolves the level-2 dynamics with two matrix-vector products of the per-site collision matrix and the local quadratic term at $1316N$ flops per step, $5.4\times10^7$ for the acoustic and $1.0\times10^{8}$ for the vortical horizon, and never touches the $81N^2$ pair array. Because collision and streaming are translation invariant, a linear state with $M_0$ Fourier modes stays on those modes, and the quadratic term generates only sums of pairs of them, so the level-2 solution of a few-mode start lives on at most $M_0 + M_0^2$ wavevectors and is evolved in Fourier space at a cost that is independent of $N$ at a fixed number of steps. For the linearized two-mode start the nine initial wavevectors, the eight of the momentum field and the rest mode, generate 35 active ones, the linear model stays on the nine, the Fourier evolution reproduces both observables at their horizons and the full state to $10^{-13}$ of the lattice evolution, and it costs $1.4\times10^{7}$ flops for the acoustic and $2.7\times10^{7}$ for the vortical horizon at any lattice size. At fixed physics the number of steps to the horizon grows with $L$, so the Fourier cost grows linearly with $L$ while the quantum count grows faster through the streaming stage and the smaller step probability. T gates and floating-point operations are different units, and their ratio is not a runtime. The comparison is between the sizes of the two counts. The quantum count of the acoustic reference case exceeds the Fourier baseline by six orders of magnitude, the vortical one by a factor of $8\times10^{5}$, and the gap widens with $L$. Accordingly, no quantum advantage is demonstrated for the tested few-mode problem with any of the encodings compared here, and the performance for more general initial data, whose Fourier support fills the lattice and whose preparation is a cost that this study does not evaluate \cite{Demirdjian2026}, remains unresolved.

Table~\ref{tab:compare} places the constructions under matched assumptions and marks the status of every entry. The sparse-access and the local population constructions are evaluated on the same collision operator and the same states, the two MRT-CLB encodings differ only in the collision block, and the history-state approach of \citet{Jennings2026} is listed with its reported numbers, since it replaces the step-by-step evolution by one linear system whose condition number grows with the Reynolds number. Its entries are statements of the quantum query complexity and of an upper bound on the possible speedup over classical evolution, not of a demonstrated speedup, and they are not reproduced here. The per-step probabilities of the population constructions are set by the Euclidean metric of the encoding and by the density of the population basis, in which relaxation is not a contraction, whereas in the weighted Hermite representation the dissipation removes the relaxed non-equilibrium norm, up to the cross term of Eq.~(\ref{eq:balance}), and the normalization is set by the block structure of Theorem~\ref{th:structure}. The coupled-block encoding changes the multi-step cost from $T^2$ to $T$ in the joint probability and from $T$ to $\sqrt T$ in the number of applications of $U$, and the rest shift removes the factor $1/\epsilon_0$ from both. Attaining the norm of the specified encoded operator is a statement about that operator, not about the optimality of the algorithm as a whole. Neither changes the comparison with the classical baselines of the few-mode problem.

\section{Generality, limitations and conclusions}\label{sec:scope}
For the D2Q9 lattice with MRT collision in the weighted Hermite basis, the cost of a coherent level-2 Carleman evolution is set by the conservation and relaxation structure of the collision. The conserved momenta feed the three relaxing second-order moments through a coupling whose rows are orthogonal, so that after a rotation of the conserved-pair slots it consists of three $2\times2$ blocks, and the norm of the scaled collision stage is $1 + c/\lambda^2 + O(\lambda^{-4})$ at fixed rate with $c = \omega/[4(2-\omega)]$. A block encoding built from three Halmos dilations attains this norm, and its step success probability, 0.985 at $\lambda = 10$ on the $4\times4$ Taylor-Green state, is verified by statevector simulation. The dissipation itself removes only the relaxed non-equilibrium norm, up to the cross term of Eq.~(\ref{eq:balance}). The step factors telescope to the exact complete-run probability, and optimizing the pair-sector scale for the specified horizon converts the normalization-selection trade-off into polynomial success scaling for the structured encodings of this representation, whereas the specified population constructions lose a constant factor per step on the same trajectory. The coupled block adds a factor linear in $T$, with its joint probability approaching $e^{-1}/(2cTN)$ at the optimal scale for large $T$ in the near-rest model, and the rest-shifted encoding multiplies the number of sites by the flow fraction $\epsilon_0$ of the norm. The same block structure holds on the D3Q19 and D3Q27 lattices, with six blocks and the same coefficient, so the statement is a property of the conserved and relaxing modes of a kinetic scheme whose coupling rows are orthogonal, which is the class the theorem covers, rather than of the two-dimensional lattice. The three-dimensional dynamics itself was not simulated.

The dynamical scope of these results is the horizon of a resolved, weakly compressible flow up to a fraction of the advective time. On the two-mode flow the level-2 velocity error is of order $\Ma^2$ at early times and grows on the advective time scale. At $\Ma = 0.087$ and $\omega = 1.5$ it is 1.4\% at half the advective time, independently of the lattice resolution at fixed Reynolds and Mach numbers, and 5.6\% after three advective times. The two nonlinearly generated modes, the acoustic mode that the vortical interaction radiates and the vortical mode that it feeds, are reproduced at their peaks within 2 and 3\% at $\Ma = 0.087$. At $\Ma = 0.173$ and $\omega = 1.9$ the truncation fails within one advective time, and at $\mathrm{Re} = 100$ the vortical peak lies beyond the window. The favorable cost and the accurate dynamics therefore coincide on a horizon on which the observables are nonlinear, but this horizon is short, and extending it requires the third Carleman level at a state dimension of order $N^3$. The probability curves of Sec.~\ref{sec:coherent} extend beyond this horizon and describe survival, not accuracy.

For the specified benchmark instances, the resource estimate provides a complete observable-level accounting of the coherent Carleman route, calibrated with the level-2 benchmark amplitudes used to set the amplitude-estimation length. Estimating the magnitude of the acoustic observable on the $32\times32$ lattice to 10\% relative precision with 95\% confidence requires $1.8\times10^{6}$ applications of a circuit of $1.5\times10^{5}$ Toffolis and $7.9\times10^{4}$ rotations on 188 logical qubits, $2.5\times10^{13}$ T gates in total under the ledger of Appendix~\ref{app:resources}, the vortical observable $2.1\times10^{13}$, and the counts grow with the lattice side. The kinetic truncation error of 2 to 3\% at the horizon is a separate, classical part of the error budget. The same few-mode problem is evolved classically in Fourier space at a cost that does not depend on the lattice size at a fixed number of steps, $1.4\times10^7$ flops for the acoustic horizon, so no quantum advantage is demonstrated for the tested few-mode problem with any of the encodings compared here, and the performance for more general initial data remains unresolved. The question of an advantage for coherent Carleman lattice Boltzmann is thus a question about initial data whose Fourier support fills the lattice, whose preparation is a cost this study has not evaluated beyond the few-mode case, and about observables that remain accurate over the truncation horizon. The present results reduce the multi-step cost by a factor that grows with the number of steps and remove the rest-state penalty, and they leave the preparation of generic data and the truncation horizon as the two open costs.

Several limitations qualify these statements. The counts exclude error correction, routing and classical control, and rest on the counting and synthesis rules stated in Appendices~\ref{app:circuit} and \ref{app:resources}, which are standard decompositions rather than constructed circuits for the operations on the moment registers and for the reflections of the estimation. The amplification of the preparation is verified as circuits, the reflections and controlled powers of the outer estimation are counted, not implemented, and the reported fidelities are the authors' statevector records on lattices of $4\times4$ sites and smaller. The estimation lengths $M$ and the run counts are instance-dependent resource estimates calibrated with the benchmark amplitude $|A|$ of the classical level-2 calculation of the same instance. No stopping rule for an output that is not known in advance has been implemented or costed, and the magnitude, not the phase, of the Fourier coefficient is estimated. The bound of Eq.~(\ref{eq:recurrence}) is loose, and a bound that uses the dissipation of the linear part is not derived. The accuracy window was established on one family of flows in two dimensions without boundaries or forcing, and the level-3 checks were limited to lattices small enough for a dense stepper. Future work should address the preparation of generic initial data, boundaries and forcing, the third Carleman level, and the hybrid combination with deterministic relaxation \cite{KhanSucciYao2026}, for which the block structure identified here provides the sparse representation of the nonlinear component. For the D2Q9 scheme considered here, the cost of the coherent Carleman route is now tied to three relaxing moments and one flow fraction, and whether an advantage over classical evolution exists for initial data beyond the few-mode family tested here is not settled by the present results.

\begin{acknowledgments}
The research is financed by the Swedish Transport Administration in the project ``GEneric Multidisciplinary optimization for sail INstallation on wInd-assisted ships'' (GEMINI, Grant No.\ TRV~2023/32107). The data that support the findings of this study, and the scripts that produce every number, table and figure, are available from the authors upon reasonable request.
\end{acknowledgments}

\appendix

\section{Closure of the level-2 moment hierarchy under streaming}\label{app:closure}
This appendix examines whether the quadratic sector of the moment hierarchy can be closed locally or requires the pair array, and records the evidence that fixed the pair-array formulation of Sec.~\ref{sec:problem}. A local formulation would keep, per site, the nine moments and the three products $q = JJ^\Tr = (J_x^2, J_xJ_y, J_y^2)$ of the conserved momentum, twelve components, because the quadratic part of the equilibrium depends on $q$ alone and $q$ is collision invariant. This construction closes exactly for the collision step, in line with the exactness of the level-2 lift for collision-only updates found by \citet{ItaniSucci2022} on the D1Q3 lattice. With $\rho = 1$ inside the quadratic term the per-site update of $(m, q)$ is the linear map with blocks $C_1$, $\Omega B$ and $I_3$, where $Bq$ is the quadratic part of the equilibrium moments, and this was verified to machine precision. After streaming, the momentum at $x$ and the quadratic sector are
\begin{equation}
\begin{aligned}
J(x,t+1) &= \sum_i c_i\, f_i'(x - c_i),\\ q(x,t+1) &= J(x,t+1)\,J(x,t+1)^\Tr,
\end{aligned}
\label{eq:qstream}
\end{equation}
a linear combination of the post-collision moments of the nine neighbors and a bilinear form in them. To identify the terms generated by streaming, $q(x,t+1)$ was expanded exactly with bookkeeping parameters for the orders in the Mach number, in the non-equilibrium moments and in the gradient. The parameters are $J = \Ma\,u$, $m_s = (Bq)_s + \Ma\,\delta\,n_s$ for the second-order moments and $m_h = \Ma\,\delta\,h$ for the ghosts, and a factor $\kappa$ for every difference between a neighbor value and the value at $x$. The result has 12314 monomials in fifteen classes according to the kinds of factors they contain. Only the class of same-site products $u_au_b$, forty monomials, lies in the span of the local state $\{m(y), q(y)\}$. Table~\ref{tab:classes} lists the classes that enter at first order in the gradient with the continuum form of their leading terms; the remaining ten classes are products of two transport terms and enter at second order.

\begin{table*}[t]\centering\small
\caption{Classes of monomials contributing to $q(x,t+1)$ through first order in the gradient, with their orders in $(\Ma, \delta)$, same-site and cross-site counts, and leading continuum forms. $\sigma'$ is the post-collision non-equilibrium stress.}\label{tab:classes}
\begin{tabular}{llccl}\toprule
class & order & same site & cross site & continuum form\\\midrule
$u\cdot u$ & $(2,0)$ & 40 & 126 & none at first order: linear transport gives $J + O(\kappa^2)$\\
$r\cdot u$ & $(3,0)$ & 28 & 164 & $2\,\mathrm{Sym}\,J\otimes(-c_s^2\nabla\rho)$, pressure work\\
$u\cdot u\cdot u$ & $(3,0)$ & 56 & 416 & $2\,\mathrm{Sym}\,J\otimes(-\nabla\!\cdot\!(JJ))$, advective flux\\
$n\cdot u$ & $(2,1)$ & 112 & 656 & $2\,\mathrm{Sym}\,J\otimes(\nabla\!\cdot\!\sigma')$, stress work\\
$h\cdot u$ & $(2,1)$ & 160 & 832 & flux of the ghost moments into $J$\\\bottomrule
\end{tabular}\end{table*}

The continuum forms are the terms of
\begin{equation}
\begin{aligned}
\frac{\mathrm{d}(JJ^\Tr)}{\mathrm{d}t} &= 2\,\mathrm{Sym}\big(J\,\partial_tJ^\Tr\big),\\
\partial_tJ &= -c_s^2\nabla\rho - \nabla\!\cdot\!(JJ) + \nabla\!\cdot\!\sigma .
\end{aligned}
\label{eq:qdot}
\end{equation}
At first order in the gradient the quadratic sector evolves by the work of the pressure gradient, of the advective flux and of the viscous stress on the momentum flux. Each of these terms is a product of $J$ at one site with a difference of a conserved or a non-conserved moment across sites, and none has an image in the local span. The transport of $q$ itself is the exception. The momentum flux of the linear equilibrium is pure pressure, so the linear transport of $J$ is $J + O(\kappa^2)$ and the $u\cdot u$ class has no first-order term. Accordingly, any local rule that reduces to the identity on uniform fields transports $q$ to second order. Two such rules were tested, the identity rule $q(x,t+1) = q(x)$ and the stencil
\begin{equation}
q(x,t+1) = \sum_i 3w_i\,\mathrm{Sym}\big(c_ic_i^\Tr\,q(x - c_i)\big),
\label{eq:stencil}
\end{equation}
the analogue of the momentum transport. On a smooth field their residuals decrease as $\kappa^{1.98}$ under lattice refinement. For lattice-scale roughness they are first order, with the leading first-order discrepancy given by the antisymmetric part of $J\otimes\nabla J$.

\begin{figure*}[t]\centering\includegraphics[width=0.9\textwidth]{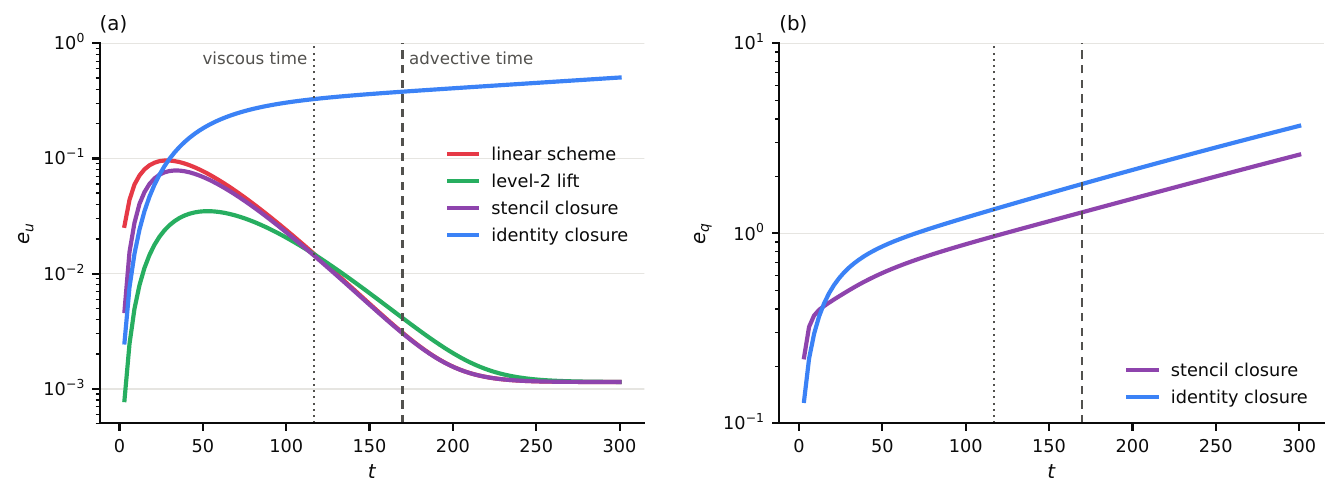}
\caption{The local closures depart from the pair-space reference on the viscous time scale. Kolmogorov flow of Eq.~(\ref{eq:kolm}) on the $32\times32$ lattice with the amplitude reduced to $\Ma = 0.052$, $\omega = 1.5$. (a) Velocity error $e_u$ against the nonlinear reference, as a function of the time step $t$, for the linear scheme, the level-2 population lift and the two local closures (identity and stencil); the viscous (dotted) and advective (dashed) times are marked in both panels. (b) Error $e_q$ of the local quadratic sector against $JJ^\Tr$ of the reference.}\label{fig:closure}\end{figure*}

Figure~\ref{fig:closure} shows the behavior of local schemes built on these rules on the Kolmogorov flow of \citet{SanavioSucci2024},
\begin{equation}
\begin{aligned}
f_i = w_i\big[1 &+ A_x\cos(2\pi k_xy/L)\,c_{i,x}\\ &+ A_y\cos(2\pi k_yx/L)\,c_{i,y}\big],
\end{aligned}
\label{eq:kolm}
\end{equation}
with $k_x = 1$, $k_y = 4$ on the $32\times32$ lattice, $A_y = 2A_x/3$, and the Mach number $\Ma = U_0/c_s$ with $U_0 = A_x/3$, so that the amplitudes $A_x = 0.3$, $A_y = 0.2$ of that paper correspond to $\Ma = 0.17$. The closure test uses the amplitude reduced to $\Ma = 0.052$ and $\omega = 1.5$. The population level-2 lift has $e_u = 8\times10^{-4}$ at $t = 3$ compared with $2.6\times10^{-2}$ for the linear scheme, and is five to thirty times more accurate over the first twenty steps. The two errors meet near the viscous time and decay together to $1.1\times10^{-3}$ after the advective time $1/(\kappa U_0) = 170$. The two local schemes have the same moment-sector update, exact for one step, and differ only in $q$. The identity closure retains the initial quadratic-stress contribution after the flow has decayed, and its velocity error reaches 0.5 by $t = 300$. The stencil rule diffuses the spatial variation of $q$ at the lattice rate $1/6$ rather than the viscous rate, so that within tens of steps only the lattice mean of $q$ remains, which the stencil conserves and which exerts no force. The scheme then reduces to the linear one, whose velocity error it reproduces to three digits at all late times. In both schemes the quadratic sector departs from $JJ^\Tr$ on the viscous time scale (Fig.~\ref{fig:closure}b), and the populations acquire a pressure contribution that departs from the reference, $e_f = 0.07$ to $0.5$ compared with $10^{-3}$ for the level-2 lift, where $e_f = \|F - F_{\rm ref}\|/\|F_{\rm ref} - w\|$ is the population error relative to the flow part. On the Taylor-Green vortex the pressure work and the advective flux cancel, and only the viscous decay of $q$ remains. The velocity is then unaffected, because the quadratic stress is a gradient, but the identity closure reaches $e_q = e^2 - 1 = 6.4$ at $t = 2t_{\rm visc}$, its undecayed value, and $e_f = 0.2$.

The cross-site pairs contain the work terms of the momentum flux, and the pair array stays. With the full pair array retained, the moment lift is the weighted population lift in another orthogonal basis, $\mathcal{T}\Cc_g\mathcal{T}^\Tr$, and a similarity transform of the plain population lift, with identical dynamics, identical truncation error and the same validity window of the level-2 truncation. For the $\omega = 1.9$, $\Ma = 0.17$ case of \citet{SanavioSucci2024}, the level-2 velocity error grows to order unity against the decaying reference over 300 steps on the $32\times32$ lattice, and the moment representation shows the same behavior, as expected from the basis equivalence. The twelve local components remain an exact collision block within the full pair representation, and this block is where the structure of Sec.~\ref{sec:structure} resides.

\section{Representations of the collision stage}\label{app:representations}
The rows of $H$ are the Gram-Schmidt orthonormalization of the monomials $1$, $c_x$, $c_y$, $c_x^2 + c_y^2$, $c_x^2 - c_y^2$, $c_xc_y$, $c_xc_y^2$, $c_x^2c_y$ and $c_x^2c_y^2$ in the weighted inner product, so that $HWH^\Tr = I$ and the first three moments are $\rho$, $\sqrt3J_x$ and $\sqrt3J_y$. With $E_1 = W(\mathbf{1}\mathbf{1}^\Tr + 3CC^\Tr)$, $C$ the matrix of lattice velocities, the linear equilibrium map is $HE_1H^{-1} = \mathrm{diag}(1,1,1,0,\dots,0)$. The quadratic part of the equilibrium, $\tfrac92(c_i\!\cdot\!J)^2 - \tfrac32J\!\cdot\!J$, projects onto the trace, the deviator and the shear moment only, which gives the coupling block of Eq.~(\ref{eq:csite}) with the three orthogonal rows $(1,1,0,0)$, $(1,-1,0,0)$ and $(0,0,1,1)$ of norm $\sqrt2$, scaled by $\omega_s/2$. Its singular value decomposition has the natural rotation of Sec.~\ref{sec:structure} as right singular vectors and the antisymmetric slot in its null space.

The diagnostics of a stage $A$ are the row and column sparsities $s_r$, $s_c$ (largest number of nonzeros per row and column, entries below $10^{-13}$ of the largest counted as zero), the largest entry $\max|a_{ij}|$, the sparse-access subnormalization $\alpha_s = \sqrt{s_rs_c}\max|a_{ij}|$ \cite{SanavioSimonSucci2025}, the extreme singular values, and the number of Pauli strings with nonzero coefficient in the decomposition of $A$ zero-padded to $2^n\times2^n$, reported as a filling fraction of $4^n$, together with the one-norm $\alpha_P$ of the coefficients, which is the subnormalization of a Pauli linear combination of unitaries. The success probability of a block encoding with subnormalization $\alpha$ on a stated state $\psi$ is $p = \|A\psi\|^2/(\alpha^2\|\psi\|^2)$, Eq.~(\ref{eq:p}).

\begin{figure*}[t]\centering\includegraphics[width=0.9\textwidth]{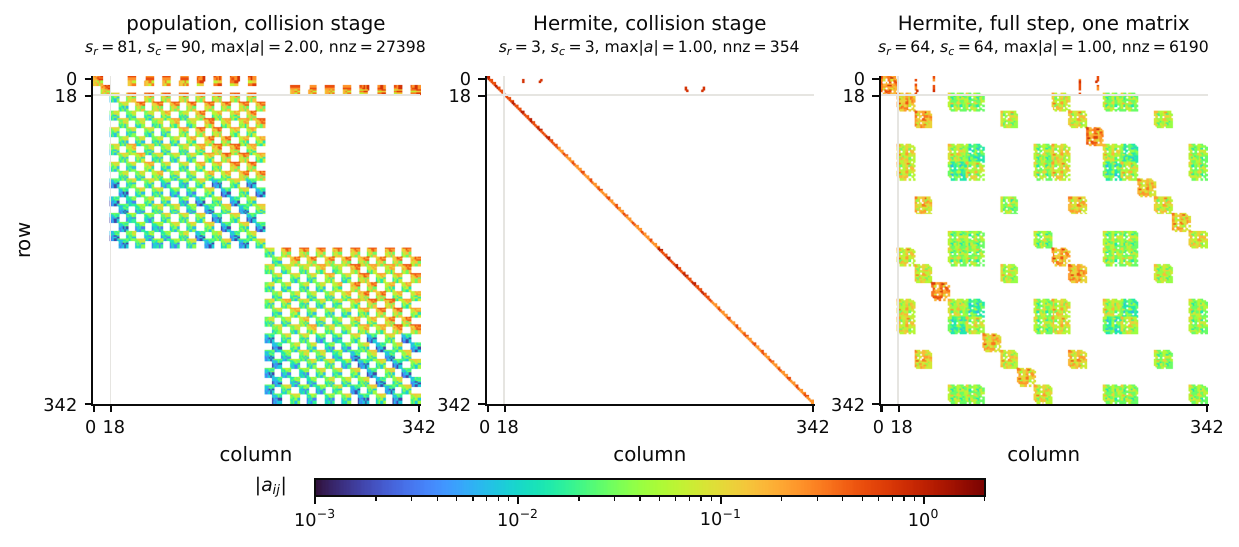}
\caption{Sparsity and magnitude of the level-2 collision stage on a two-site lattice (dimension 342, BGK $\omega = 1.5$) in the population representation with plain encoding and in the weighted Hermite moment representation, and of the assembled full-step matrix.}\label{fig:sparsity}\end{figure*}

Table~\ref{tab:structure} and Fig.~\ref{fig:sparsity} give the level-2 collision stage $\Kc$ of the Hermite MRT scheme in these four representations on a two-site lattice. The sparsities, the entries and the singular values are independent of the lattice size, because the stage is block diagonal over site pairs. The Pauli filling and the one-norm $\alpha_P$ depend on the binary embedding and the padding and are reported for the two-site instance. In the population representation with plain encoding, the construction of \citet{SanavioSimonSucci2025}, the level-1 block $L$ is dense with nine nonzeros per row, the stage has sparsities 81 and 90, the coupling $Q$ has entries up to 2.0 at $\omega = 1.5$, and the singular values span 0.05 to 9.4 because the Euclidean metric on populations is not the one in which relaxation is a contraction. The weighted encoding $f/\sqrt w$ alone moves the singular values into $[0.25, 1.5]$ and reduces $\alpha_s$ from 171 to 28.5, a factor of six that is due to the metric. The Hermite basis reduces the sparsity to three in rows and columns and the largest entry to one, $\alpha_s = 3$, and its Pauli decomposition fills 2.5\% of all strings on two sites compared with 89\% in the population representation, with a one-norm of the coefficients of 16 compared with 401. The full step written as one matrix is denser in moment space than in population space, because streaming couples each moment to the nine moments of nine neighbors, with $s = 59^2$ on $3\times3$ and $4\times4$ compared with 81 and 90 in population space. The gain is confined to the factorized step of Eq.~(\ref{eq:lift}), permutation streaming, orthogonal transform and sparse collision.

\begin{table}[t]\centering\scriptsize\setlength{\tabcolsep}{2.5pt}
\caption{Level-2 collision stage of the Hermite MRT scheme for BGK relaxation at $\omega = 1.5$ in four representations, with the row and column sparsities, the largest entry, the sparse-access subnormalization $\alpha_s$, the extreme singular values, and the Pauli filling and one-norm $\alpha_P$ of the Pauli coefficients on the two-site lattice. The representations are the population basis with plain and weighted encoding and the Euclidean and Hermite moment bases with plain and weighted encoding.}\label{tab:structure}
\begin{tabular}{lcccccccc}\toprule
repr. & $s_r$ & $s_c$ & $\max|a|$ & $\alpha_s$ & $\sigma_{\max}$ & $\sigma_{\min}$ & fill. & $\alpha_P$\\\midrule
pop., plain & 81 & 90 & 2.00 & 171 & 9.39 & 0.052 & 0.89 & 401\\
pop., weighted & 81 & 90 & 0.33 & 28.5 & 1.50 & 0.25 & 0.68 & 112\\
mom., Eucl., plain & 4 & 9 & 4.50 & 27 & 9.39 & 0.052 & 0.18 & 190\\
mom., Hermite, wtd. & 3 & 3 & 1.00 & 3 & 1.50 & 0.25 & 0.025 & 15.8\\\bottomrule
\end{tabular}\end{table}

\section{Proof of Theorem~\ref{th:structure} and extensions}\label{app:proof}
Under assumption (i) the level-1 block $C_1$ is diagonal in the metric with entries in $[-1,1]$, strictly inside $(-1,1)$ on the relaxing moments coupled by the quadratic term, and the pair block $C_1\otimes C_1$ is diagonal with products of these entries. Under (ii) the quadratic coupling of the lifted collision acts from the same-site products of conserved moments, which in relative coordinates occupy the slice $r = 0$ of the pair register, into the coupled relaxing moments, and from nowhere else. The conserved-pair slots are products of two conserved moments and are therefore left unchanged by $C_1\otimes C_1$. On the subspace spanned by the coupled moments $s$ and the conserved-pair slots the stage is thus the block upper triangular matrix with $\mathrm{diag}(1-\omega_s)$ in the upper left, the identity in the lower right and the coupling $C_{\rm site}/\lambda$ in the upper right. In general the singular value decomposition $C_{\rm site} = U_{\rm out}\Sigma V^\Tr$ rotates the output space as well, and $U_{\rm out}$ mixes relaxing moments with different rates, so that no rotation of the pair slots alone produces independent blocks. Assumption (iii) states that the rows of $C_{\rm site}$ are orthogonal, which is the same as $C_{\rm site}C_{\rm site}^\Tr$ being diagonal, so that $U_{\rm out}$ is the identity up to signs and the decomposition reads $C_{\rm site} = \Sigma V^\Tr$ with $V$ orthogonal on the pair slots and $\Sigma$ the matrix of singular values $\sigma_s$ on the diagonal. Rotating the pair slots by $W_n = V^\Tr$ leaves the identity block unchanged and turns the coupling into $\Sigma/\lambda$, so that on this subspace the matrix is the direct sum of the blocks $B_s$ of Eq.~(\ref{eq:block}) over the coupled moments and of the identity on the remaining conserved-pair slots. On the rest of the lifted state the stage remains the diagonal of assumption (i), with entries $1 - \omega_s$ and their products, not the identity. For D2Q9 the singular vectors are the natural rotation stated after the theorem, because the three rows of Eq.~(\ref{eq:csite}) are orthogonal with norm $\omega_s/\sqrt2$, and the antisymmetric slot spans the null space. Assumption (iv) is used only to make the norm ratio of a step equal to that of the collision stage, so that Eq.~(\ref{eq:total}) telescopes. The norm of a direct sum is the largest norm of its summands, which proves $\|\Kc_\lambda\|_2 = \max_s\sigma_{\max}(B_s)$ since the diagonal entries have modulus at most one and $\sigma_{\max}(B_s)\ge1$.

With $p = 1 - \omega_s$ and $q = \sigma_s/\lambda$ the largest singular value of $B_s$ is
\begin{equation}
\sigma_{\max}^2 = \frac{1 + p^2 + q^2 + \sqrt{(1 + p^2 + q^2)^2 - 4p^2}}{2},
\label{eq:sigma2}
\end{equation}
whose series in $q$ at fixed $|p| < 1$ is even, $\sigma_{\max} = 1 + q^2/[2(1-p^2)] + q^4c_4'(p) + O(q^6)$. This gives $c_s = \sigma_s^2/[2\omega_s(2-\omega_s)]$, which for $\sigma_s = \omega_s/\sqrt2$ is Eq.~(\ref{eq:c}), and substituting $p = 1 - \omega$ and $q^2 = \omega^2/(2\lambda^2)$ in the next term gives the fourth-order coefficient $c_4 = -\omega(3\omega^2 - 6\omega + 4)/[32(2-\omega)^3]$ of $\|\Kc_\lambda\|_2 = 1 + c/\lambda^2 + c_4/\lambda^4 + O(\lambda^{-6})$. The expansion to second order matches the exact norm to $6\times10^{-5}$ at $\lambda = 10$ and $7\times10^{-9}$ at $\lambda = 100$ for $\omega = 1.5$. At $\lambda = 10$ it gives 1.048 against the exact 1.036 for $\omega = 1.9$, and 1.098 against 1.051 for $\omega = 1.95$, because $c$ diverges as $\omega\to2$. At the endpoint $p = -1$, that is $\omega_s = 2$, the square root in Eq.~(\ref{eq:sigma2}) equals $q\sqrt{4 + q^2}$ and is not analytic in $q^2$, so that $\sigma_{\max} = 1 + q/2 + O(q^2)$ and the largest singular value acquires an $O(1/\lambda)$ correction. The expansion of Eq.~(\ref{eq:sigma}) therefore requires the strict contractivity $|1 - \omega_s| < 1$ of assumption (i) on the coupled relaxing moments and is not uniform as $\omega_s\to2^-$. All costs of the main text use the exact singular value. The quadratic order of the correction does not depend on assumption (iii). For a strictly contractive diagonal relaxing block $D$ and an arbitrary coupling $C/\lambda$ from the conserved slots, the largest singular value of the stage is $1 + \|C^\Tr(I - D^2)^{-1}C\|_2/(2\lambda^2) + O(\lambda^{-4})$, of which the expansion above is the scalar case, verified numerically for random contractive $D$ and non-orthogonal $C$. Assumption (iii) turns this generic spectral-gap scaling into the exact direct sum, the attained norm and the three-dilation circuit.

For the three-dimensional lattices the same construction from the monomials up to fourth order gives six second-order moments $r^2$, $3x^2 - r^2$, $y^2 - z^2$, $xy$, $yz$ and $xz$, and the coupling matrix from the six symmetric products of the momentum components, taken in the orthonormal directions $\ket{aa}$ and $(\ket{ab} + \ket{ba})/\sqrt2$ on which a product state has coordinates $m_a^2$ and $\sqrt2\,m_am_b$, into these moments has six orthogonal rows of norm $\omega_s/\sqrt2$ in the ordered pair space, verified numerically for D3Q19 and D3Q27. The coupling from products involving a non-conserved moment vanishes to $1.5\times10^{-15}$ on D3Q19 and $3\times10^{-16}$ on D3Q27. Assumption (iii) therefore holds there, the corollary follows from the same argument, and the norm of the scaled stage, $1.504861$, $1.076738$, $1.007436$, $1.000833$ and $1.000075$ at $\lambda = 1$, 3, 10, 30 and 100 for $\omega = 1.5$, coincides on the three lattices to six digits.

\section{Quantum construction}\label{app:circuit}
\begin{figure}[t]\centering\includegraphics[width=0.92\columnwidth]{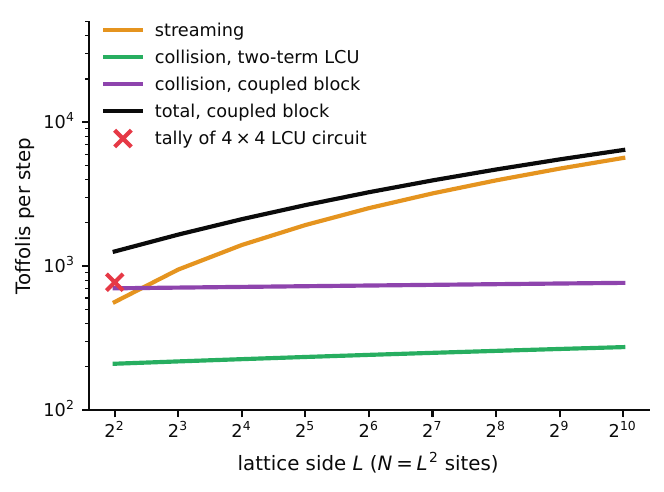}
\caption{Toffolis per step against the lattice side for the streaming stage, the two-term LCU collision stage of the earlier circuit, the coupled-block collision stage and the total of the coupled-block step. The cross is the tally of the complete $4\times4$ LCU circuit, 774 Toffolis, which equals the sum of its streaming and collision counts. Streaming dominates from $L = 8$ onward and sets the $O(\log^2N)$ scaling.}\label{fig:cost}\end{figure}

\emph{Registers and stages.} The register layout follows the relative-coordinate encoding of \citet{BastidaZamora2026}. The pair sector is stored as $\psi[x, i, r, j] = \lambda X[(x,i),(x+r,j)]$ and the level-1 sector in the padding slot $j = 15$ of the second moment register with $r = 0$, so that no level qubit is needed. The registers are the site $x$ and the relative position $r$, $2n$ qubits each, two moment registers of four qubits and the ancillas. One step consists of the orthogonal transform $(U\oplus I_7)\otimes(U\oplus I_7)$ on the moment registers, the collision stage, the inverse transform and the streaming $x \leftarrow x + c_i$, $r \leftarrow r + c_j - c_i$ by controlled incrementers, at $36n^2 + 204n + 12$ Toffolis. Algorithm~\ref{alg:step} lists the stages of the coherent protocol, and Fig.~\ref{fig:cost} shows the Toffolis per step against the lattice side for the streaming stage, the two collision stages and the coupled-block total, with the tally of the complete $4\times4$ LCU circuit, 774 Toffolis, reproduced by the sum of the streaming and collision formulas. Streaming dominates from $L = 8$ onward and sets the $O(\log^2N)$ scaling of the step.

\begin{algorithm}[t]
\caption{Coherent MRT-CLB protocol on the site register $x$, the relative-position register $r$, the moment registers $m_1$ and $m_2$, the preparation ancillas and the encoding ancillas $(a_1, a_2, b)$ of each step. Gate counts (right) are Toffolis for $L = 2^n$ unless stated otherwise; $k$ is the number of amplification rounds of the preparation and $N_U$ the number of uses of $U$.}\label{alg:step}
\SetAlgoLined\DontPrintSemicolon\SetKwComment{Comment}{}{}\SetCommentSty{textrm}
\textbf{Preparation} $U_{\rm prep}$: $\psi[x,i,0,15] \leftarrow g_0[x,i]$ by a linear combination of $M$ unitaries; $\psi[x,i,r,j] \leftarrow \lambda g_0[x,i]\,g_0[x+r,j]$ by a second copy on the pair branch; exact amplification with $2k+1$ preparation circuits\Comment*[r]{$4.7\times10^4$ for the task}
\For{$t = 1$ \KwTo $T$}{
\textbf{Transform} $(U\oplus I_7)\otimes(U\oplus I_7)$ on $m_1$, $m_2$\Comment*[r]{200 CNOT}
\textbf{Collision} flags $f$ (slice $r = 0$), $g$, $h_1$, $h_2$; rotation $W_n$ of the conserved-pair slots; three Halmos dilations of $B_s/\alpha$ with the fresh ancilla $b_t$; diagonal on $(a_{1,t}, a_{2,t})$; $W_n^\Tr$\Comment*[r]{$8n + 686$}
\textbf{Inverse transform} on $m_1$, $m_2$\Comment*[r]{200 CNOT}
\textbf{Streaming} $x \leftarrow x + c_i$, $r \leftarrow r + c_j - c_i$\Comment*[r]{$36n^2 + 204n + 12$}
}
\textbf{Output recovery} amplitude estimation of $|A|$ with $Q = -US_0U^\dagger S_\chi$, $S_\chi = VS_0V^\dagger$ with $V$ the reference-state preparation, both reflections controlled by the phase register; median of $r$ runs\Comment*[r]{$N_U = r(2M-1)$ applications of $U$, $U^\dagger$}
\end{algorithm}

\emph{Collision stage.} The coupled-block collision stage consists of the flag $f$ of the slice $r = 0$, computed and uncomputed by a $2n$-controlled X; three flags $g$, $h_1$, $h_2$ that mark the padding slot and the conserved moments of the two registers, computed from the moment registers by two-level operations and uncomputed before any operation that moves amplitude between moments, since a flag computed from a register that is later modified would otherwise remain entangled with the wrong branch; two uniformly controlled rotations of the ancillas $a_1$ and $a_2$ that block encode the diagonal entries of modulus at most one; the rotation $W_n$ on the four conserved-pair slots of the slice, controlled by $f$; three Halmos dilations of $B_s/\alpha$ on the pairs $\{(s,15), u_s\}$ of the two moment registers with the ancilla $b$, controlled by $f$; a rotation of the ancilla for the decoupled antisymmetric slot; and $W_n^\Tr$. The uniform normalization of Eq.~(\ref{eq:assembled}) is realized in the implemented circuit as follows. The rotation of $a_1$ is controlled by $f$, $g$, $h_2$ and the first moment register, that of $a_2$ by $f$, $h_1$ and the second moment register. On a pair slot $(i, j)$ that no block handles, the two rotations leave the amplitudes $d_i/\sqrt\alpha$ and $d_j/\sqrt\alpha$ on $a_1 = 0$ and $a_2 = 0$, with $d_i$ the entry of $C_1$ for moment $i$, so that the slot carries $d_id_j/\alpha$; this includes the conserved-pair slots at $r \ne 0$, where $d_i = d_j = 1$ gives $1/\alpha$, and the six unused moment values of each register, where $d_i = 0$. On the level-1 slots $(i, 15)$ the rotation of $a_1$ leaves $d_i/\alpha$ and that of $a_2$ leaves the amplitude unchanged, so that the level-1 sector carries $C_1/\alpha$ in one factor. The coupled output slots $(s, 15)$ of the three coupled moments and the four conserved-pair slots of the slice $r = 0$ are left with amplitude one by both rotations, so that they are not rescaled a second time, and their normalization is applied once, by the dilations of $B_s/\alpha$ on the pairs $\{(s, 15), u_s\}$ and by the rotation of $b$ by $2\arccos(1/\alpha)$ on the decoupled antisymmetric slot $u_{\rm a}$ (the slots $(s, 15)$ at $r \ne 0$, which the state never occupies, keep amplitude one). Every sector of the stage therefore carries the single factor $1/\alpha$, and the block of the assembled stage on $a_1 = a_2 = b = 0$ is $\Kc_\lambda/\alpha$ on the encoded physical subspace, the subspace occupied by the state. The count is $4(2n-1)$ Toffolis for the flag $f$, 60 for the moment flags, 72 for $W_n$ and its inverse, at most 540 for the three dilations at ten two-level operations each and 18 for the antisymmetric slot, in total $8n + 686$ Toffolis under the counting rules below, $192$ CNOT from the two uniformly controlled rotations and $218$ rotations, 197 about a fixed axis and 21 generic single-qubit gates, that is 260 $z$-rotations, against $8n + 194$ Toffolis for the LCU stage. The qubit count of the step register is $4n + 15$.

\emph{Preparation.} The preparation of $\psi_\lambda(0)$ for a few-mode field proceeds as follows. The coefficients of the level-1 state are $c_{k,a} = \sqrt{3N}\,\hat{J}_a(k)$ on the $M$ nonzero wavevectors and velocity components, PREP loads $\sqrt{|c_m|/\|c\|_1}$ on the ancillas, SELECT writes the basis state $\ket{k_x, k_y}$ into the site register by multi-controlled X gates, the phase of $c_m$ by a multi-controlled phase on the ancillas and the velocity vector $\ket{v_a}$ on the moment register by a state preparation controlled on the component ancilla, two quantum Fourier transforms turn the basis states into plane waves, and PREP$^\dagger$ followed by postselection gives $\ket{g}$ with probability $p_1 = (\|c\|_2/\|c\|_1)^2$. A branch ancilla rotated by $\theta = 2\arctan[\lambda\|g\|_g/\sqrt{p_1}]$ selects the pair branch, on which a second copy of the preparation writes $\ket{g}$ into the second register in absolute coordinates and a modular subtraction $r \leftarrow r - x$ brings it into relative coordinates, while the level-1 branch sets the second register to $(r = 0, j = 15)$; the branch ancilla is uncomputed from the padding flag. The compensation of $\theta$ by $1/\sqrt{p_1}$ makes the postselected state exactly $\psi_\lambda(0)$ with success $P_{\rm prep} = p_1(\|g\|^2 + \lambda^2\|g\|^4)/(\|g\|^2 + \lambda^2\|g\|^4/p_1)$. The second copy uses its own set of $n_a = \lceil\log_2M\rceil$ LCU ancillas, because after the first copy the shared ancillas are entangled with its failure branch.

\emph{Exact amplification of the preparation.} With $\sin^2\theta_{\rm p} = P_{\rm prep}$ the good weight after $k$ rounds of $Q = -U_{\rm prep}S_0U_{\rm prep}^\dagger S_{\rm good}$ is $\sin^2[(2k+1)\theta_{\rm p}]$ \cite{BrassardHoyer2002}. Exactness is obtained with one auxiliary qubit rotated by $\beta$ with $\cos(\beta/2) = \sqrt{P'/P_{\rm prep}}$ before the preparation, which reduces the good weight to $P' = \sin^2\theta'$, $\theta' = \pi/[2(2k+1)]$, for the smallest $k$ with $\theta'\le\theta_{\rm p}$, that is $k = \lceil\pi/(4\theta_{\rm p}) - 1/2\rceil$. The good subspace includes the auxiliary qubit in $\ket0$, and after $k$ rounds the good weight is exactly one and the good branch is exactly $\psi_\lambda(0)$. On the $2\times2$ lattice ($P_{\rm prep} = 0.4295$, $k = 1$, $P' = 0.250$) the statevector after one round has good weight $1 - 4\times10^{-14}$ and fidelity one with $\psi_\lambda(0)$ when the reflections are applied to the exact statevector, and good weight $1 - 2\times10^{-13}$ at fidelity one when the two reflections are built as multi-controlled phase gates on the 16-qubit preparation register and the whole amplified preparation is run as one circuit (Supplemental Material). For the task, $P_{\rm prep} = 4.3\times10^{-3}$ gives $k = 12$, 25 preparation circuits and 12 pairs of reflections on the preparation register of $4n + 8 + 2n_a + 2$ qubits, $4.7\times10^4$ Toffolis and $1.1\times10^{4}$ $z$-rotations against 1734 and 429 for one preparation. The number of applications of $U$ then depends on the combined amplitude alone, and without this amplification it would be larger by $1/\sqrt{P_{\rm prep}} = 15$.

\emph{Registers of the estimation, reflections and readout.} In the coherent protocol each step uses a fresh triple $(a_1, a_2, b)$, so that $3(T-1)$ ancillas are added to the step register, and the preparation adds the branch ancilla, the two LCU ancilla sets and the auxiliary qubit, so that $U$ acts on $q = 4n + 15 + 3(T-1) + 2n_a + 2$ qubits, 162 for the acoustic task on $32\times32$. The multi-controlled gates of the step and of the preparation are counted with an AND chain into clean workspace qubits. The flag of the slice $r = 0$ on $2n$ controls needs $2n - 1$, and the two-level operations and the controlled rotation of the collision stage, on nine controls, need eight. The latter act while the three moment flags are uncomputed, so that a pool of $2n - 1$ clean qubits together with these flags supplies the workspace for every lattice of the resource tables, and this pool is part of the register. The amplitude estimation adds $m = \log_2M$ phase qubits, so that the estimation runs on $q + 2n - 1 + m$ qubits, 188 for the acoustic and 298 for the vortical task. The reflection about $\ket0$ of $Q$ acts on the $q$ qubits of $U$ and carries the control of the phase register. It is a $q$-controlled Z, counted with one workspace qubit in any state, taken from the pool, at $8(q-3)$ Toffolis. The reflection about the reference product state is $S_\chi = VS_0V^\dagger$, where $V$ writes the wavevector into the site register by X gates, applies the two $n$-qubit quantum Fourier transforms that turn it into the plane wave $e^{ik\cdot x}$, prepares the real velocity vector $\ket{v_a}$ on the moment register and sets the level-1 slot $(r = 0, j = 15)$ by X gates; $V$ and $V^\dagger$ need no control because they cancel on the branch that skips the reflection. The transverse observable uses the combination of the two reference states as its $\ket{v_a}$. The readout is the estimate of $|A|$ of Eq.~(\ref{eq:A}) and the classical recovery $|\hat J| = c_s\alpha^T\|\psi_\lambda(0)\|_g|A|/\sqrt N$. The two-step protocol on the $2\times2$ lattice at $\lambda = 3$ uses 25 qubits and is verified in the same way as the one-step protocol of Sec.~\ref{sec:coherent}, and the one-step protocol with the amplified preparation built as circuits uses 23 qubits. The numbers are listed in the Supplemental Material. The two-step evolution factor $P_{\rm evol} = 0.2045$ is small because the $2\times2$ lattice at $\kappa = \pi$ generates non-equilibrium norm at the lattice scale in every step.

\emph{Counting rules.} The counts rest on standard decompositions rather than on circuits constructed gate by gate, and they are constructive upper bounds under the stated decompositions rather than minimal gate counts, applied with one convention to every stage. A $k$-controlled X or phase gate with $k \ge 2$ is $2(k-1)$ Toffolis of an AND chain into $k - 1$ clean workspace qubits plus the singly controlled base gate, a CNOT or a controlled phase at 2 CNOT and 3 $z$-rotations \cite{Barenco1995}; where the register does not supply the clean workspace, the reflections on the full register, a $k$-controlled gate is $8(k-3)$ Toffolis with one workspace qubit in any state, Corollary 7.4 of \citet{Barenco1995}. A two-level operation on the eight moment qubits controlled by one flag is budgeted at 18 Toffolis, an upper bound under the AND-chain rule; a uniformly controlled rotation on $m$ controls is $2^m$ CNOT and $2^m$ rotations, and a real state on $m$ qubits is prepared with $2^m - 2$ CNOT and $2^m - 1$ rotations \cite{Mottonen2004}; a generic four-qubit unitary is at most 100 CNOT and 120 generic single-qubit gates and the Halmos dilation of a $2\times2$ block on two moment basis states and one ancilla at most ten two-level operations, three CNOT-type and seven of generic single-qubit type \cite{Shende2006}; a controlled generic single-qubit gate is 2 CNOT, three generic single-qubit gates and one fixed-axis rotation \cite{Barenco1995}; an $n$-bit modular subtractor is $2n$ Toffolis \cite{Cuccaro2004}, and its form controlled by the branch ancilla, $r \leftarrow r - (a\wedge x)$, computes $a\wedge x$ into $n$ pool qubits and uncomputes it, $4n$ Toffolis. Rotations are counted as $z$-rotations to be synthesized: a rotation about a fixed axis, a controlled-phase angle or a uniformly controlled Ry is one, and a generic single-qubit gate is three, its Euler decomposition with the Y rotation Clifford-conjugate to a Z rotation. With these rules the transforms of one step cost 400 CNOT and 480 generic single-qubit gates, 1440 $z$-rotations, the coupled-block collision stage $8n + 686$ Toffolis, 192 CNOT and 260 $z$-rotations, the two-term LCU stage $8n + 194$ Toffolis, 66 CNOT and 69 $z$-rotations, and one preparation of $\psi_\lambda(0)$ for the 16-term field on $32\times32$ costs 1734 Toffolis, 296 CNOT and 429 $z$-rotations: the 92 set bits of the wavevectors of the two copies written by 4- and 5-controlled X gates, the 16 phases, the two PREP pairs, the two velocity-vector preparations controlled on the component ancilla and, in the second copy, on the branch ancilla, the two quantum Fourier transforms per copy, the two controlled subtractors and the branch logic. The CNOT figures count the uniformly controlled rotations, the state preparations and the controlled phases; the base CNOT of each multi-controlled X gate and the basis changes of the two-level operations are Clifford gates that are not tallied and do not enter the T-gate total. The velocity-vector preparation is counted as the controlled real-state preparation of $\ket{v_a}$, 14 CNOT and 15 Ry with every gate controlled; the statevector checks implement it as a unitary completed from that first column, whose action on $\ket0$ is the same. The statevector simulations verify the operators the circuits implement, not these counts, which apply the rules to the operator list of each stage.

\section{Validation and error analysis}\label{app:validation}
\emph{Error definitions.} Errors of a truncated evolution against the nonlinear LB reference started from the same state are measured on the velocity by $e_u$ of Eq.~(\ref{eq:eu}), on the populations by $e_f = \|F - F_{\rm ref}\|/\|F_{\rm ref} - w\|$ relative to the flow part, on individual Fourier modes by the relative error of the mode at its peak (Sec.~\ref{sec:accuracy}), and, for the comparison with \citet{SanavioSucci2024}, by their metric $\langle\mathrm{RMSE}\rangle = \tfrac19\sum_q\epsilon_q$ with $\epsilon_q = [N^{-1}\sum_x((f_q^{\rm LB} - f_q^{\rm CL})/f_q^{\rm LB})^2]^{1/2}$, a relative population error averaged over the nine velocities. All norms are Euclidean over the lattice.

\begin{figure*}[t]\centering\includegraphics[width=0.9\textwidth]{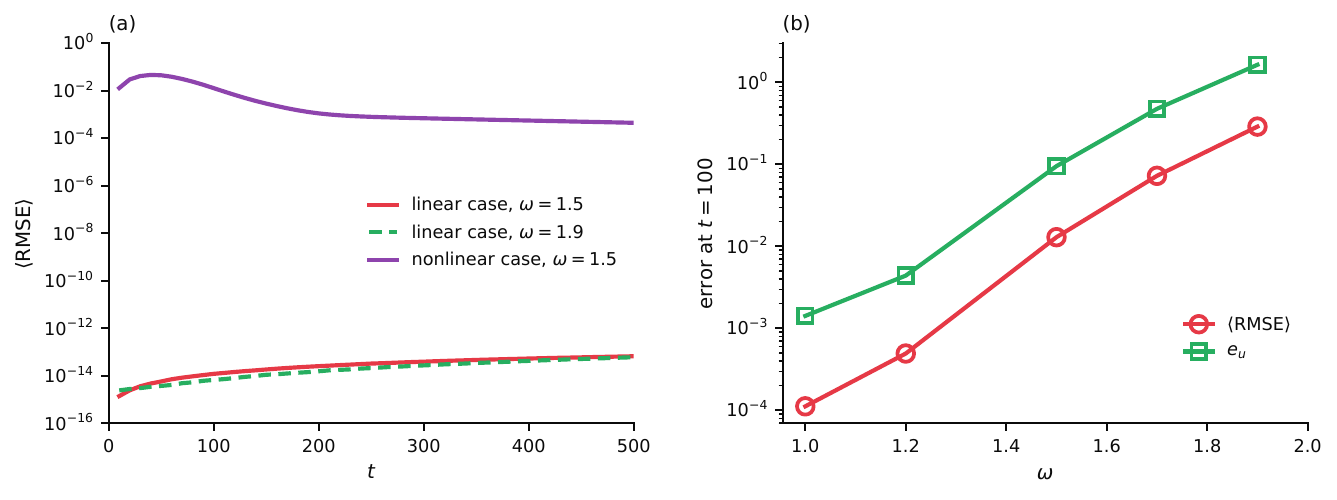}
\caption{Level-2 lift against the LB reference of \citet{SanavioSucci2024} (density-divided equilibrium) on their Kolmogorov cases on the $32\times32$ lattice. (a) $\langle\mathrm{RMSE}\rangle$ against the time step for the linear case at $\omega = 1.5$ and $1.9$ and the nonlinear case at $\omega = 1.5$. (b) $\langle\mathrm{RMSE}\rangle$ and $e_u$ at $t = 100$ against $\omega$ for the nonlinear case.}\label{fig:calibration}\end{figure*}

\emph{Calibration against the published level-2 results.} Figure~\ref{fig:calibration} compares the level-2 lift with the LB reference on the Kolmogorov cases of \citet{SanavioSucci2024} on the $32\times32$ lattice. The reference implements the equilibrium of that work, which divides by the local density, rather than the quadratic equilibrium of Eq.~(\ref{eq:feq}) used elsewhere in this paper, and the lift is built from the exactly quadratic map. For the shear wave $A_x = 0.3$, $A_y = 0$ (their Fig.~2) at $\omega = 1.5$ and $1.9$ the lift reproduces the reference to machine precision, $\langle\mathrm{RMSE}\rangle$ below $10^{-13}$ over 500 steps, because a shear wave keeps the density uniform and the streamed quadratic term carries no momentum. For the nonlinear case $A_x = 0.3$, $A_y = 0.2$, $k_y = 4$ (their Fig.~3) at $\omega = 1.5$, $\langle\mathrm{RMSE}\rangle$ rises to $4.7\times10^{-2}$ at $t = 40$, is $1.3\times10^{-2}$ at $t = 100$ and decays to $4.5\times10^{-4}$ at $t = 500$ as the flow decays. At $t = 100$ the error grows with $\omega$, $\langle\mathrm{RMSE}\rangle = 1.1\times10^{-4}$, $4.9\times10^{-4}$, $1.3\times10^{-2}$, $7.3\times10^{-2}$ and $0.29$ at $\omega = 1.0$, 1.2, 1.5, 1.7 and 1.9, with $e_u = 1.4\times10^{-3}$, $4.4\times10^{-3}$, $9.5\times10^{-2}$, $0.48$ and $1.6$, the last value being the loss of accuracy at $\omega = 1.9$ and $\Ma = 0.17$ noted in Appendix~\ref{app:closure}.

\begin{figure*}[t]\centering\includegraphics[width=0.9\textwidth]{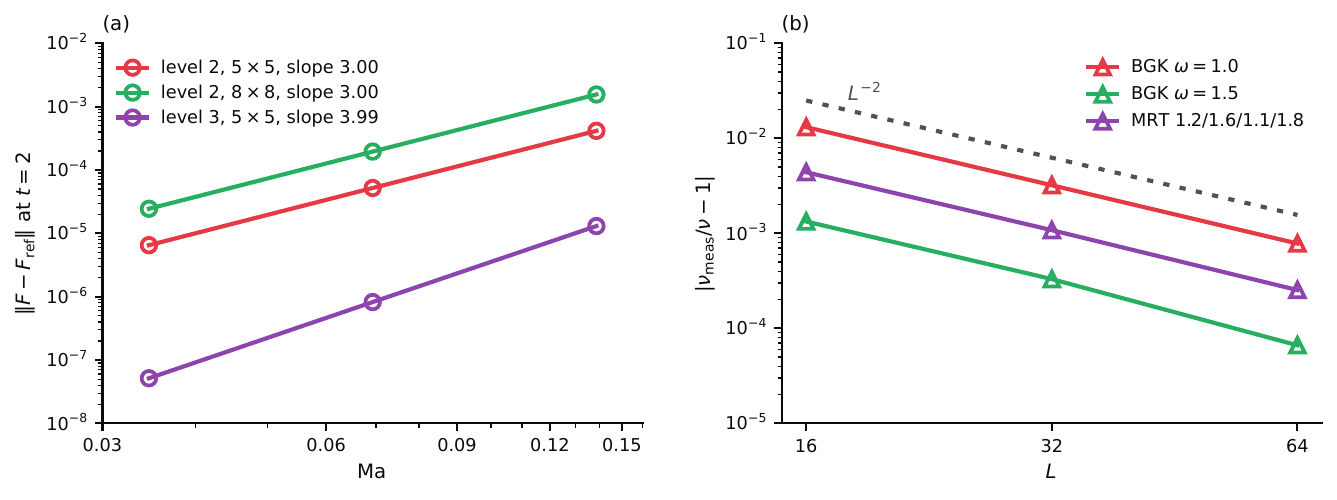}
\caption{(a) Absolute population error after two steps of the Taylor-Green vortex against the Mach number for the MRT rates $\omega_\nu/\omega_e/\omega_q/\omega_\varepsilon = 1.3/1.6/1.1/1.8$: level 2 on $5\times5$ and $8\times8$ and level 3 on $5\times5$, with the fitted slopes. (b) Relative error of the viscosity measured from the decay of the Taylor-Green vortex against the lattice side for BGK at $\omega = 1.0$ and $1.5$ and for the MRT rates $\omega_\nu/\omega_e/\omega_q/\omega_\varepsilon = 1.2/1.6/1.1/1.8$; the dotted line is $L^{-2}$.}\label{fig:truncation}\end{figure*}

\emph{Truncation order and hydrodynamic discretization error.} Two errors of different origin enter the accuracy statements of Sec.~\ref{sec:accuracy}. The first is the truncation error of the Carleman lift against the kinetic reference, the LB scheme itself, which is the error the quantum evolution carries. Figure~\ref{fig:truncation}(a) shows the absolute population error $\|F - F_{\rm ref}\|$ after two steps of the Taylor-Green vortex against the Mach number on $5\times5$ and $8\times8$ lattices for the MRT rates $1.3/1.6/1.1/1.8$, with fitted slopes $3.00$ and $3.00$ at level 2 and $3.99$ at level 3, so that relative to the flow amplitude, itself of order $\Ma$, the early-time error is of order $\Ma^2$ at level 2 and $\Ma^3$ at level 3 on these lattices, in agreement with the $\Ma^{2.0}$ scaling of the relative error along the two-mode runs found below and with the level-3 check on the $6\times6$ lattice of Sec.~\ref{sec:accuracy}. The $4\times4$ check of that section shows the steeper $\Ma^{4}$ scaling of the velocity error at level 3. The second is the discretization error of the LB reference itself against the Navier-Stokes solution, which the quantum algorithm inherits from the scheme and cannot reduce. Figure~\ref{fig:truncation}(b) shows the relative error of the measured viscosity, from the decay of the kinetic energy of the Taylor-Green vortex at $U_0 = 0.01$ as $\exp(-4\nu\kappa^2t)$, against the lattice side: $1.3\times10^{-2}$, $3.2\times10^{-3}$ and $7.8\times10^{-4}$ at $\omega = 1.0$, $1.3\times10^{-3}$, $3.3\times10^{-4}$ and $6.6\times10^{-5}$ at $\omega = 1.5$, and $4.4\times10^{-3}$, $1.1\times10^{-3}$ and $2.5\times10^{-4}$ for the MRT rates $1.2/1.6/1.1/1.8$ on $L = 16$, 32 and 64, second order in the lattice spacing and identical in the Gram-Schmidt and Hermite bases to a relative difference below $10^{-7}$. On the $32\times32$ lattice at $\omega = 1.5$ this hydrodynamic error, $3\times10^{-4}$, is far below the level-2 truncation error of the observables at the horizon, 1 to 3\%, so the accuracy window of Sec.~\ref{sec:accuracy} is set by the Carleman truncation and not by the lattice resolution.

\emph{Error recurrence along the reference run.} Equation~(\ref{eq:recurrence}) was evaluated along the two-mode runs on $32\times32$ at $\omega = 1.5$ and the three Mach numbers for 300 steps, with the quadratic-equilibrium start of Sec.~\ref{sec:accuracy}. The two parts of the source, $\Qc(e_t\otimes(F_{\rm ref}-w)) + \Qc((G-w)\otimes e_t)$ from the error itself and the same expression with $d_t$ from the nonlinear correction, add up to $S_t$ to $10^{-12}$. At $\Ma = 0.087$ the part from $d_t$ exceeds the part from $e_t$ by a factor of 18 at $t = 2$, 3.2 at $t = 10$, 5.9 at $t = 40$ and 4.5 at $t = 100$, at $\Ma = 0.020$ by 77 to 14, and at $\Ma = 0.173$ by 8.9 to 1.7 over the same steps. The smallest ratios over the full runs are 2.0, 8.2 and 1.1. The relative error $\|e_t\|_g/\|F_{\rm ref} - w\|_g$ is $3.6\times10^{-3}$, $1.2\times10^{-2}$, $1.7\times10^{-2}$ and $5.6\times10^{-2}$ at $t = 2$, 10, 40 and 300 for $\Ma = 0.087$, and scales as $\Ma^{2.0}$ between the three Mach numbers at $t = 2$. The summed bound $\sum_{k<t}\|S_k\|_g$ exceeds $\|e_t\|_g$ by 5.5, 24, 42 and 120 at $t = 10$, 40, 100 and 300, at every Mach number alike, so the triangle inequality on the homogeneous part discards the dissipation that the actual error experiences. The ratio $\|S_t\|_g/(\|\delta_t\|_g\|{\rm flow}\|_g)$ lies between 0.010 and 0.068 on all three runs, against the generic bound $2\|Q\|_g = 2.12$, because the momentum products that $Q$ maps into the second-order moments are a small part of the populations. The step-by-step values are tabulated in the Supplemental Material.

\section{Resource derivations and classical comparisons}\label{app:resources}
\emph{Ledger.} Table~\ref{tab:ledger} lists the ledger of the two reference cases. For each case of Table~\ref{tab:resources} the horizon $T$ is the specified step, the scale $\lambda$ maximizes the near-rest model with $N\epsilon_0$ in place of $N$, $\alpha$ is the exact norm of the scaled collision stage at that $\lambda$, the joint probability is $P = (\|g(T)\|_g^2/\|g_0\|_g^2)\,\alpha^{-2T}/(1 + \lambda^2N\epsilon_0)$ with the norm ratio from the shifted trajectory started from the prepared linearized state, and the combined amplitude $|A|$ follows from Eq.~(\ref{eq:A}) with $\|g_0\|_g^2 = N\epsilon_0$ and $|\hat J|$ from the level-2 lift at the horizon. The two routes $\sqrt N|\hat J|/(c_s\alpha^T\|\psi_\lambda(0)\|_g)$ and $\sqrt P|a|$ agree to $10^{-16}$. The resolution parameter $M$ of a run is the smallest power of two satisfying Eq.~(\ref{eq:M}), the number of runs is $r = 7$, and each run applies $U$ once and $Q$ a further $M - 1$ times, with the Fourier transform pair on the phase register, $m(m-1)$ controlled phases. The preparation success $P_{\rm prep}$ of Appendix~\ref{app:circuit} with $p_1 = 0.0658$ for the 16-term field is amplified exactly in $k = \lceil\pi/(4\theta_{\rm p}) - 1/2\rceil$ rounds inside every application of $U$. The categories of the ledger are I for operations implemented as circuits and verified by statevector simulation, on lattices of $4\times4$ sites and smaller, A for operations counted analytically from their operator lists with the rules of Appendix~\ref{app:circuit}, and S for the leading-order synthesis estimate. The T-gate total is seven per Toffoli plus the $z$-rotations of the $r$ runs times $3\log_2(1/\delta_r)$ \cite{RossSelinger2016}, with $\delta_r = 0.1\epsilon_{\rm rel}|A|/R_{\rm run}$ and $R_{\rm run}$ the rotations of one run, so that the synthesis errors of one run, added in operator norm, are a tenth of the amplitude precision. The conditional repeat-until-success line uses $1.15\log_2(1/\delta_r) + 9.2$ T gates per rotation in expectation \cite{Bocharov2015} and requires one ancilla and a mid-circuit measurement with feed-forward per rotation inside the coherent estimation circuit. The incoherent alternative measures the projector onto the reference product state after each $U\ket0$, a Bernoulli variable with mean $|A|^2$, and needs $1.96^2(1 - |A|^2)/(4\epsilon_{\rm rel}^2|A|^2)$ samples for the same precision and confidence in the normal approximation of the binomial sample. In both estimators $M$, $r$ and the sample size are set from the benchmark amplitude $|A|$ of the classical level-2 calculation of the instance; the ledger is an instance-dependent resource estimate calibrated with that amplitude, and no stopping rule for an unknown output is costed. The counts exclude error correction, routing and classical control.

\begin{table*}[t]\centering
\caption{Resource ledger of the two $32\times32$ reference cases at $\Ma = 0.087$, $\omega = 1.5$, coupled-block encoding, $\epsilon_{\rm rel} = 0.1$, 95\% confidence. Category I: implemented as circuits and verified by statevector simulation on lattices up to $4\times4$; A: counted analytically from the operator list with the rules of Appendix~\ref{app:circuit}; S: leading-order synthesis estimate. Rotations are $z$-rotations to be synthesized.}\label{tab:ledger}
\scriptsize\setlength{\tabcolsep}{3pt}\begin{tabular}{p{52mm}cp{40mm}p{40mm}}\toprule
item & category & acoustic, $T = 40$ & vortical, $T = 77$\\\midrule
\multicolumn{4}{l}{\emph{inputs}}\\
horizon $T$, scale $\lambda$, $\alpha$ &  & 40, 7.59, 1.01281 & 77, 10.64, 1.00658\\
$\|\psi_\lambda(0)\|_g$, $|\hat J|$ at $T$, $|A|$ &  & 55.5, $6.29\times10^{-4}$, $3.78\times10^{-4}$ & 77.7, $2.07\times10^{-3}$, $8.93\times10^{-4}$\\
$M$ (Eq.~\ref{eq:M}; unrounded), $m$, runs $r$ &  & $2^{17}$ ($9.1\times10^{4}$), 17, 7 & $2^{16}$ ($3.9\times10^{4}$), 16, 7\\
qubits: register, workspace, phase, total &  & 162, 9, 17, 188 & 273, 9, 16, 298\\
\multicolumn{4}{l}{\emph{per application of $U$ (Toffoli, $z$-rotations)}}\\
$T$ steps: streaming, collision, transforms & I, A & 77280, 29040, 16000 CNOT; 68000 rotations & 148764, 55902, 30800 CNOT; 130900 rotations\\
one preparation $U_{\rm prep}$ & I, A & 1734, 429 & 1734, 429\\
amplified preparation ($2k+1$ circuits, $k$ reflection pairs) & I, A & 46806, 10750 & 46806, 10750\\
total per application of $U$ & A & 153126, 78750 & 251472, 141650\\
\multicolumn{4}{l}{\emph{per application of $Q$}}\\
reflection about $\ket0$ on the register, controlled (each of two) & A & 1272, 0 & 2160, 0\\
reference-state preparation $V$ (each of two) & A & 0, 75 & 0, 75\\
total per application of $Q$ & A & 308796, 157650 & 507264, 283450\\
\multicolumn{4}{l}{\emph{totals}}\\
per run ($U$, $M - 1$ applications of $Q$, QFT pair) & A & $4.05\times10^{10}$, $2.07\times10^{10}$ & $3.32\times10^{10}$, $1.86\times10^{10}$\\
over $r$ runs: Toffolis, $z$-rotations, applications of $U$ or $U^\dagger$ & A & $2.83\times10^{11}$, $1.45\times10^{11}$, $1.84\times10^{6}$ & $2.33\times10^{11}$, $1.30\times10^{11}$, $9.17\times10^{5}$\\
synthesis: $\delta_r$, T per rotation (RS) & S & $1.8\times10^{-16}$, 157 & $4.8\times10^{-16}$, 153\\
T gates: from Toffolis, from rotations, total & S & $2.0\times10^{12}$, $2.3\times10^{13}$, $2.5\times10^{13}$ & $1.6\times10^{12}$, $2.0\times10^{13}$, $2.1\times10^{13}$\\
conditional RUS synthesis: T per rotation, total & S, cond. & 69, $1.2\times10^{13}$ & 68, $1.0\times10^{13}$\\
incoherent repetitions; applications without amplified preparation & A & $6.7\times10^{8}$; $2.8\times10^{7}$ & $1.2\times10^{8}$; $1.4\times10^{7}$\\
classical flops to the horizon: lattice, Fourier (support) & A & $5.4\times10^{7}$, $1.4\times10^{7}$ (35) & $1.0\times10^{8}$, $2.7\times10^{7}$ (35)\\
\bottomrule\end{tabular}

\end{table*}

\emph{Classical comparisons.} The factorized level-2 evolution keeps the pair sector as the product $G(t)\otimes G(t)$ of Eq.~(\ref{eq:product}) and advances $F$ and $G$ with two matrix-vector products of the per-site $9\times9$ collision matrix, the local quadratic term and the streaming permutation. Its operation count is derived from the implemented stepper with the structural zeros of the collision matrices skipped: per step $N(2\cdot81 + 496)$ real multiply-adds, the 81 nonzeros of the collision matrix in each of the two products and the 496 nonzeros of the $9\times81$ quadratic map, that is $1316N$ flops at two flops per multiply-add (the dense implementation executes $729$ multiply-adds per site for the quadratic term instead of 496). In Fourier space every wavevector evolves independently under the $9\times9$ matrix of the per-site collision followed by the streaming phase, and the quadratic term at wavevector $k$ is the sum over pairs of modes of $G$ with $k_1 + k_2 = k$; per step the implemented stepper performs $|G|^2\cdot496$ complex multiply-adds for the quadratic term and $81 + 9$ per active mode of $F$ and of $G$ for the collision and the phase, at eight flops per complex multiply-add. From the linearized two-mode start the nine initial wavevectors, the eight of the momentum field and the rest mode, stay the support of $G$, and $F$ occupies 35 wavevectors from the first step onward. The Fourier evolution reproduces $|\hat J_x(1,0)|$ at $T = 40$ and $|\hat J_\perp(1,2)|$ at $T = 77$ to $5\times10^{-15}$ of the lattice values, which are the values of Table~\ref{tab:resources}, and the full state to $4\times10^{-14}$ of the flow norm at $T = 77$. The flop counts of Table~\ref{tab:resources} and Fig.~\ref{fig:resources} are $1316NT$ for the lattice evolution and the summed per-step count of the Fourier evolution over the same $T$, $5.4\times10^{7}$ and $1.4\times10^{7}$ for the acoustic reference case and $1.0\times10^{8}$ and $2.7\times10^{7}$ for the vortical one. From the quadratic-equilibrium start of the accuracy study the initial support is 35 wavevectors and the active support 137, and the Fourier evolution agrees with the lattice evolution to $3\times10^{-13}$ of the flow norm over the 300, 600 and 1200 steps of the runs on $32^2$, $64^2$ and $128^2$. The Fourier cost is independent of the lattice size at a fixed number of steps and a fixed Fourier support. At fixed physics the horizon grows with $L$, so the Fourier cost grows linearly with $L$. T gates and floating-point operations are different units, and the comparison states the two counts side by side without a conversion.

\bibliographystyle{apsrev4-2}
\bibliography{refs}

\clearpage
\setcounter{section}{0}\setcounter{table}{0}\setcounter{figure}{0}\setcounter{equation}{0}
\renewcommand{\thesection}{S\arabic{section}}\renewcommand{\thetable}{S\arabic{table}}\renewcommand{\thefigure}{S\arabic{figure}}\renewcommand{\theequation}{S\arabic{equation}}
\makeatletter\def\@hangfrom@section#1#2#3{\@hangfrom{#1#2}\MakeTextUppercase{#3}}\def\@sectioncntformat#1{\csname the#1\endcsname.\quad}\makeatother
\begin{center}\textbf{\large Supplemental Material}\end{center}

This Supplemental Material collects the catalogs, records and sweeps behind the figures and tables of the main text: the parameters and the accuracy window of every two-mode run, the statevector records of every circuit check, the step-by-step statistics of the error recurrence, the full resource sweep, and the manifest that links every figure and table to the script and data file that produce it. Definitions, derivations and counting rules are in the appendices of the main text.

\section{Parameters of the two-mode runs}
Table~\ref{tab:parameters} lists every run of the accuracy study of Sec.~V of the main text (quadratic-equilibrium start): the lattice side, the velocity amplitude $U_0$, the Mach number, the relaxation rates, the viscosity, the Reynolds number $\mathrm{Re} = U_0L/\nu$, the wavenumber $\kappa = 2\pi/L$, the advective time $t_{\rm adv} = 1/(\kappa U_0)$, the viscous time $t_{\rm visc} = 1/(2\nu\kappa^2)$ of the $(1,\pm1)$ modes, the number of steps and the initial excess norm $\epsilon_0 = \|F_0\|_g^2/N - 1$. The three fixed-physics series keep $\Ma = 0.087$ and $\mathrm{Re} = 28.8$ or $100$ while the lattice side doubles, with the rate chosen so that $\nu\propto L$. The seven resource cases of Sec.~VI of the main text are the runs ref28\_L16, ref28\_L32, ref28\_L64, ref100\_L32, ref100\_L64, L32\_U0.0115\_w1.5 and L32\_U0.1\_w1.5 started from the linearized equilibrium of Sec.~II of the main text instead, with the same parameters.
\begin{table*}[t]\centering
\caption{Parameters of the two-mode runs.}\label{tab:parameters}
\small\setlength{\tabcolsep}{4pt}\begin{tabular}{lccccccccccc}\toprule
run & $L$ & $U_0$ & Ma & rates $\omega_\nu/\omega_e/\omega_q/\omega_\varepsilon$ & $\nu$ & Re & $\kappa$ & $t_{\rm adv}$ & $t_{\rm visc}$ & steps & $\epsilon_0$\\\midrule
L32\_U0.0115\_w1 & 32 & 0.0115 & 0.020 & 1 & 0.1667 & 2.2 & 0.1963 & 443 & 78 & 300 & $3.8\times10^{-4}$\\
L32\_U0.0115\_w1.5 & 32 & 0.0115 & 0.020 & 1.5 & 0.0556 & 6.6 & 0.1963 & 443 & 233 & 300 & $3.8\times10^{-4}$\\
L32\_U0.0115\_w1.9 & 32 & 0.0115 & 0.020 & 1.9 & 0.0088 & 42.0 & 0.1963 & 443 & 1478 & 300 & $3.8\times10^{-4}$\\
L32\_U0.05\_w1 & 32 & 0.05 & 0.087 & 1 & 0.1667 & 9.6 & 0.1963 & 102 & 78 & 300 & $7.2\times10^{-3}$\\
L32\_U0.05\_w1.5 & 32 & 0.05 & 0.087 & 1.5 & 0.0556 & 28.8 & 0.1963 & 102 & 233 & 300 & $7.2\times10^{-3}$\\
L32\_U0.05\_w1.9 & 32 & 0.05 & 0.087 & 1.9 & 0.0088 & 182.4 & 0.1963 & 102 & 1478 & 300 & $7.2\times10^{-3}$\\
L32\_U0.05\_MRT & 32 & 0.05 & 0.087 & 1.3/1.6/1.1/1.8 & 0.0897 & 17.8 & 0.1963 & 102 & 145 & 300 & $7.2\times10^{-3}$\\
L32\_U0.1\_w1 & 32 & 0.1 & 0.173 & 1 & 0.1667 & 19.2 & 0.1963 & 51 & 78 & 300 & $2.9\times10^{-2}$\\
L32\_U0.1\_w1.5 & 32 & 0.1 & 0.173 & 1.5 & 0.0556 & 57.6 & 0.1963 & 51 & 233 & 300 & $2.9\times10^{-2}$\\
L32\_U0.1\_w1.9 & 32 & 0.1 & 0.173 & 1.9 & 0.0088 & 364.8 & 0.1963 & 51 & 1478 & 300 & $2.9\times10^{-2}$\\
ref28\_L16 & 16 & 0.05 & 0.087 & 1.71 & 0.0278 & 28.8 & 0.3927 & 51 & 117 & 150 & $7.2\times10^{-3}$\\
ref28\_L32 & 32 & 0.05 & 0.087 & 1.5 & 0.0556 & 28.8 & 0.1963 & 102 & 233 & 300 & $7.2\times10^{-3}$\\
ref28\_L64 & 64 & 0.05 & 0.087 & 1.2 & 0.1111 & 28.8 & 0.0982 & 204 & 467 & 600 & $7.2\times10^{-3}$\\
ref28\_L128 & 128 & 0.05 & 0.087 & 0.857 & 0.2222 & 28.8 & 0.0491 & 407 & 934 & 1200 & $7.2\times10^{-3}$\\
ref100\_L32 & 32 & 0.05 & 0.087 & 1.82 & 0.0160 & 100.0 & 0.1963 & 102 & 811 & 300 & $7.2\times10^{-3}$\\
ref100\_L64 & 64 & 0.05 & 0.087 & 1.68 & 0.0320 & 100.0 & 0.0982 & 204 & 1621 & 600 & $7.2\times10^{-3}$\\
ref100\_L128 & 128 & 0.05 & 0.087 & 1.45 & 0.0640 & 100.0 & 0.0491 & 407 & 3242 & 1200 & $7.2\times10^{-3}$\\
\bottomrule\end{tabular}

\end{table*}

\section{Accuracy window of the two-mode runs}
Table~\ref{tab:window} lists, for every run of Table~\ref{tab:parameters}, the first step at which the level-2 velocity error exceeds $10^{-2}$, the error at a quarter and a half of the advective time, and the relative error of $|\hat{J}(1,0)|$ at the peak of the reference. The level-3 spot checks on $4\times4$ and $6\times6$ use a dense stepper of the full three-level lift at $\omega = 1.5$; the level-2 error at $t = 2$ scales as $\Ma^{2.00}$ and $\Ma^{2.02}$ and the level-3 error as $\Ma^{4.00}$ and $\Ma^{3.01}$ on the two lattices, and at $\Ma = 0.173$ the level-2 error first exceeds $10^{-2}$ at $t = 4$ and $t = 2$ while the level-3 error stays below $10^{-2}$ over the 30 and 40 steps of the runs.
\begin{table*}[t]\centering
\caption{Accuracy window of the level-2 lift on the two-mode flow. $t(e_u > 10^{-2})$ is the first step at which the velocity error exceeds $10^{-2}$ (a dash means never within the run), and $e_j$ the relative error of $|\hat{J}(1,0)|$ at its peak.}\label{tab:window}
\small\begin{tabular}{lccccccccc}\toprule
run & $L$ & Ma & $\omega_\nu$ & Re & $t_{\rm adv}$ & $t(e_u > 10^{-2})$ & $e_u(t_{\rm adv}/4)$ & $e_u(t_{\rm adv}/2)$ & $e_j$ at the peak\\\midrule
L32\_U0.0115\_w1 & 32 & 0.020 & 1.000 & 2.2 & 443 & -- & $3.4\times10^{-4}$ & $6.4\times10^{-4}$ & $5.3\times10^{-4}$\\
L32\_U0.0115\_w1.5 & 32 & 0.020 & 1.500 & 6.6 & 443 & -- & $1.9\times10^{-3}$ & $2.9\times10^{-3}$ & $9.9\times10^{-4}$\\
L32\_U0.0115\_w1.9 & 32 & 0.020 & 1.900 & 42.0 & 443 & 168 & $4.9\times10^{-3}$ & $1.6\times10^{-2}$ & $1.2\times10^{-3}$\\
L32\_U0.05\_w1 & 32 & 0.087 & 1.000 & 9.6 & 102 & 194 & $6.7\times10^{-3}$ & $4.7\times10^{-3}$ & $1.0\times10^{-2}$\\
L32\_U0.05\_w1.5 & 32 & 0.087 & 1.500 & 28.8 & 102 & 10 & $1.0\times10^{-2}$ & $1.4\times10^{-2}$ & $1.9\times10^{-2}$\\
L32\_U0.05\_w1.9 & 32 & 0.087 & 1.900 & 182.4 & 102 & 8 & $1.3\times10^{-2}$ & $2.2\times10^{-2}$ & $2.2\times10^{-2}$\\
L32\_U0.05\_MRT & 32 & 0.087 & 1.300 & 17.8 & 102 & 11 & $9.3\times10^{-3}$ & $9.9\times10^{-3}$ & $1.6\times10^{-2}$\\
L32\_U0.1\_w1 & 32 & 0.173 & 1.000 & 19.2 & 51 & 4 & $3.2\times10^{-2}$ & $2.7\times10^{-2}$ & $4.0\times10^{-2}$\\
L32\_U0.1\_w1.5 & 32 & 0.173 & 1.500 & 57.6 & 51 & 4 & $4.6\times10^{-2}$ & $4.0\times10^{-2}$ & $7.1\times10^{-2}$\\
L32\_U0.1\_w1.9 & 32 & 0.173 & 1.900 & 364.8 & 51 & 4 & $5.5\times10^{-2}$ & $5.0\times10^{-2}$ & $4.3\times10^{-2}$\\
ref28\_L16 & 16 & 0.087 & 1.714 & 28.8 & 51 & 6 & $9.0\times10^{-3}$ & $1.3\times10^{-2}$ & $1.8\times10^{-2}$\\
ref28\_L32 & 32 & 0.087 & 1.500 & 28.8 & 102 & 10 & $1.0\times10^{-2}$ & $1.4\times10^{-2}$ & $1.9\times10^{-2}$\\
ref28\_L64 & 64 & 0.087 & 1.200 & 28.8 & 204 & 19 & $1.1\times10^{-2}$ & $1.4\times10^{-2}$ & $1.9\times10^{-2}$\\
ref28\_L128 & 128 & 0.087 & 0.857 & 28.8 & 407 & 37 & $1.1\times10^{-2}$ & $1.4\times10^{-2}$ & $1.9\times10^{-2}$\\
ref100\_L32 & 32 & 0.087 & 1.825 & 100.0 & 102 & 8 & $1.2\times10^{-2}$ & $2.1\times10^{-2}$ & $2.2\times10^{-2}$\\
ref100\_L64 & 64 & 0.087 & 1.678 & 100.0 & 204 & 16 & $1.3\times10^{-2}$ & $2.1\times10^{-2}$ & $2.2\times10^{-2}$\\
ref100\_L128 & 128 & 0.087 & 1.445 & 100.0 & 407 & 32 & $1.3\times10^{-2}$ & $2.1\times10^{-2}$ & $2.2\times10^{-2}$\\
\bottomrule\end{tabular}

\end{table*}

\section{Statevector records}
Table~\ref{tab:circuit} lists the statevector checks of the coupled-block collision circuit on the $4\times4$ lattice, of the preparation circuit of the few-mode state, of its exact amplification, of the coherent protocol on the $2\times2$ lattice, and of the norm-free readout identity of Sec.~IV of the main text. The predicted step probability is $r_\lambda/\alpha^2$ from the classical scaled state, the predicted preparation success is $(\|c\|_2/\|c\|_1)^2$ for the level-1 state and $p_1(\|g\|^2 + \lambda^2\|g\|^4)/(\|g\|^2 + \lambda^2\|g\|^4/p_1)$ for the full state with the compensated branch angle, the predicted weights of the protocol are $P_{\rm prep}P_{\rm evol}$ with $P_{\rm evol} = \prod_tr_t/\alpha^{2T}$ and $\sin^23\theta$ after one round with $\sin\theta = \sqrt{P_{\rm prep}P_{\rm evol}}$. The predicted readout amplitude, the overlap of $U\ket0$ with the product of the reference state $\ket{\phi_a(k)}$, the level-1 slot and the zero ancillas, is $\sqrt{P_{\rm prep}P_{\rm evol}}\sqrt N\hat J_a(k)/(c_s\|\psi_\lambda(T)\|_g)$ from the classical state, which equals the norm-free form $\sqrt{P_{\rm prep}}\sqrt N\hat J_a(k;T)/(c_s\alpha^T\|\psi_\lambda(0)\|_g)$ of the main text to $3\times10^{-15}$ on all eight recorded amplitudes (the rows with the unamplified preparation). The last four rows are the run of the exactly amplified preparation, with the auxiliary rotation and the two reflections of the amplification round built as multi-controlled phase gates, followed by one coupled-block step on 23 qubits: the amplified preparation alone has good weight $1 - 2\times10^{-13}$ at fidelity one with $\psi_\lambda(0)$, the chain $U\ket0$ has the good weight $P_{\rm evol}$ with no preparation factor, and its amplitudes on the reference product states equal $\sqrt N\hat J_a(k;1)/(c_s\alpha\|\psi_\lambda(0)\|_g)$ to $10^{-14}$, up to the global sign of the amplification round, which amplitude estimation does not see. On the $2\times2$ lattice the product wavevector $(1,1)$ carries no momentum because the discrete divergence of a stress mode at the Nyquist wavevector vanishes, so the two initial wavevectors are used. The fidelities and amplitudes are the authors' statevector records (qiskit-aer, exact statevector method; the step unitaries are applied as native operators and only the preparation is decomposed into gates). On the $2\times2$ lattice the two-mode field of the main text vanishes ($\kappa = \pi$), and the protocol uses the field $J_x = U_0\cos\pi y$, $J_y = 0.6U_0\cos\pi x$ with $U_0 = 0.05$ instead.
\begin{table*}[t]\centering
\caption{Statevector checks of the coupled-block circuit, the preparation, its amplification, the coherent protocol and the norm-free readout identity. $F$ denotes the fidelity of the postselected or good-branch state with the classical target.}\label{tab:circuit}
\footnotesize\setlength{\tabcolsep}{4pt}\begin{tabular}{lcccccc}\toprule
check & qubits & $\lambda$ & quantity & measured & predicted & max.\ error\\\midrule
collision stage, $4\times4$, BGK 1.5, step 1 & 23 & 10 & $p$ & 0.984965 & 0.984965 & $5.4\times10^{-15}$\\
collision stage, $4\times4$, BGK 1.5, step 2 & 23 & 10 & $p$ & 0.965744 & 0.965744 & $6.5\times10^{-15}$\\
collision stage, $4\times4$, BGK 1.5, step 3 & 23 & 10 & $p$ & 0.980375 & 0.980375 & $6.5\times10^{-15}$\\
collision stage, $4\times4$, BGK 1.5, step 1 & 23 & 100 & $p$ & 0.999518 & 0.999518 & $1.1\times10^{-14}$\\
collision stage, $4\times4$, BGK 1.5, step 1 & 23 & 3 & $p$ & 0.862257 & 0.862257 & $2.1\times10^{-14}$\\
collision stage, $4\times4$, MRT 1.3/1.6/1.1/1.8, step 1 & 23 & 10 & $p$ & 0.980263 & 0.980263 & $1.1\times10^{-14}$\\
level-1 preparation, $L = 4$, 12 terms & 12 & -- & success & 0.1062 & 0.1062 & $1 - F = 10^{-16}$\\
level-1 preparation, $L = 8$, 16 terms & 14 & -- & success & 0.0658 & 0.0658 & $1 - F = 5\times10^{-16}$\\
$\psi_\lambda(0)$ preparation, $L = 2$ & 15 & 3 & success & 0.4295 & 0.4295 & $1 - F = 10^{-16}$\\
$\psi_\lambda(0)$ preparation, $L = 4$ & 25 & 10 & success & 0.0120 & 0.0120 & $1 - F = 5\times10^{-15}$\\
exact amplification of the $\psi_\lambda(0)$ preparation, $L = 2$, 1 round & 16 & 3 & good weight & 1.000000 & 1 & $1 - F = 2\times10^{-16}$\\
protocol $U = U_{\rm evol}U_{\rm prep}$, $2\times2$, $T = 1$ & 22 & 3 & good weight & 0.373547 & 0.373547 & $1 - F = 3\times10^{-16}$\\
one amplification round, $T = 1$ & 22 & 3 & good weight & 0.847005 & 0.847005 & $1 - F = 2\times10^{-16}$\\
overlap readout, $T = 1$, $\hat{J}_{y}(1,0)$ & 22 & 3 & amplitude & +0.089269 & +0.089269 & $2\times10^{-15}$\\
overlap readout, $T = 1$, $\hat{J}_{x}(0,1)$ & 22 & 3 & amplitude & +0.148782 & +0.148782 & $3\times10^{-15}$\\
protocol $U = U_{\rm evol}U_{\rm prep}$, $2\times2$, $T = 2$ & 25 & 3 & good weight & 0.087842 & 0.087842 & $1 - F = 10^{-16}$\\
one amplification round, $T = 2$ & 25 & 3 & good weight & 0.616231 & 0.616231 & $1 - F = 10^{-16}$\\
overlap readout, $T = 2$, $\hat{J}_{y}(1,0)$ & 25 & 3 & amplitude & -0.082907 & -0.082907 & $7\times10^{-16}$\\
overlap readout, $T = 2$, $\hat{J}_{x}(0,1)$ & 25 & 3 & amplitude & -0.138179 & -0.138179 & $10^{-15}$\\
amplified preparation as circuits, $L = 2$, 1 round & 23 & 3 & good weight & 1.000000 & 1 & $1 - F = 10^{-16}$\\
protocol with amplified preparation, $2\times2$, $T = 1$ & 23 & 3 & good weight & 0.869781 & 0.869781 & $1 - F = 3\times10^{-16}$\\
norm-free readout identity, $T = 1$, $\hat{J}_{y}(1,0)$ & 23 & 3 & $|A|$ & 0.136218 & 0.136218 & $10^{-14}$\\
norm-free readout identity, $T = 1$, $\hat{J}_{x}(0,1)$ & 23 & 3 & $|A|$ & 0.227030 & 0.227030 & $8\times10^{-15}$\\
\bottomrule\end{tabular}

\end{table*}

Table~\ref{tab:verify_lcu} reproduces the verification record of the two-term LCU circuit of the preliminary version of this work (Yao and Succi, Zenodo, doi 10.5281/zenodo.22714041) on the same $4\times4$ lattice (22 qubits, Taylor-Green flow at $\Ma = 0.17$). Its values coincide with five of the six crosses of Fig.~5(a) of the main text (the $\lambda = 3$ value there is computed from the classical state only) and with the LCU rows of Fig.~2 of the main text. After $k$ steps the postselected state equals $\Cc_\lambda^k\psi/\alpha^k$ to $10^{-13}$ in both sectors, and the measured success of each step equals $r_\lambda/\alpha^2$ from the classical state to four digits. The value 0.80 at the second step has a physical origin. On this coarse lattice, $\kappa = \pi/2$, the linear evolution generates 1.3\% of non-equilibrium norm in one step, of which 1.0\% is in the deviator; at $\omega = 1.5$ the level-1 sector relaxes three quarters of this norm and the pair sector twice that, 2.0\%, which with $\alpha = 1.106$ gives the measured 0.80. The coupled-block circuit on the same states reaches 0.985 and 0.966 at $\lambda = 10$ (Table~\ref{tab:circuit}), because its subnormalization is $1 + c/\lambda^2$ rather than $1 + a/\lambda$.
\begin{table}[t]\centering\scriptsize\setlength{\tabcolsep}{2.5pt}
\caption{Statevector verification of the two-term LCU circuit of the preliminary version on $4\times4$ for the Taylor-Green flow at $\Ma = 0.17$. MRT rates are listed as $\omega_\nu/\omega_e/\omega_q/\omega_\varepsilon$.}\label{tab:verify_lcu}
\begin{tabular}{lccccc}\toprule
rates & $\lambda$ & step & rel.\ error & $p$ measured & $r_\lambda/\alpha^2$\\\midrule
BGK 1.5 & 10 & 1 & $3.2\times10^{-14}$ & 0.8171 & 0.8171\\
BGK 1.5 & 10 & 2 & $6.1\times10^{-14}$ & 0.8012 & 0.8012\\
BGK 1.5 & 10 & 3 & $9.1\times10^{-14}$ & 0.8133 & 0.8133\\
BGK 1.5 & 100 & 1 & $3.3\times10^{-14}$ & 0.9788 & 0.9788\\
MRT 1.3/1.6/1.1/1.8 & 10 & 1 & $3.2\times10^{-14}$ & 0.8068 & 0.8068\\\bottomrule
\end{tabular}\end{table}

\section{Error recurrence}
Table~\ref{tab:recurrence} lists the quantities of the error recurrence $e_{t+1} = \mathcal L e_t + S_t$ of Sec.~V of the main text along the three two-mode runs on $32\times32$ at $\omega = 1.5$ (quadratic-equilibrium start, 300 steps): the relative error $\|e_t\|_g/\|F_{\rm ref} - w\|_g$, the ratio of the summed bound $\sum_{k<t}\|S_k\|_g$ to $\|e_t\|_g$, the ratio of the two parts of the source, and the range of $\|S_t\|_g/(\|\delta_t\|_g\|{\rm flow}\|_g)$ against the generic bound $2\|Q\|_g = 2.12$ in the weighted norm. The full series along the runs are in the data file listed in Table~\ref{tab:manifest}.
\begin{table*}[t]\centering
\caption{Error recurrence along the two-mode runs on $32\times32$ at $\omega = 1.5$. $e$ is the relative error $\|e_t\|_g/\|F_{\rm ref} - w\|_g$, $B/e$ the ratio of the summed source bound to the error, $S_d/S_e$ the ratio of the source part from the nonlinear correction to the part from the error, and $\sigma$ the range of $\|S_t\|_g/(\|\delta_t\|_g\|{\rm flow}\|_g)$ over the run.}\label{tab:recurrence}
\scriptsize\setlength{\tabcolsep}{2.5pt}\begin{tabular}{lccccccccccc}\toprule
run & Ma & $e(2)$ & $e(10)$ & $e(40)$ & $e(300)$ & $B/e(10)$ & $B/e(40)$ & $B/e(300)$ & $S_d/S_e(2)$ & $S_d/S_e(40)$ & $\sigma$\\\midrule
ref28\_L32 & 0.087 & $3.6\times10^{-3}$ & $1.2\times10^{-2}$ & $1.7\times10^{-2}$ & $5.6\times10^{-2}$ & 5.5 & 23.8 & 122 & 17.7 & 5.9 & 0.011 to 0.066\\
L32\_U0.0115\_w1.5 & 0.020 & $1.9\times10^{-4}$ & $6.5\times10^{-4}$ & $9.1\times10^{-4}$ & $3.1\times10^{-3}$ & 5.5 & 23.7 & 117 & 77.0 & 25.2 & 0.011 to 0.068\\
L32\_U0.1\_w1.5 & 0.173 & $1.4\times10^{-2}$ & $4.8\times10^{-2}$ & $6.6\times10^{-2}$ & $1.9\times10^{-1}$ & 5.6 & 24.0 & 136 & 8.9 & 3.1 & 0.010 to 0.063\\
\bottomrule\end{tabular}

\end{table*}

\section{Resource sweep}
Table~\ref{tab:resources_full} lists every case of the resource accounting of Sec.~VI and Appendix~F of the main text: the seven cost cases, the acoustic observable $|\hat J_x(1,0)|$ at its peak and, for the three cases of the $\mathrm{Re} = 28.8$ series, the vortical observable $|\hat J_\perp(1,2)|$ at its peak, each at the estimation precisions $\epsilon_{\rm rel} = 0.1$ and $0.02$ with 95\% confidence (the median of seven independent runs), for the coupled-block and the two-term LCU encodings, with the T-gate totals under the ancilla-free Ross-Selinger rule (RS) and, as a conditional alternative, the repeat-until-success rule of Bocharov, Roetteler and Svore (RUS), which counts $1.15\log_2(1/\delta_r) + 9.2$ T gates per rotation in expectation and requires one ancilla and a mid-circuit measurement with classical feed-forward per rotation inside the coherent estimation circuit. The RUS column is valid only where that capability is available and is not used in the main text. The columns $P_{\rm prep}$ and $k$ are the preparation success and the rounds of its exact amplification, $|A|$ the combined amplitude, $M$ the resolution parameter of the estimation, with $M - 1$ applications of $Q$ per run, and the uses of $U$ the applications of $U$ or $U^\dagger$ over the seven runs. Over the sweep at $\epsilon_{\rm rel} = 0.1$ with the coupled block the T-gate count ranges from $5.2\times10^{12}$ to $2.1\times10^{14}$ and the qubit count from 121 to 539.
\begin{turnpage}\begin{table*}[p]\centering
\caption{Resource sweep of the observable task.}\label{tab:resources_full}
\scriptsize\setlength{\tabcolsep}{2pt}\begin{tabular}{llccccccccccccc}\toprule
case & observable & $T$ & $\lambda$ & $e_{\rm trunc}$ & $\epsilon_{\rm rel}$ & $P_{\rm prep}$ & $k$ & $P$ & $|A|$ & $M$ & uses of $U$ & qubits & T gates (RS) & T gates (RUS, cond.)\\\midrule
Re 28.8, $16^2$ & $|\hat J_x(1,0)|$ & 20 & 7.2 & $1.8\times10^{-2}$ & 0.1 & $4.4\times10^{-3}$ & 12 & $2.0\times10^{-3}$ & $7.2\times10^{-4}$ & $2^{16}$ & $9.2\times10^{5}$ & 121 & $6.5\times10^{12}$ & $3.2\times10^{12}$\\
Re 28.8, $16^2$ & $|\hat J_x(1,0)|$ & 20 & 7.2 & $1.8\times10^{-2}$ & 0.02 & $4.4\times10^{-3}$ & 12 & $2.0\times10^{-3}$ & $7.2\times10^{-4}$ & $2^{18}$ & $3.7\times10^{6}$ & 123 & $2.8\times10^{13}$ & $1.4\times10^{13}$\\
Re 28.8, $16^2$ (LCU) & $|\hat J_x(1,0)|$ & 20 & 23.1 & $1.8\times10^{-2}$ & 0.1 & $4.3\times10^{-3}$ & 12 & $7.5\times10^{-5}$ & $1.4\times10^{-4}$ & $2^{18}$ & $3.7\times10^{6}$ & 123 & $2.5\times10^{13}$ & $1.2\times10^{13}$\\
Re 28.8, $16^2$ (LCU) & $|\hat J_x(1,0)|$ & 20 & 23.1 & $1.8\times10^{-2}$ & 0.02 & $4.3\times10^{-3}$ & 12 & $7.5\times10^{-5}$ & $1.4\times10^{-4}$ & $2^{21}$ & $2.9\times10^{7}$ & 126 & $2.2\times10^{14}$ & $1.1\times10^{14}$\\
Re 28.8, $16^2$ & $|\hat J_\perp(1,2)|$ & 37 & 10.1 & $2.3\times10^{-2}$ & 0.1 & $4.3\times10^{-3}$ & 12 & $7.0\times10^{-4}$ & $1.6\times10^{-3}$ & $2^{15}$ & $4.6\times10^{5}$ & 171 & $5.2\times10^{12}$ & $2.5\times10^{12}$\\
Re 28.8, $16^2$ & $|\hat J_\perp(1,2)|$ & 37 & 10.1 & $2.3\times10^{-2}$ & 0.02 & $4.3\times10^{-3}$ & 12 & $7.0\times10^{-4}$ & $1.6\times10^{-3}$ & $2^{17}$ & $1.8\times10^{6}$ & 173 & $2.2\times10^{13}$ & $1.1\times10^{13}$\\
Re 28.8, $16^2$ (LCU) & $|\hat J_\perp(1,2)|$ & 37 & 43.7 & $2.3\times10^{-2}$ & 0.1 & $4.3\times10^{-3}$ & 12 & $1.4\times10^{-5}$ & $2.3\times10^{-4}$ & $2^{18}$ & $3.7\times10^{6}$ & 174 & $4.1\times10^{13}$ & $2.0\times10^{13}$\\
Re 28.8, $16^2$ (LCU) & $|\hat J_\perp(1,2)|$ & 37 & 43.7 & $2.3\times10^{-2}$ & 0.02 & $4.3\times10^{-3}$ & 12 & $1.4\times10^{-5}$ & $2.3\times10^{-4}$ & $2^{20}$ & $1.5\times10^{7}$ & 176 & $1.8\times10^{14}$ & $8.3\times10^{13}$\\
Re 28.8, $32^2$ & $|\hat J_x(1,0)|$ & 40 & 7.6 & $1.9\times10^{-2}$ & 0.1 & $4.3\times10^{-3}$ & 12 & $5.0\times10^{-4}$ & $3.8\times10^{-4}$ & $2^{17}$ & $1.8\times10^{6}$ & 188 & $2.5\times10^{13}$ & $1.2\times10^{13}$\\
Re 28.8, $32^2$ & $|\hat J_x(1,0)|$ & 40 & 7.6 & $1.9\times10^{-2}$ & 0.02 & $4.3\times10^{-3}$ & 12 & $5.0\times10^{-4}$ & $3.8\times10^{-4}$ & $2^{19}$ & $7.3\times10^{6}$ & 190 & $1.1\times10^{14}$ & $5.1\times10^{13}$\\
Re 28.8, $32^2$ (LCU) & $|\hat J_x(1,0)|$ & 40 & 41.4 & $1.9\times10^{-2}$ & 0.1 & $4.3\times10^{-3}$ & 12 & $6.2\times10^{-6}$ & $4.2\times10^{-5}$ & $2^{20}$ & $1.5\times10^{7}$ & 191 & $2.0\times10^{14}$ & $9.4\times10^{13}$\\
Re 28.8, $32^2$ (LCU) & $|\hat J_x(1,0)|$ & 40 & 41.4 & $1.9\times10^{-2}$ & 0.02 & $4.3\times10^{-3}$ & 12 & $6.2\times10^{-6}$ & $4.2\times10^{-5}$ & $2^{22}$ & $5.9\times10^{7}$ & 193 & $8.4\times10^{14}$ & $3.9\times10^{14}$\\
Re 28.8, $32^2$ & $|\hat J_\perp(1,2)|$ & 77 & 10.6 & $2.6\times10^{-2}$ & 0.1 & $4.3\times10^{-3}$ & 12 & $1.6\times10^{-4}$ & $8.9\times10^{-4}$ & $2^{16}$ & $9.2\times10^{5}$ & 298 & $2.1\times10^{13}$ & $1.0\times10^{13}$\\
Re 28.8, $32^2$ & $|\hat J_\perp(1,2)|$ & 77 & 10.6 & $2.6\times10^{-2}$ & 0.02 & $4.3\times10^{-3}$ & 12 & $1.6\times10^{-4}$ & $8.9\times10^{-4}$ & $2^{18}$ & $3.7\times10^{6}$ & 300 & $9.3\times10^{13}$ & $4.4\times10^{13}$\\
Re 28.8, $32^2$ (LCU) & $|\hat J_\perp(1,2)|$ & 77 & 80.6 & $2.6\times10^{-2}$ & 0.1 & $4.3\times10^{-3}$ & 12 & $1.0\times10^{-6}$ & $7.1\times10^{-5}$ & $2^{19}$ & $7.3\times10^{6}$ & 301 & $1.7\times10^{14}$ & $8.1\times10^{13}$\\
Re 28.8, $32^2$ (LCU) & $|\hat J_\perp(1,2)|$ & 77 & 80.6 & $2.6\times10^{-2}$ & 0.02 & $4.3\times10^{-3}$ & 12 & $1.0\times10^{-6}$ & $7.1\times10^{-5}$ & $2^{22}$ & $5.9\times10^{7}$ & 304 & $1.5\times10^{15}$ & $7.0\times10^{14}$\\
Re 28.8, $64^2$ & $|\hat J_x(1,0)|$ & 80 & 7.7 & $1.9\times10^{-2}$ & 0.1 & $4.3\times10^{-3}$ & 12 & $1.2\times10^{-4}$ & $1.9\times10^{-4}$ & $2^{18}$ & $3.7\times10^{6}$ & 315 & $9.8\times10^{13}$ & $4.8\times10^{13}$\\
Re 28.8, $64^2$ & $|\hat J_x(1,0)|$ & 80 & 7.7 & $1.9\times10^{-2}$ & 0.02 & $4.3\times10^{-3}$ & 12 & $1.2\times10^{-4}$ & $1.9\times10^{-4}$ & $2^{20}$ & $1.5\times10^{7}$ & 317 & $4.2\times10^{14}$ & $2.0\times10^{14}$\\
Re 28.8, $64^2$ (LCU) & $|\hat J_x(1,0)|$ & 80 & 67.0 & $1.9\times10^{-2}$ & 0.1 & $4.3\times10^{-3}$ & 12 & $6.0\times10^{-7}$ & $1.3\times10^{-5}$ & $2^{22}$ & $5.9\times10^{7}$ & 319 & $1.6\times10^{15}$ & $7.5\times10^{14}$\\
Re 28.8, $64^2$ (LCU) & $|\hat J_x(1,0)|$ & 80 & 67.0 & $1.9\times10^{-2}$ & 0.02 & $4.3\times10^{-3}$ & 12 & $6.0\times10^{-7}$ & $1.3\times10^{-5}$ & $2^{24}$ & $2.3\times10^{8}$ & 321 & $6.8\times10^{15}$ & $3.2\times10^{15}$\\
Re 28.8, $64^2$ & $|\hat J_\perp(1,2)|$ & 155 & 10.7 & $2.7\times10^{-2}$ & 0.1 & $4.3\times10^{-3}$ & 12 & $4.0\times10^{-5}$ & $4.6\times10^{-4}$ & $2^{17}$ & $1.8\times10^{6}$ & 539 & $8.9\times10^{13}$ & $4.3\times10^{13}$\\
Re 28.8, $64^2$ & $|\hat J_\perp(1,2)|$ & 155 & 10.7 & $2.7\times10^{-2}$ & 0.02 & $4.3\times10^{-3}$ & 12 & $4.0\times10^{-5}$ & $4.6\times10^{-4}$ & $2^{19}$ & $7.3\times10^{6}$ & 541 & $3.8\times10^{14}$ & $1.8\times10^{14}$\\
Re 28.8, $64^2$ (LCU) & $|\hat J_\perp(1,2)|$ & 155 & 130.7 & $2.7\times10^{-2}$ & 0.1 & $4.3\times10^{-3}$ & 12 & $1.0\times10^{-7}$ & $2.3\times10^{-5}$ & $2^{21}$ & $2.9\times10^{7}$ & 543 & $1.4\times10^{15}$ & $6.8\times10^{14}$\\
Re 28.8, $64^2$ (LCU) & $|\hat J_\perp(1,2)|$ & 155 & 130.7 & $2.7\times10^{-2}$ & 0.02 & $4.3\times10^{-3}$ & 12 & $1.0\times10^{-7}$ & $2.3\times10^{-5}$ & $2^{23}$ & $1.2\times10^{8}$ & 545 & $6.2\times10^{15}$ & $2.9\times10^{15}$\\
Re 100, $32^2$ & $|\hat J_x(1,0)|$ & 41 & 13.7 & $2.2\times10^{-2}$ & 0.1 & $4.3\times10^{-3}$ & 12 & $2.2\times10^{-4}$ & $2.1\times10^{-4}$ & $2^{18}$ & $3.7\times10^{6}$ & 192 & $5.2\times10^{13}$ & $2.5\times10^{13}$\\
Re 100, $32^2$ & $|\hat J_x(1,0)|$ & 41 & 13.7 & $2.2\times10^{-2}$ & 0.02 & $4.3\times10^{-3}$ & 12 & $2.2\times10^{-4}$ & $2.1\times10^{-4}$ & $2^{20}$ & $1.5\times10^{7}$ & 194 & $2.2\times10^{14}$ & $1.1\times10^{14}$\\
Re 100, $32^2$ (LCU) & $|\hat J_x(1,0)|$ & 41 & 51.6 & $2.2\times10^{-2}$ & 0.1 & $4.3\times10^{-3}$ & 12 & $5.8\times10^{-6}$ & $3.4\times10^{-5}$ & $2^{20}$ & $1.5\times10^{7}$ & 194 & $2.0\times10^{14}$ & $9.6\times10^{13}$\\
Re 100, $32^2$ (LCU) & $|\hat J_x(1,0)|$ & 41 & 51.6 & $2.2\times10^{-2}$ & 0.02 & $4.3\times10^{-3}$ & 12 & $5.8\times10^{-6}$ & $3.4\times10^{-5}$ & $2^{23}$ & $1.2\times10^{8}$ & 197 & $1.8\times10^{15}$ & $8.2\times10^{14}$\\
Re 100, $64^2$ & $|\hat J_x(1,0)|$ & 82 & 14.4 & $2.2\times10^{-2}$ & 0.1 & $4.3\times10^{-3}$ & 12 & $5.2\times10^{-5}$ & $1.0\times10^{-4}$ & $2^{19}$ & $7.3\times10^{6}$ & 322 & $2.1\times10^{14}$ & $1.0\times10^{14}$\\
Re 100, $64^2$ & $|\hat J_x(1,0)|$ & 82 & 14.4 & $2.2\times10^{-2}$ & 0.02 & $4.3\times10^{-3}$ & 12 & $5.2\times10^{-5}$ & $1.0\times10^{-4}$ & $2^{21}$ & $2.9\times10^{7}$ & 324 & $8.9\times10^{14}$ & $4.2\times10^{14}$\\
Re 100, $64^2$ (LCU) & $|\hat J_x(1,0)|$ & 82 & 96.1 & $2.2\times10^{-2}$ & 0.1 & $4.3\times10^{-3}$ & 12 & $4.3\times10^{-7}$ & $9.4\times10^{-6}$ & $2^{22}$ & $5.9\times10^{7}$ & 325 & $1.6\times10^{15}$ & $7.7\times10^{14}$\\
Re 100, $64^2$ (LCU) & $|\hat J_x(1,0)|$ & 82 & 96.1 & $2.2\times10^{-2}$ & 0.02 & $4.3\times10^{-3}$ & 12 & $4.3\times10^{-7}$ & $9.4\times10^{-6}$ & $2^{25}$ & $4.7\times10^{8}$ & 328 & $1.4\times10^{16}$ & $6.6\times10^{15}$\\
Ma 0.020, $32^2$ & $|\hat J_x(1,0)|$ & 40 & 7.8 & $1.0\times10^{-3}$ & 0.1 & $4.5\times10^{-3}$ & 12 & $8.9\times10^{-3}$ & $3.7\times10^{-4}$ & $2^{17}$ & $1.8\times10^{6}$ & 188 & $2.5\times10^{13}$ & $1.2\times10^{13}$\\
Ma 0.020, $32^2$ & $|\hat J_x(1,0)|$ & 40 & 7.8 & $1.0\times10^{-3}$ & 0.02 & $4.5\times10^{-3}$ & 12 & $8.9\times10^{-3}$ & $3.7\times10^{-4}$ & $2^{19}$ & $7.3\times10^{6}$ & 190 & $1.1\times10^{14}$ & $5.1\times10^{13}$\\
Ma 0.020, $32^2$ (LCU) & $|\hat J_x(1,0)|$ & 40 & 41.4 & $1.0\times10^{-3}$ & 0.1 & $4.3\times10^{-3}$ & 12 & $1.1\times10^{-4}$ & $4.2\times10^{-5}$ & $2^{20}$ & $1.5\times10^{7}$ & 191 & $2.0\times10^{14}$ & $9.4\times10^{13}$\\
Ma 0.020, $32^2$ (LCU) & $|\hat J_x(1,0)|$ & 40 & 41.4 & $1.0\times10^{-3}$ & 0.02 & $4.3\times10^{-3}$ & 12 & $1.1\times10^{-4}$ & $4.2\times10^{-5}$ & $2^{22}$ & $5.9\times10^{7}$ & 193 & $8.4\times10^{14}$ & $3.9\times10^{14}$\\
Ma 0.173, $32^2$ & $|\hat J_x(1,0)|$ & 40 & 7.6 & $7.2\times10^{-2}$ & 0.1 & $4.3\times10^{-3}$ & 12 & $1.3\times10^{-4}$ & $3.8\times10^{-4}$ & $2^{17}$ & $1.8\times10^{6}$ & 188 & $2.5\times10^{13}$ & $1.2\times10^{13}$\\
Ma 0.173, $32^2$ & $|\hat J_x(1,0)|$ & 40 & 7.6 & $7.2\times10^{-2}$ & 0.02 & $4.3\times10^{-3}$ & 12 & $1.3\times10^{-4}$ & $3.8\times10^{-4}$ & $2^{19}$ & $7.3\times10^{6}$ & 190 & $1.1\times10^{14}$ & $5.1\times10^{13}$\\
Ma 0.173, $32^2$ (LCU) & $|\hat J_x(1,0)|$ & 40 & 41.4 & $7.2\times10^{-2}$ & 0.1 & $4.3\times10^{-3}$ & 12 & $1.6\times10^{-6}$ & $4.2\times10^{-5}$ & $2^{20}$ & $1.5\times10^{7}$ & 191 & $2.0\times10^{14}$ & $9.4\times10^{13}$\\
Ma 0.173, $32^2$ (LCU) & $|\hat J_x(1,0)|$ & 40 & 41.4 & $7.2\times10^{-2}$ & 0.02 & $4.3\times10^{-3}$ & 12 & $1.6\times10^{-6}$ & $4.2\times10^{-5}$ & $2^{22}$ & $5.9\times10^{7}$ & 193 & $8.4\times10^{14}$ & $3.9\times10^{14}$\\
\bottomrule\end{tabular}

\end{table*}\end{turnpage}

\section{Reproducibility manifest}
Every number, figure and table of the paper is produced by a script that writes its parameters, the script name and the date into a JSON data file, from which the figures and tables are generated without hand-typed numbers. Table~\ref{tab:manifest} lists the chain script $\to$ data $\to$ figure or table for the main text and the appendices. The scripts share the reference lattice Boltzmann solver, the Carleman lift and the circuit modules of the project (lbm\_ref, carleman\_lift, mtr\_clb\_circuit, coupled\_block\_circuit); the circuit checks use qiskit 2.5 and qiskit-aer with the exact statevector method.
\begin{table*}[t]\centering\scriptsize\setlength{\tabcolsep}{3pt}
\caption{Reproducibility manifest: scripts, data files and the figures and tables they produce. Paths are relative to the project's code/prxq, data/prxq and figures/prxq directories unless stated otherwise.}\label{tab:manifest}
\raggedcells\begin{tabular}{p{34mm}p{46mm}p{92mm}}\toprule
script & data & figure or table\\\midrule
A1\_coupled\_block.py & A1\_coupled\_block.json & Fig.~3 (G1\_figures.py figB), the norm values of Sec.~III; Fig.~1(b,c) is drawn by G1 fig1\_overview directly from the collision matrices\\
A2\_verify\_coupled.py, coupled\_block\_circuit.py & A2\_circuit\_check.json & Fig.~5(a) and Table~\ref{tab:circuit} collision rows (G1 fig5\_protocol, G2\_tables.py tab\_circuit)\\
A3\_joint\_success.py & A3\_joint\_success.json & Fig.~4(b) optimum curves; the near-rest model of Eqs.~(21)--(23)\\
A4\_matched\_joint.py & A4\_matched\_joint.json & Fig.~4(a) matched joint-success curves (G1 fig4\_run\_success)\\
B1\_multimode.py, B1\_campaign.sh & B1/*.json (17 runs) & Fig.~7(a,b), Table~\ref{tab:parameters}, the norm-balance statements of Sec.~III\\
B2\_level3.py & B2\_level3\_L4.json, B2\_level3\_L6.json & Fig.~7(c), level-3 statements of Sec.~V\\
B3\_summary.py & B3\_summary.json & Table~\ref{tab:window} (G2 tab\_window), Sec.~V window statements\\
B4\_error\_recurrence.py & B4\_error\_recurrence.json & Appendix~E recurrence numbers, Table~\ref{tab:recurrence} (G2 tab\_recurrence)\\
O2\_general\_gap.py & O2\_general\_gap.json & the general spectral-gap remark of Appendix~C (random contractive $D$, non-orthogonal $C$)\\
C1\_observable.py (linear) & C1\_observable\_linear.json & shifted norms, overlaps and trajectory values of Fig.~4(c), Fig.~6(c), Table~II of the main text; C1\_observable.json (quadratic start) for the comparison of the two starts in Sec.~II\\
C2\_observable\_physics.py & C2\_observable\_physics.json & Fig.~6(a,b), peaks and errors of the observables in Secs.~V and VI\\
D1\_prepare.py & D1\_prepare.json & preparation rows of Table~\ref{tab:circuit}, $p_1$ and $P_{\rm prep}$\\
D2\_coherent.py & D2\_coherent\_T1.json, D2\_coherent\_T2.json & Fig.~5(b), protocol and readout rows of Table~\ref{tab:circuit}\\
D3\_resources.py & D3\_resources.json & Table~II of the main text (G2 tab\_resources), Table~\ref{tab:resources_full} (tab\_resources\_full), the ledger Table~V of Appendix~F (tab\_ledger), Fig.~8 (G1 fig8\_resources), Secs.~VI and VII numbers\\
D5\_readout\_identity.py & D5\_readout\_identity.json & norm-free readout identity of Eq.~(25) of the main text: check against the D2 records and the 32x32 cases, end-to-end run with the amplified preparation as circuits (last rows of Table~\ref{tab:circuit})\\
D4\_prep\_amplification.py & D4\_prep\_amplification.json & amplification row of Table~\ref{tab:circuit}, Fig.~5(b) amplified weight, the schedule $k = 12$ of Table~\ref{tab:resources_full}\\
E1\_fourier\_baseline.py (linear, quad) & E1\_fourier\_baseline.json, E1\_fourier\_baseline\_quad.json & Fourier support, observable check and operation counts of the classical baselines of Appendix~F and Fig.~8(c); the quadratic-start record\\
F1\_lattices.py & F1\_lattices.json & D3Q19 and D3Q27 checks of Corollary~1 and Appendix~C\\
G3\_success\_ladder.py (from fig3\_success.py of the preliminary version) & data/C2\_success.json & Fig.~2\\
G4\_cost\_decomposition.py & data/D2\_gate\_counts.json & Fig.~11\\
G5\_validation\_figures.py & data/calibration\_ss2024.json, data/carleman\_unit\_tests.json, data/taylor\_green\_viscosity.json & Figs.~12 and 13\\
closure\_symbolic.py, closure\_refinement\_check.py (recorded in the data files) & data/B1\_closure.json, data/B1\_refinement.json & Table~III of the main text and the class counts of Appendix~A\\
fig2\_closure.py (preliminary version) & data/B3\_runs.json & Fig.~9 (arxiv\_fig2\_closure)\\
fig1\_sparsity.py (preliminary version), C1\_structure.py & computed directly; data/C1\_structure.json & Fig.~10 (arxiv\_fig1\_sparsity), Table~IV of the main text\\
\bottomrule\end{tabular}
\end{table*}

\end{document}